\documentclass[a4paper,fleqn,usenatbib]{mnras}

\usepackage{newtxtext,newtxmath}

\usepackage[T1]{fontenc}
\usepackage{ae,aecompl}

\usepackage{graphicx}	
\usepackage{natbib}
\usepackage{soul}
\bibpunct{(}{)}{;}{a}{}{,}
\usepackage{graphicx}
\usepackage{multirow}

\usepackage{txfonts}
\usepackage{natbib}
\usepackage{xspace}
\usepackage{dcolumn}
\usepackage{longtable}
\usepackage{lscape}
\usepackage{soul}        

\usepackage{rotating}
\usepackage{afterpage}

\newcommand{\MJup}{M$_{\mathrm{Jup}}$\xspace}

\newcommand{\mic}{$\mu$m\xspace}

\title[The effect of stellar multiplicity on discs.]{The Ophiuchus DIsc Survey Employing ALMA (ODISEA). II. The effect of stellar multiplicity on disc properties\thanks{Based on ESO observations (programs number 099.C-0465, 0103.C-0466, 089.D-0199, 075.C-0042, 60.A-9800,079.C-0307), Keck observations (program number GN-2017A-Q-29), ALMA observations (program number 2016.1.00545.S).}}

\author[A. Zurlo et al.]{Alice Zurlo$^{1,2}$\thanks{E-mail: alice.zurlo@mail.udp.cl}, Lucas A. Cieza$^{1}$, Sebasti\'an P\'erez$^{3}$, Valentin Christiaens$^{4}$, \newauthor Jonathan P. Williams$^{5}$, Greta Guidi$^{6}$, Hector C\'anovas$^{7}$, Simon Casassus$^{8}$, \newauthor Antonio Hales$^{9}$, David A. Principe$^{10}$, Dary Ru\'iz-Rodr\'iguez$^{11}$, Antonia Fernandez-Figueroa$^{8}$ \\
$^{1}$N\'ucleo de Astronom\'ia, Facultad de Ingenier\'ia, Universidad Diego Portales, Av. Ejercito 441, Santiago, Chile\\
  $^{2}$Escuela de Ingenier\'ia Industrial, Facultad de Ingenier\'ia y Ciencias, Universidad Diego Portales, Av. Ejercito 441, Santiago, Chile \\
   $^{3}$Universidad de Santiago de Chile, Av. Libertador Bernardo O'Higgins 3363, Estaci\'on Central, Santiago, Chile \\
  $^{4}$School of Physics and Astronomy, Monash University, VIC 3800, Australia \\
  $^{5}$Institute for Astronomy, University of Hawaii at Manoa, Honolulu, HI, 96822, USA\\
  $^{6}$ETH Z\"urich, Institute for Particle Physics and Astrophysics, Wolfgang-Pauli-Str. 27, 8093 Z\"urich, Switzerland \\
$^{7}$Aurora Technology B.V. for ESA, ESA-ESAC, Camino Bajo del Castillo s/n, 28692 Villanueva de la Ca\~nada, Madrid, Spain \\
$^{8}$Universidad de Chile, Camino el Observatorio 1515, Santiago, Chile \\
  $^{9}$Atacama Large Millimeter/Submillimeter Array, Joint ALMA Observatory, Alonso de C\'ordova 3107, Vitacura 763-0355, Santiago, Chile\\
$^{10}$Kavli Institute for Astrophysics and Space Research, Cambridge, MA, USA \\
  $^{11}$Chester F. Carlson Center for Imaging Science, School of Physics \& Astronomy, and Laboratory for Multiwavelength Astrophysics,\\
  Rochester Institute of Technology, 54 Lomb Memorial Drive, Rochester NY 14623 USA}

\date{Accepted 2020 June 25. Received 2020 June 2; in original form 2020 March 26 }

\pubyear{2020}

\begin{document}
\label{firstpage}
\pagerange{\pageref{firstpage}--\pageref{lastpage}}
\maketitle

\begin{abstract}
  We present Adaptive Optics (AO) near infrared (NIR) observations using VLT/NACO and Keck/NIRC2 of ODISEA targets. ODISEA is an ALMA survey of the entire population of circumstellar discs in the Ophiuchus molecular cloud.  From the whole sample of ODISEA we select all the discs that are not already observed in the NIR with AO and that are observable with NACO or NIRC2.  The NIR-ODISEA survey consists of 147 stars observed in NIR AO imaging for the first time, as well as revisiting almost all the binary systems of Ophiuchus present in the literature (20 out of 21). In total, we detect 20 new binary systems and one triple system. For each of them we calculate the projected separation and position angle of the companion, as well as their NIR and millimeter flux ratios. From the NIR contrast we derived the masses of the secondaries, finding that 9 of them are in the sub-stellar regime (30-50 \MJup). Discs in multiple systems reach a maximum total dust mass of $\sim$ 50 M$_{\oplus}$\xspace, while discs in single stars can reach a dust mass of 200 M$_{\oplus}$\xspace. Discs with masses above 10 M$_{\oplus}$\xspace are found only around binaries with projected separations larger than $\sim$ 110 au. The maximum disc size is also larger around single star than binaries.  However, since most discs in Ophiuchus are very small and low-mass, the effect of visual binaries is relatively weak in the general disc population. 
  
\end{abstract}




\section{Introduction}

 The very high incidence of extrasolar planets \citep{2013Sci...340..572H, 2015ApJ...809....8B} suggests that most of the circumstellar discs we see in star-forming regions should form planetary systems.  The occurrence of exoplanets is particularly high for low-mass planets around M-dwarfs, where the statistics are robust. \citet{2016MNRAS.457.2877G} estimate that M-type main-sequence stars host an average of 2.2 $\pm$ 0.3 planets with radii of 1-4 R$_{\oplus}$. The incidence of extrasolar planets at larger distances from their hosts is more poorly constrained, but microlensing studies also indicate that most stars in the Milky Way might harbor ice or gas giants at 5-10 au separations \citep{2012Natur.481..167C}.  \newpage

In this context, studying full populations of protoplanetary discs can help linking disc properties to the properties of the planets detected so far.  ALMA's unprecedented sensitivity provides the opportunity to study nearly complete and unbiased samples of discs at sub-arcsecond resolution in the (sub)mm regime \citep[e.g.,][]{2016ApJ...828...46A, 2016ApJ...831..125P, 2018MNRAS.478.3674R}. Circumstellar discs may become optically thin at these wavelengths and therefore, resolved disc images inform on the spatial distribution of mass. 

We have recently conducted a survey of the entire population of circumstellar discs ($\sim$290) identified by \emph{Spitzer} in the Ophiuchus molecular cloud to study both their gas and dust components: the Ophiuchus Disk Survey Employing ALMA \citep[ODISEA; Paper I,][]{2019MNRAS.482..698C}.  At a distance of 140$\pm$10 pc \citep{2017ApJ...834..141O,2019A&A...626A..80C}, Ophiuchus is the closest of the major star-forming regions in the solar neighborhood, and ODISEA is the largest ALMA survey of its kind thus far. 

One of the main objectives of ODISEA is to study the dependence of disc properties on stellar multiplicity. While most stars form in multiple systems \citep[see, e.g.,][]{2013ARA&A..51..269D}, the effects that (sub)stellar companions have on planet formation are not well understood. Observational studies show that discs are less frequent \citep{2009ApJ...696L..84C,2012ApJ...745...19K} and less massive \citep{2005ApJ...619L.175A,2017ApJ...851...83C} in close binary systems (separations $<100$~au) and that much wider binaries have little effects on discs. 
From extra-solar planet studies, it is also clear that planets can form around very tight ($<1$~au) stellar binary systems \citep{2011Sci...333.1602D} as well as around individual stars in wide projected separation ($>100$~au) binaries \citep{2014A&A...569A.120L}.
Medium-separation binaries (e.g., $\sim$10-50~au), like those resolvable with the near-IR (NIR) Adaptive Optics (AO) observations presented here, might result in compact discs which might not have enough mass to form giant planets and/or might not survive long enough to form rocky planets \citep[e.g.,][]{2009ApJ...696L..84C}. However, these hypotheses still need to be observationally verified. 

Here we present observations and analysis of the NIR-AO follow-up of 164 objects of the ODISEA survey conducted with the NACO instrument at the VLT and the NIRC2 instrument at Keck. We supplement these observations with multiplicity information from the literature and public archives. The sample selection of this NIR survey is presented in Sec.~\ref{s:sample}. The observations and data reduction of the NIR and mm data are described in Sec.~\ref{s:obs}. Our results are presented and discussed in Sec.~\ref{s:res}. We close the article with a summary of our conclusions in Sec.~\ref{s:con}.

\section{The ODISEA Sample and literature data}
\label{s:sample}

The population of circumstellar discs in the Ophiuchus molecular cloud is composed of 289 objects. The ODISEA ALMA survey has been divided into two samples, A and B. Sample A includes 147 objects of Class I, Flat spectrum, and bright (K $\le$ 10 mag) Class II sources. Sample B includes a total of 142 sources composed of the fainter Class II objects, and all Class III objects. Many of these targets were previously observed in the near infrared, providing the information of the multiplicity.

\citet{2009ApJ...696L..84C} collected multiplicity information for 349 stars in nearby star-forming regions, including 73 objects that are part of the ODISEA sample. In \citet{2013ApJ...762..100C} all 34 objects presented are in common with the sample presented here. In \citet{2015ApJ...813...83C} 50 of 114 stars presented are in common with this sample. \citet{2017ApJ...851...83C} presented a 870 \mic survey of 49 objects in $\rho$ Ophiuchus, all of them but one are in common with our sample. \citet{2018AJ....155..109S} recently presented orbital motion of 8 binaries in common with our survey. As some of the objects are in common in between these works, we found a total number of 109 (of which 21 are binaries) stars in common with our survey. 

The ESO and Keck archives were searched for the presence of unpublished NIR public data useful for determining multiplicity. We found three objects in this sample observed with the NACO and SOFI instruments on the VLT and seven observed with the NIRC and NIRC2 instruments on Keck.

Excluding the objects presented in the literature, or available in the archives, we were left with 178 objects. For these objects we planned VLT/NACO and Keck/NIRC2 observations.

The NIR wave front sensor (WFS) on NACO has a K-band magnitude limit where sources need to be brighter than K=12.  70 stars in our sample met this requirement and were observed with NACO.  All the other stars were included in the NIRC2 sample. For most of them we required the use of the laser guide star (LGS) using the target itself as natural guide star (NGS) for the laser tip-tilt correction, or a close-by bright star. The stars observable with NIRC2 were 77 in total. Unfortunately, the 29 remaining objects are too faint and do no have any close NGS for the tip-tilt correction.  Therefore, they are not observable in NIR with the current AO-imaging instruments from the ground. {These objects were excluded from the sample in the statistical analysis, the complete list of them is available in Table~\ref{t:non_o}.}

\section{Observations and data reduction}
\label{s:obs}

NIR observations exploiting VLT/NACO and Keck/NIRC2 were performed to study multiplicity in Ophiuchus. These two instruments were chosen for their efficiency in taking short exposures with small overheads. The observing strategy was the same for both instruments: very short L$^{\prime}$ filter integrations with jittering. For each star we took two or three integrations where the star fell in different quadrants of the detector. This is because the sky background, which is bright at these wavelengths, is recorded quasi-simultaneously.

\subsection{VLT/NACO}
\label{s:obsnaco}

The observations with NACO were carried out from April 8 to April 11, 2017 (4 half nights), program 099.C-0465 (PI: Zurlo). The atmospheric conditions were in general favorable, especially during the last night, with stable seeing between 0\farcs5 to 0\farcs7, 5 ms coherence time and constant wind. Additional data were taken during the nights 22 to 24 August 2019 of program 0103.C-0466 (PI: Zurlo) as backup targets. The conditions of those nights were also optimal.

We adopted the technique of "star-hopping" which permits switching from one target to another, if they are close enough, by applying an offset to the telescope (hence saving on acquisition overheads).  In practice, the AO loop is opened before the ``hop'' to the second star, and closed again with the same settings as the first star after the hop. To be able to apply this technique, the stars have to be separated by $<$ 900\arcsec and have a similar magnitude (1-2 mag of difference). We divided our sample of 70 stars into 10 groups of 7 stars each, with ranges of magnitudes with a difference of maximum 1 mag and close coordinates on the sky.

Given the faintness of our targets, we chose the AO configuration N90C10 which provides NAOS with 90\% of the stars light and the detector with the remaining 10\%.  In this way the AO was stable during the exposure. We chose the L$^{\prime}$ filter given the red colors of the stars of the sample. In this configuration the exposure time is very short, 0.16-0.17 s, to avoid saturation of the background. For each star we took two exposures (of 200 NDIT\footnote{Number of frames per dithering position} each), where the star fell in two different quadrants of the detector to permit a quasi-simultaneous subtraction of the background. The pixel scale of the NACO detector is 0.02712 arcsec/pixel. In total we observed 88 stars.

The data were reduced with the python Vortex Image Processing ({\tt VIP}) package \citep{2017AJ....154....7G}. The raw data, two exposures per star, were background subtracted one from the other, flat-fielded, then recentered to have the target in the center of the detector. A bad-pixel removal has been applied. Finally, in order to reduce the effects of atmospheric turbulences on the images, a high-pass spatial filter has been used. Each final image is the mean of the two recentered frames.

To calibrate the astrometric position of the central star we adopted the coordinates given by the ALMA centroid of the primary star, which are more precise than the coordinates given in the header. The primary was assumed to be the disc with highest flux. The scale and true north calibration has been applied as listed in the NACO user manual.  

The reduced NACO images of the stars with companions are shown in Figures~\ref{f:mos}, \ref{f:mos2}, \ref{f:mos3}, and \ref{f:mos4}. The median value of the NACO contrast curves is shown in Fig.~\ref{f:contr}, in general companions with a flux 50 times fainter than the primary are detected. Ten new binaries have been found. The properties of the systems are listed in Table~\ref{t:bin}. Note that ISO-Oph204 (ODISEA ID: C4\_134) seems to be a triple system, but it is very likely that the companion appears double because of a shift of the detector during the acquisition. Since the ghost is present in both the images taken, we can speculate that the primary is in fact a very close binary (it also appears elongated in the image) and the AO is tilting in between the two tight stars, causing the small shift of the detector. With the data available we cannot conclude on the nature of this system. This system was already identified as a wide binary by \citet{2009ApJ...696L..84C}.  

\subsection{Keck/NIRC2}

Observations with Keck/NIRC2 were carried out during the nights from the 8 to the 10 June 2017 (3 half nights), program GN-2017A-Q-29 (PI: Williams). The atmospheric conditions at Mauna Kea were exquisite: seeing from 0\farcs3 to 0\farcs4. During the first night we observed all the targets bright enough to be used as NGS for the AO. During the two other nights we observed very faint stars using the LGS, we implied the target itself for tip-tilt correction when possible, otherwise a close-by bright star. In total 51 stars have been observed. Some of the targets foreseen were not visible in the detector. 

The same method exploited for NACO has been followed: for each star we took a short L$^{\prime}$ exposure, jittering the star to place it in three different detector quadrants at each time. The pixel scale of the NIRC2 detector is 0.009942 arcsec/pixel and the true North correction is listed in the Keck header. As for NACO, the astrometric position of the central star are the coordinates given by the ALMA centroid of the primary star.  

\begin{figure}
  \begin{center}

    \includegraphics[width=0.5\textwidth]{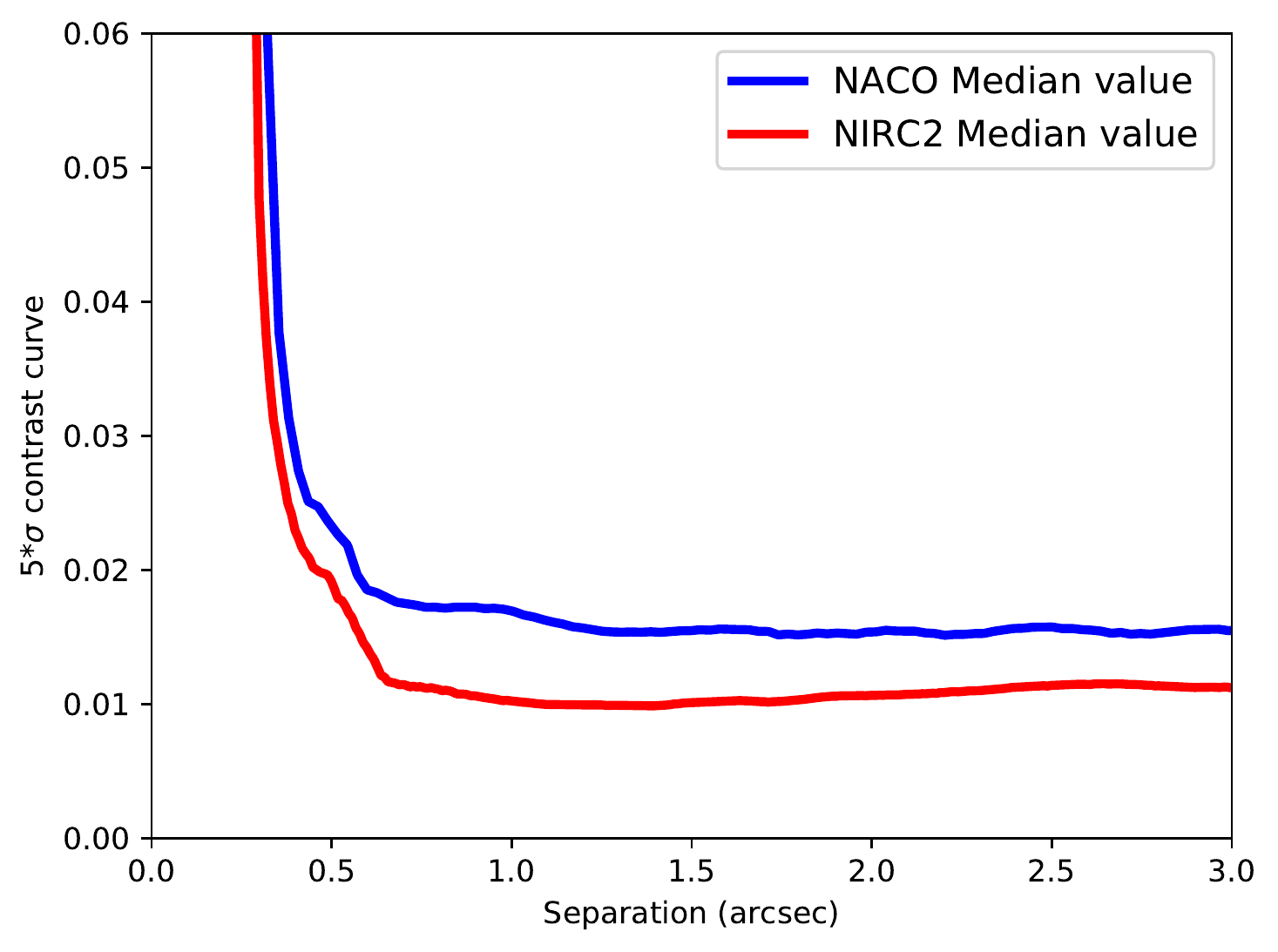}

\caption{Median value of the contrast curves obtained with the instruments NACO and NIRC2.}
\label{f:contr}
\end{center}
\end{figure}

The NIRC2 data have been reduced the same way as NACO data as described in Sec.~\ref{s:obsnaco}. The reduced NIRC2 images of the stars with companions are shown in Figures~\ref{f:mos}, \ref{f:mos2}, \ref{f:mos3}, and \ref{f:mos4}. The median value of the NIRC2 contrast curves is shown in Fig.~\ref{f:contr}, which is deeper than NACO, allowing the detection of companions 100 times fainter than the primary. Nine new binaries have been found, and we followed-up the already known binary system around V*V2131Oph (C4\_123). The properties of the systems are listed in Table~\ref{t:bin}.

\subsection{Archive data}
We found 13 objects of our sample in the ESO and Keck archives. Two objects have been observed with NACO: GY5, during the night 2005-05-28 (program 075.C-0042) and CRBR-L, during the night 2009-04-25 (program 60.A-9800). Both of them have been imaged with the jittering technique. The first target has been detected in K$_s$ filter, while the second one in L$^{\prime}$. Both appear as single stars in the detector. The images have been reduced with the same technique described in Sec.~\ref{s:obsnaco}.

{The object BKLTJ162538-242238, also known as Oph-2, which was presented as a binary system in \citet{2005A&A...437..611R}, has public NACO data in the archive under program 079.C-0307(A). We reduced and analyzed the data, with the jittering technique and K$_s$ filter, the final image is shown in Fig.~\ref{f:onlyalma} (left panel).  }

Another object, GY264, has been observed during the night 2012-04-16, with SOFI. The images, taken with the jittering method, have been median combined after recentering. No calibrations (dark, flat-field) have been found for the dataset. Nevertheless, the star is clearly visible at the center of the detector, and it appears isolated. Other stars appear in the FoV, with projected separations greater than 1.4 arcmin.

In the Keck archive, data of 7 objects of our ALMA survey are available from the instruments NIRC and NIRC2. All the available data have been taken in K$_s$ filter. For three stars sparse aperture masking (SAM) data are available. Four stars appear to be single stars. In conclusion, all the stars with archival data are single stars.

\subsection{ALMA 1.3 mm data}
For the ALMA data reduction and presentation we refer the reader to paper I \citep{2019MNRAS.482..698C} for Sample A, which consists of 147 targets observed under the Cycle-4 ALMA programme 2016.1.00545.S, with the antennas configuration C40-5. {For this sample the typical rms is of $\sim 0.15 $ mJy/beam with a synthesized beam of 0\farcs28$\times$0\farcs19. In this first observation block 120 sources are detected and 27 are non-detections.}  For Sample B we refer the reader to \cite{2019ApJ...875L...9W}, these 142 other objects were observed with a different antennas configuration (C40-3 array), under the same program. Among the 289 objects \cite{2019ApJ...875L...9W} identified 23 spurious objects that are not part of Ophiuchus. {For sample B, the sensitivity is $\sim 0.2 $ mJy/beam, the beam size is $\sim$ 0\farcs98$\times$0\farcs74. The number of detections and non-detections in this sample are 66 and 76, respectively. Note that the difference between the two ODISEA ALMA samples is the K-band magnitude, which is intrinsically correlated with the mass of the stars. The sample with lower resolution is composed of stars with magnitudes fainter than K-band = 10 and contain the lower mass stars, which explains the much lower mm detection rate given the dependance of disc masses on stellar mass \citep{2013ApJ...771..129A, 2016ApJ...831..125P}.}

\begin{figure*}
\begin{center}
  \includegraphics[height=0.3\textwidth]{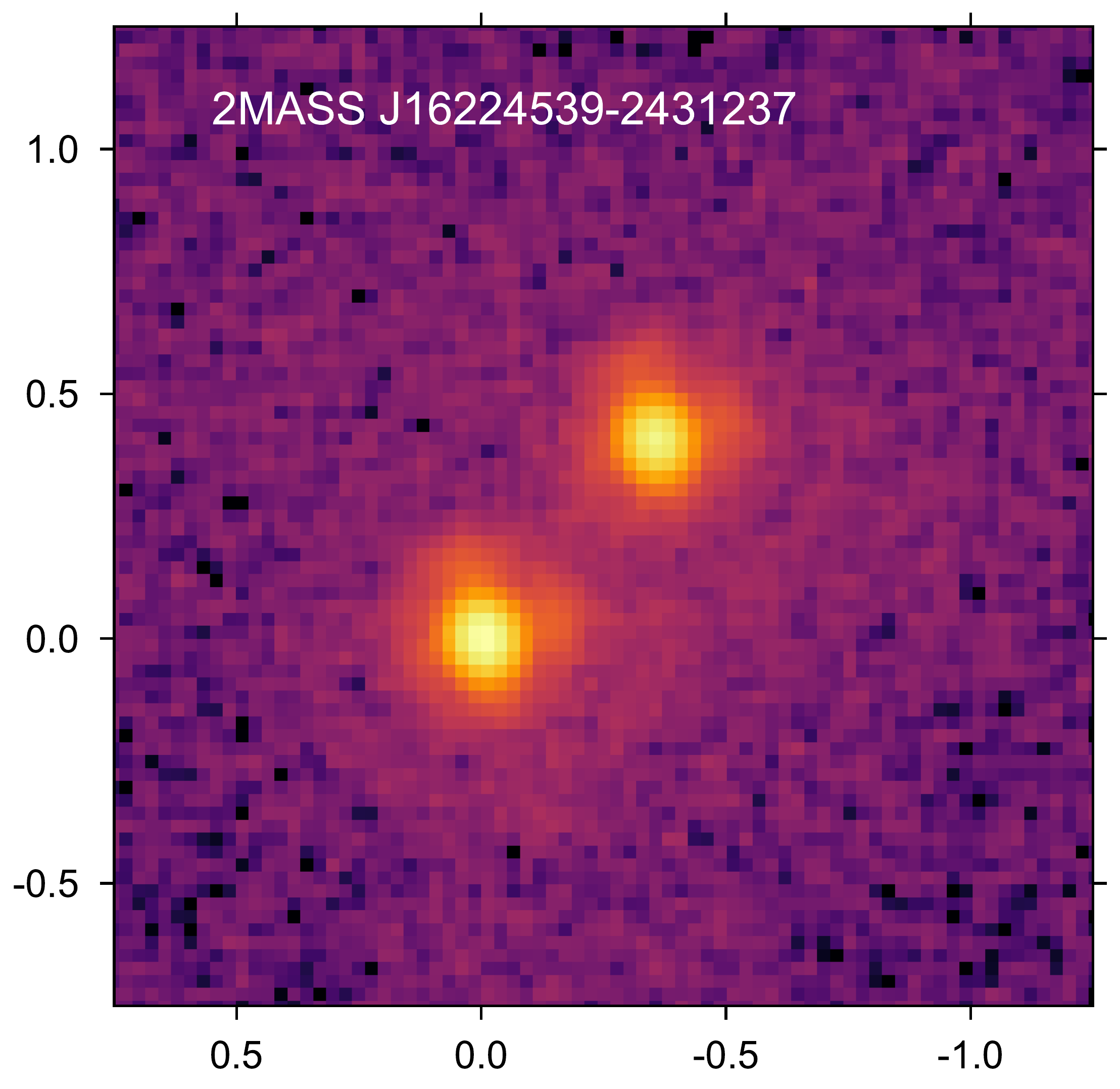}  \hfill
  \includegraphics[height=0.3\textwidth]{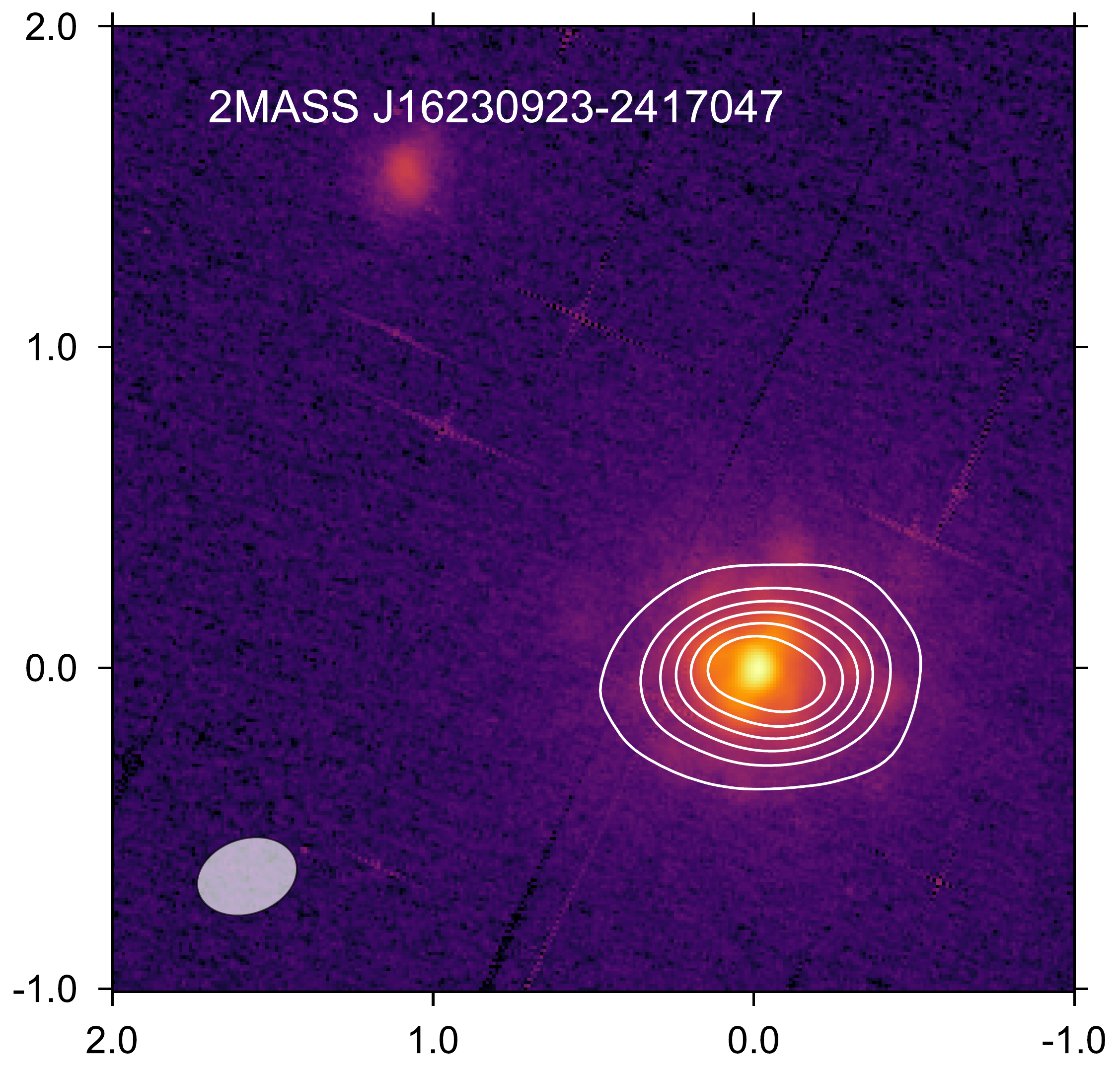}  \hfill
  \includegraphics[height=0.3\textwidth]{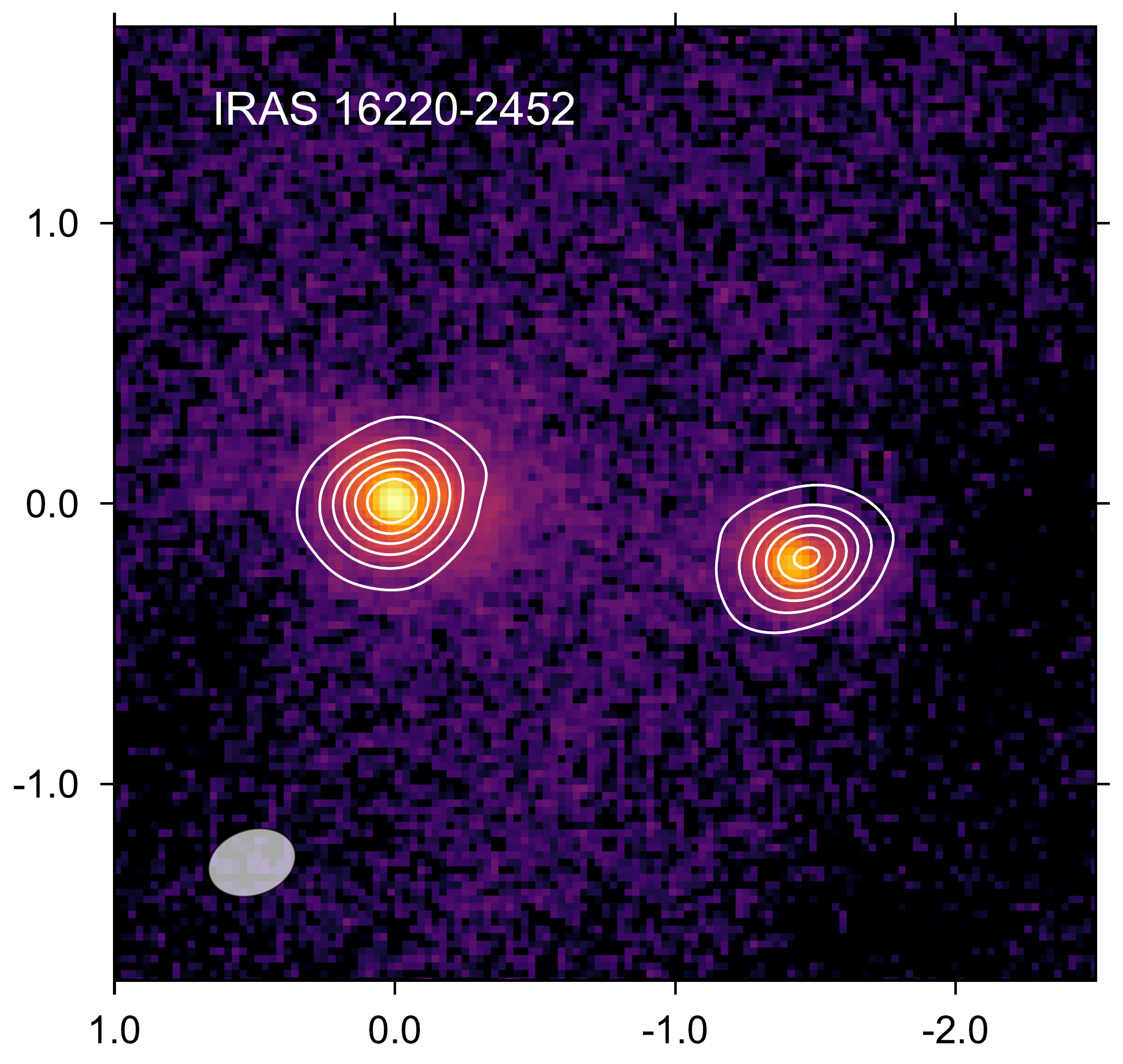}  \\    
  \includegraphics[height=0.3\textwidth]{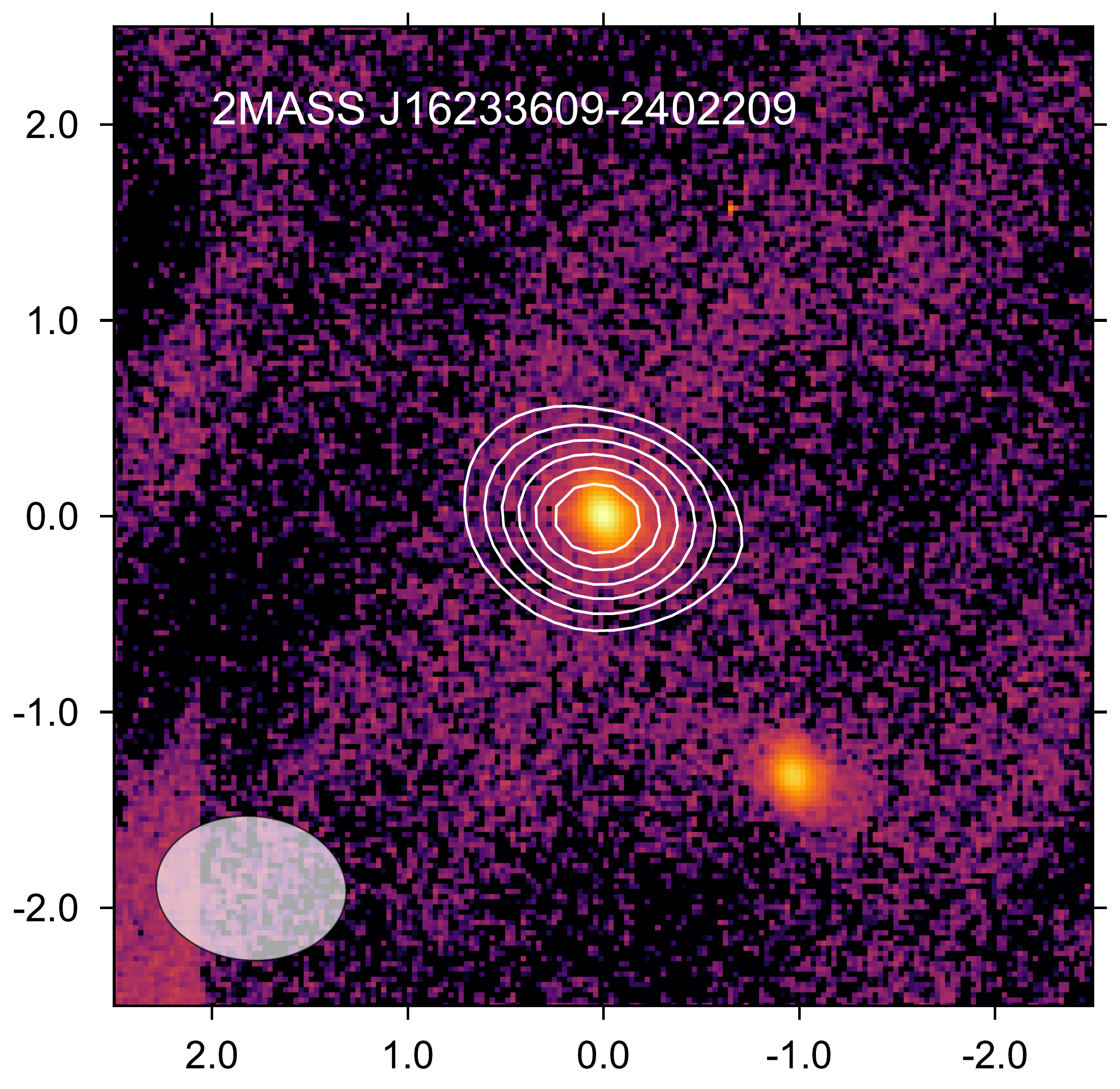}  \hfill
  \includegraphics[height=0.3\textwidth]{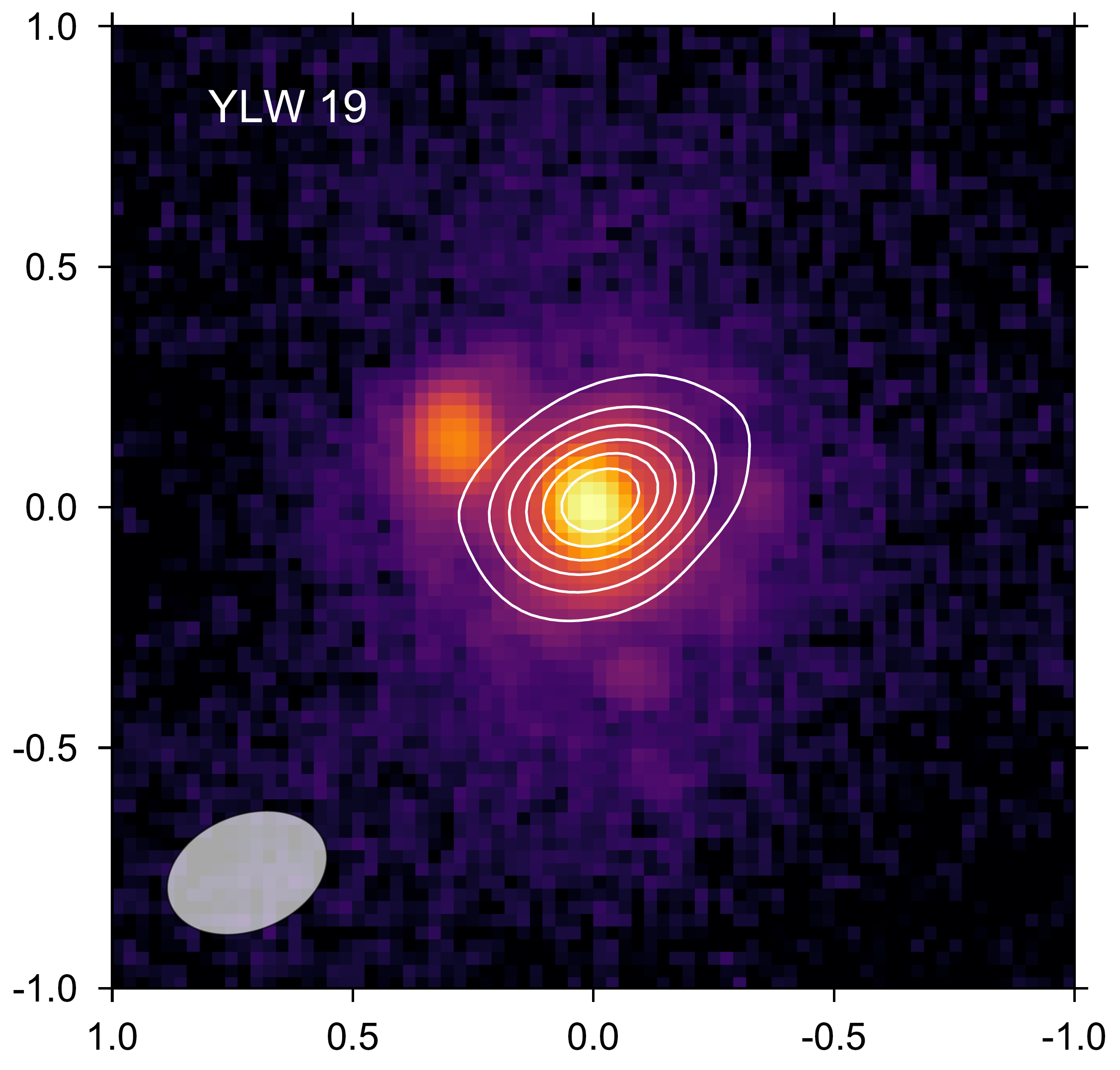}  \hfill
  \includegraphics[height=0.3\textwidth]{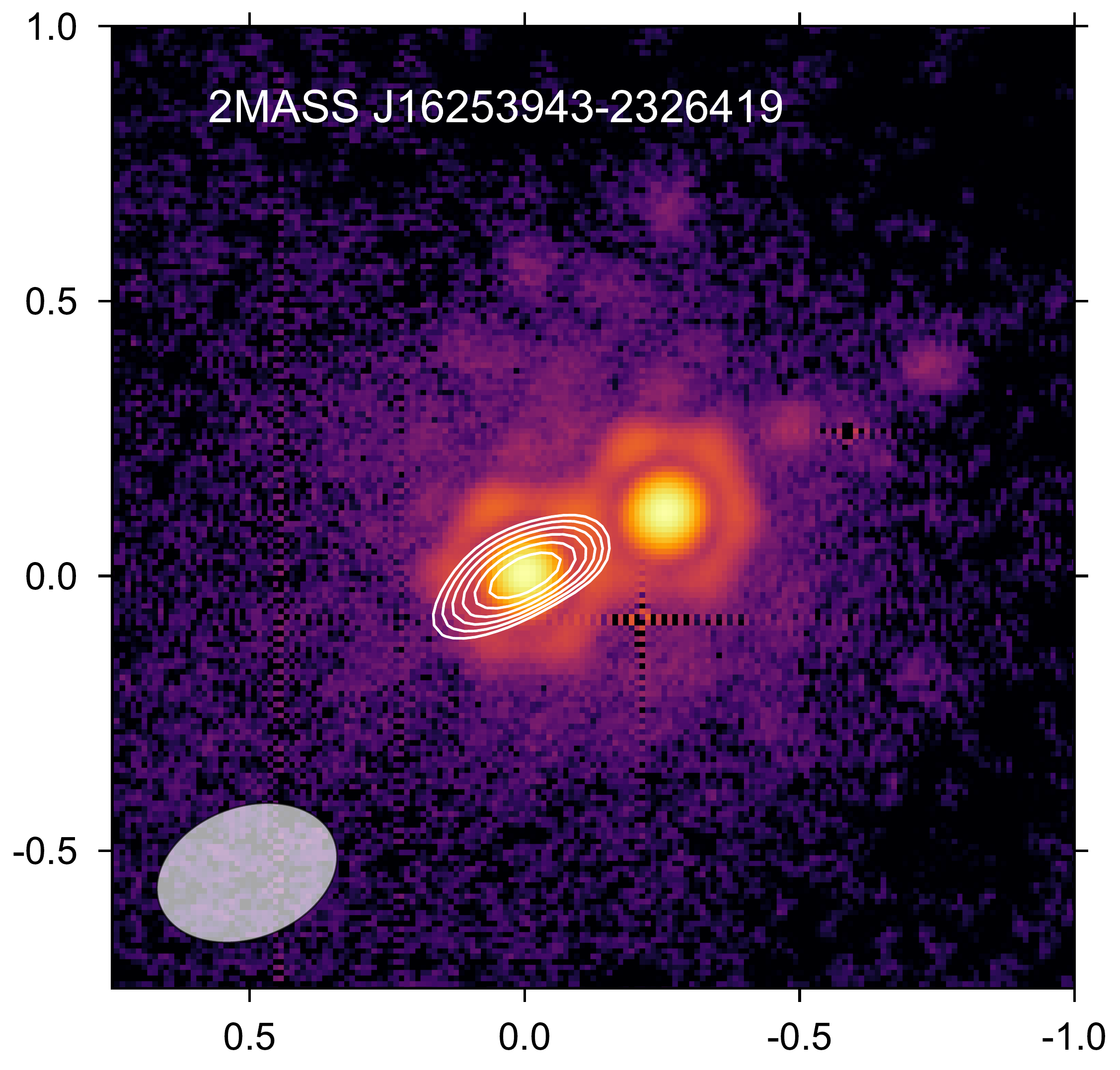}  \\    
  \includegraphics[height=0.3\textwidth]{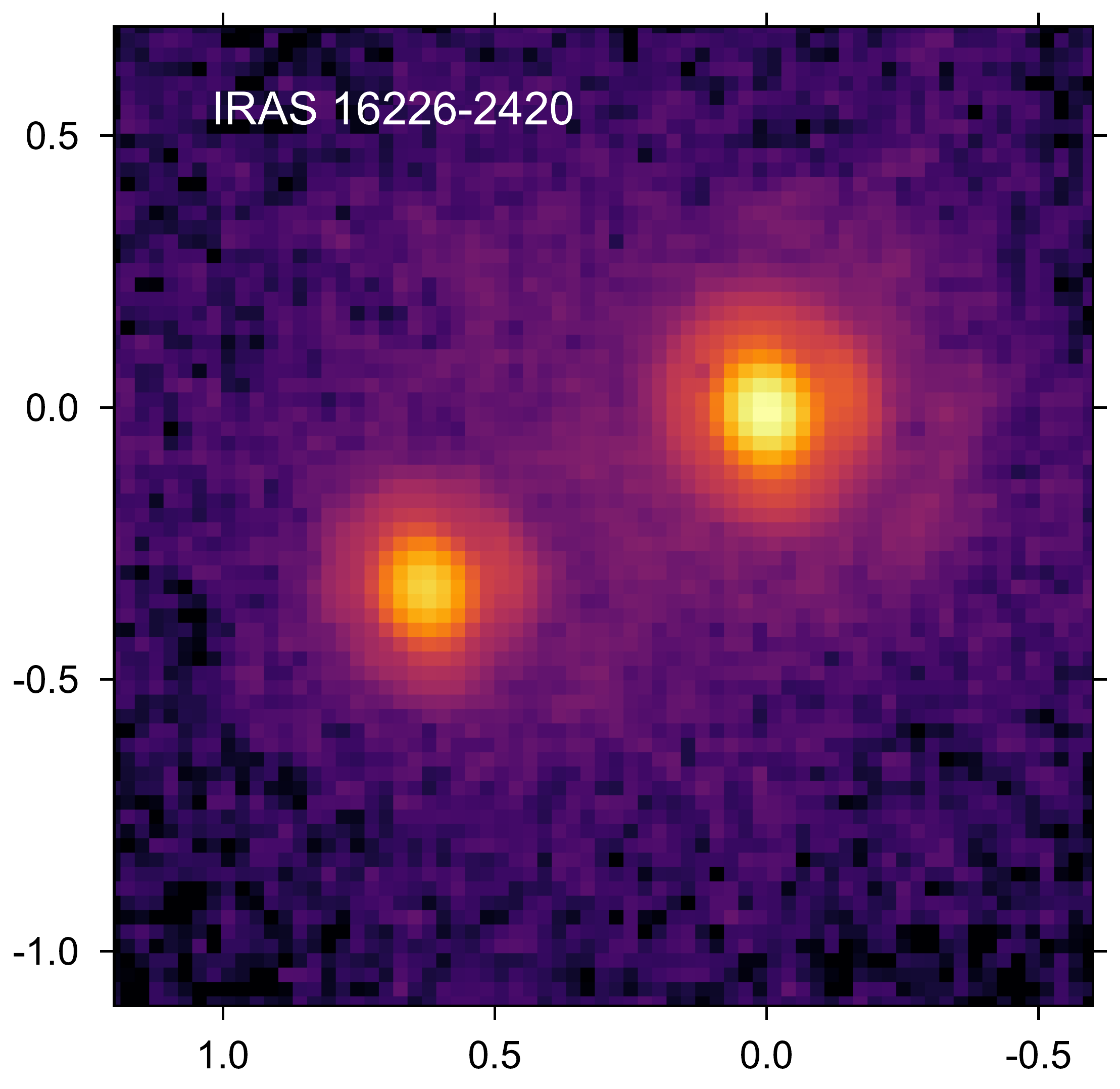}  \hfill
  \includegraphics[height=0.3\textwidth]{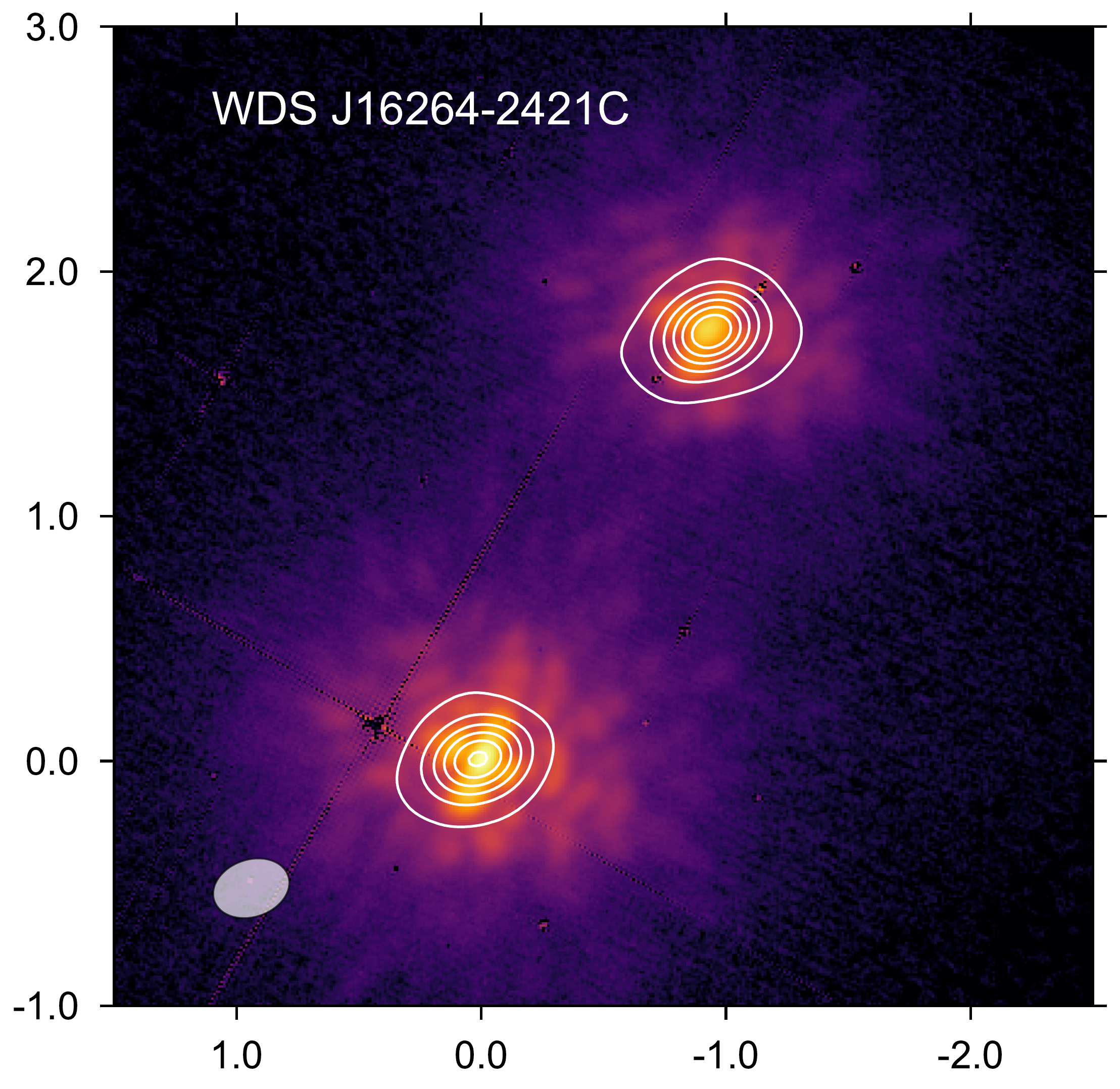}  \hfill
  \includegraphics[height=0.3\textwidth]{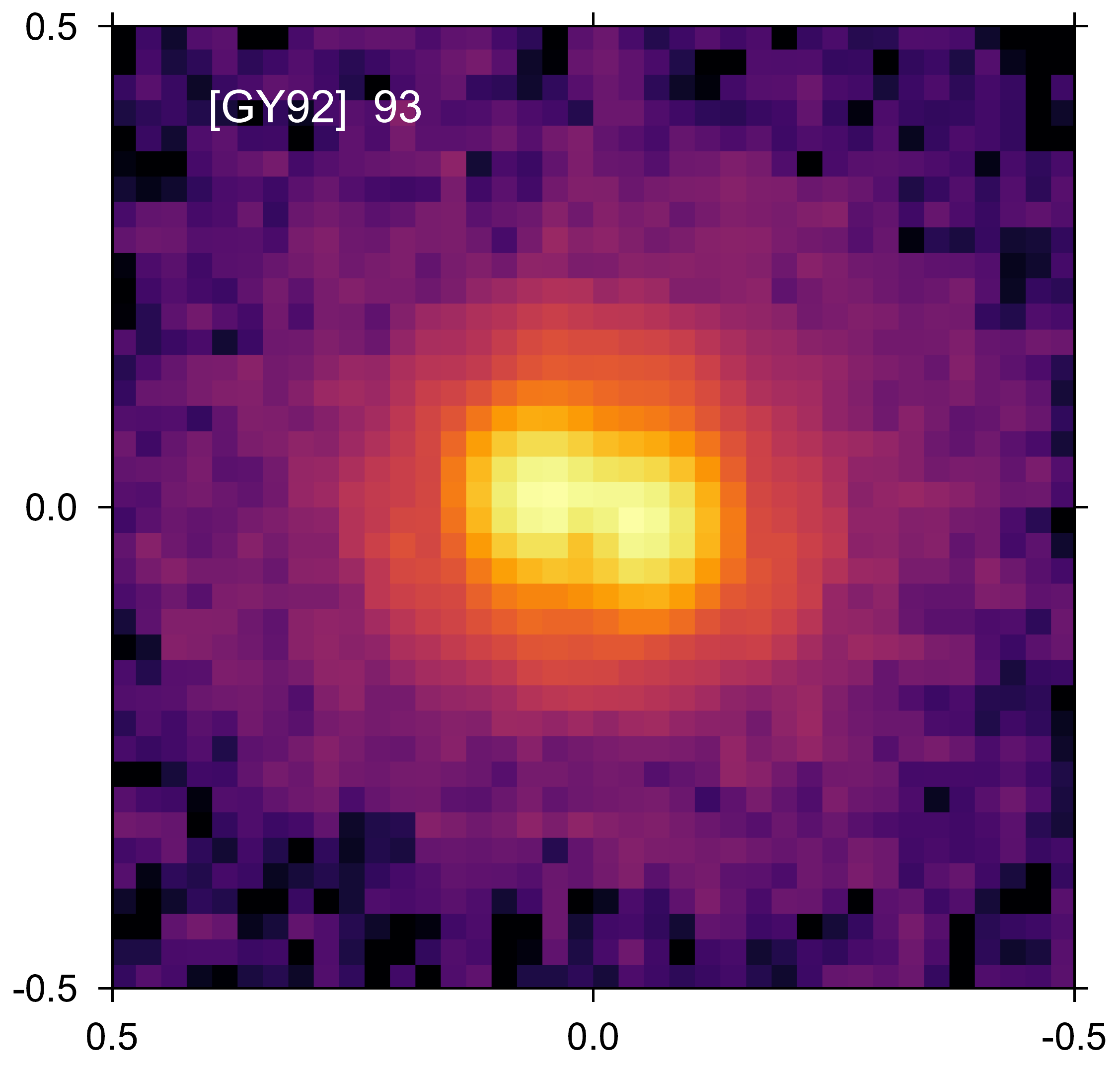}  \\    
  \includegraphics[height=0.3\textwidth]{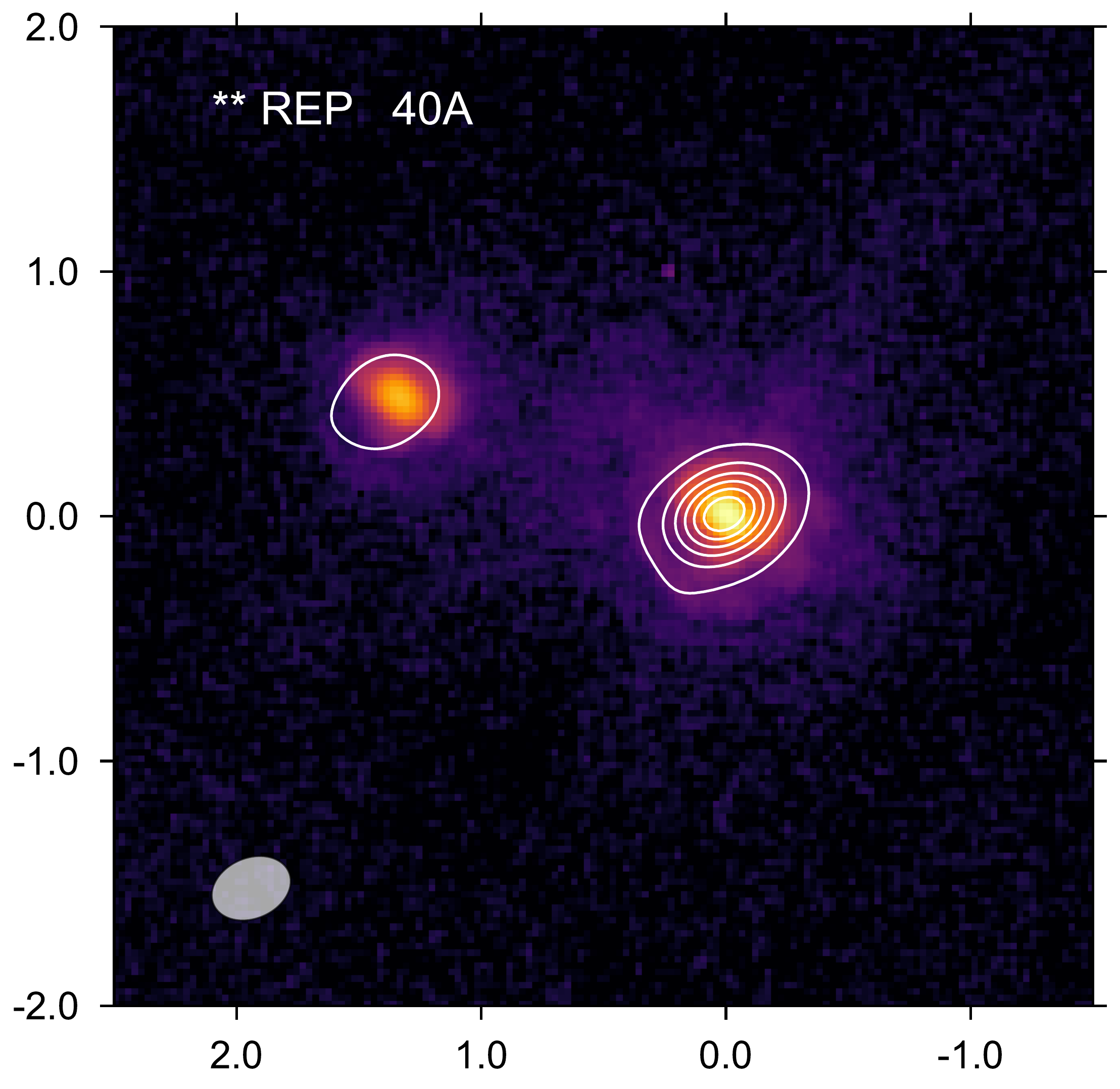}  \hfill
  \includegraphics[height=0.3\textwidth]{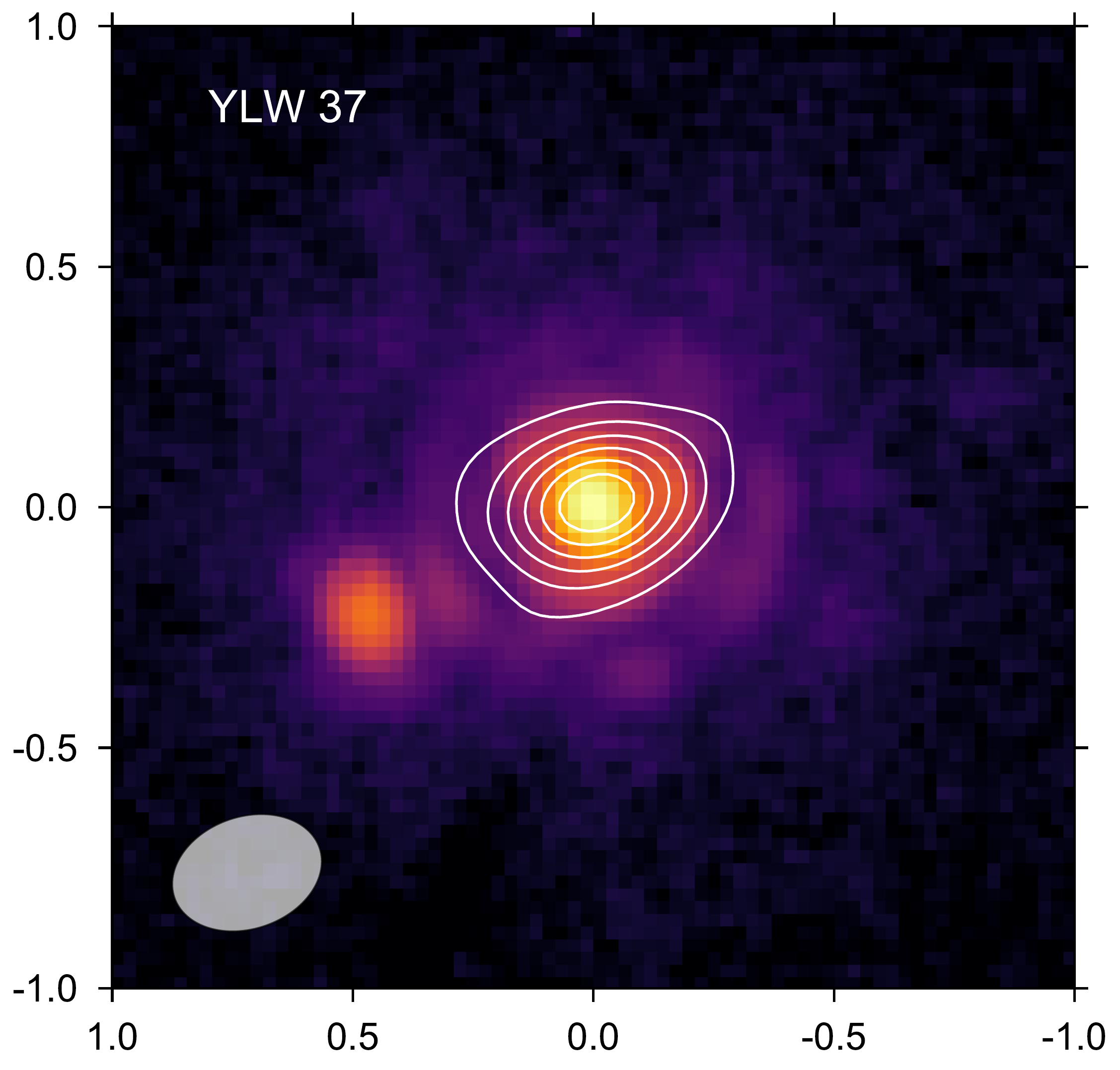}  \hfill
  \includegraphics[height=0.3\textwidth]{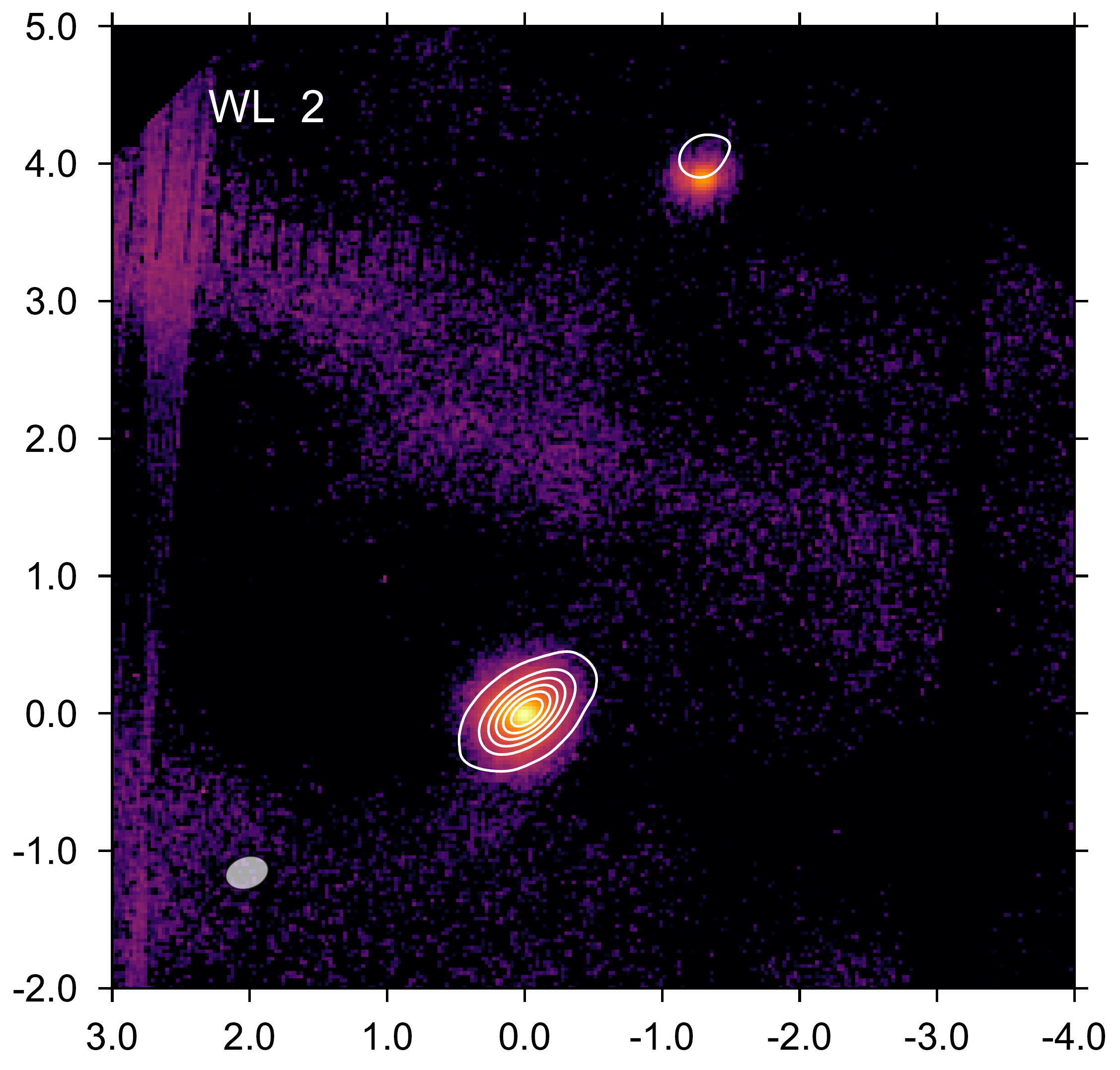}

  \caption{A gallery including all the {detected multiple systems} of the sample. For each object, the NIRC2 or NACO image is shown with a logarithmic colour stretch. The millimeter ALMA counterpart 1.3 mm emission is shown in white contours levels, ranging from 5 times the RMS noise (normally the RMS noise is 0.15-0.2 mJy in each map) to the peak emission, when detected. The ALMA synthesized beam is shown in the left bottom corner. North is up, East is left. }


\label{f:mos}
\end{center}
\end{figure*}

\begin{figure*}
  \begin{center}
  \includegraphics[height=0.3\textwidth]{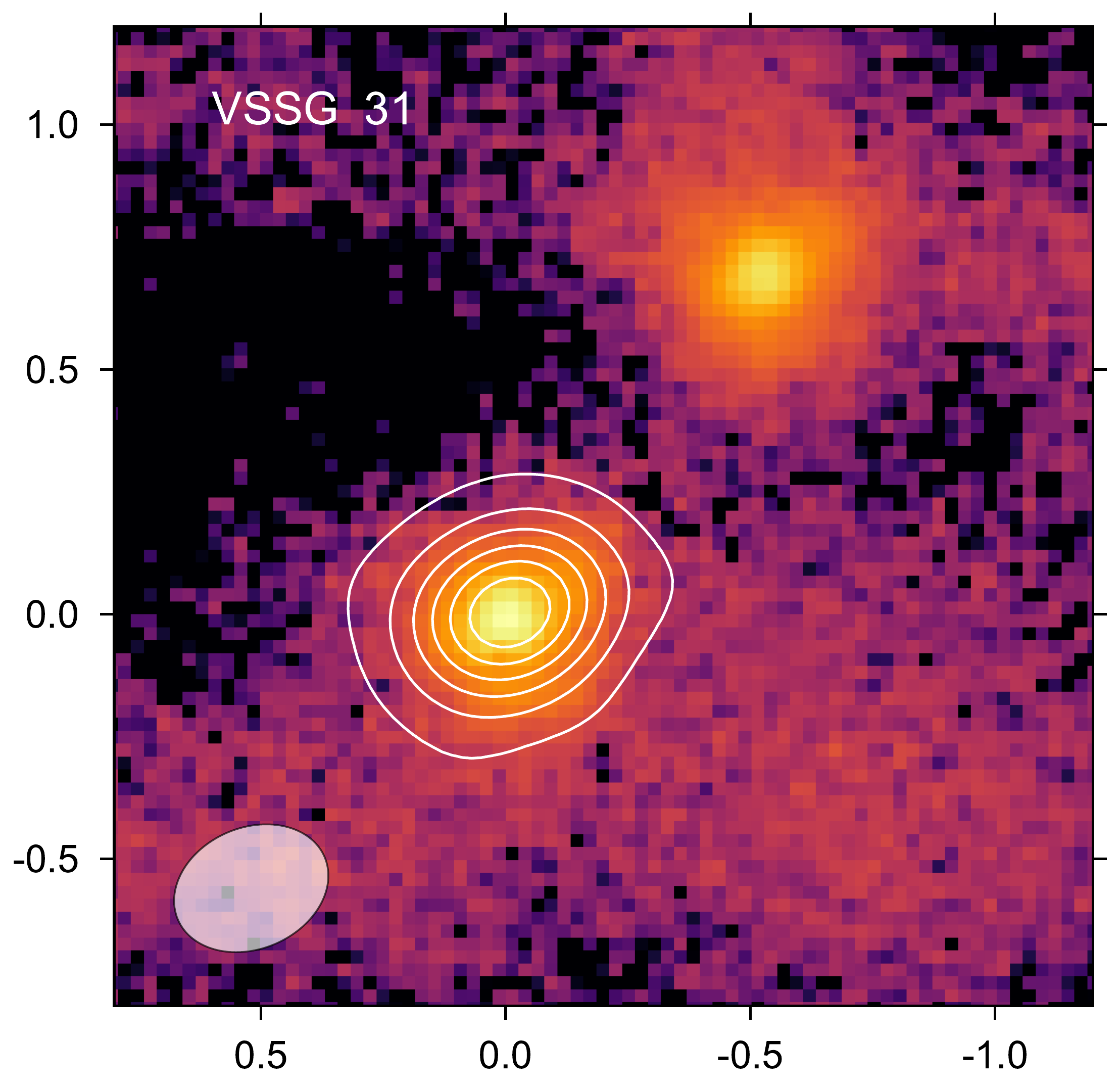}  \hfill
  \includegraphics[height=0.3\textwidth]{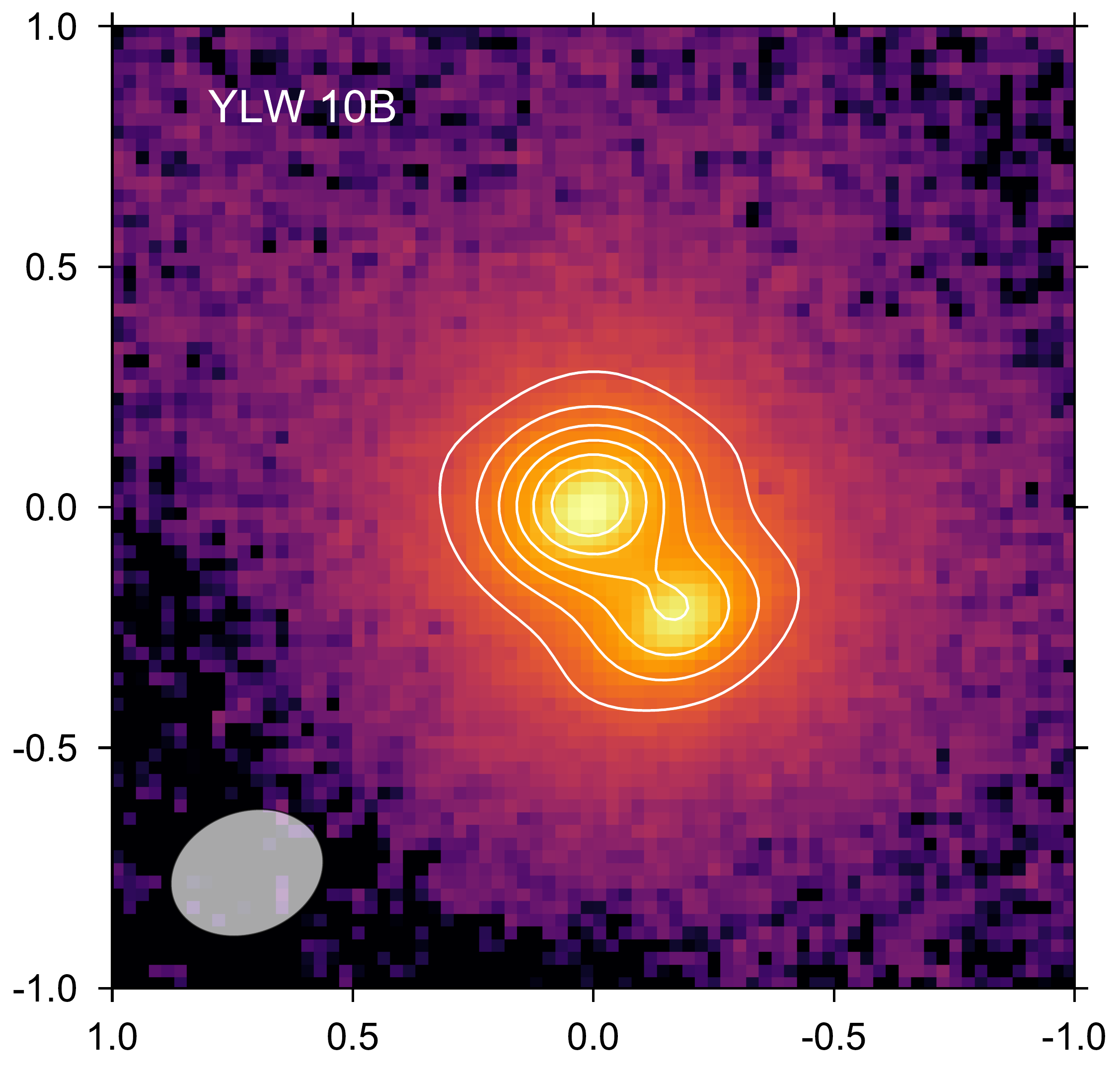}  \hfill
  \includegraphics[height=0.3\textwidth]{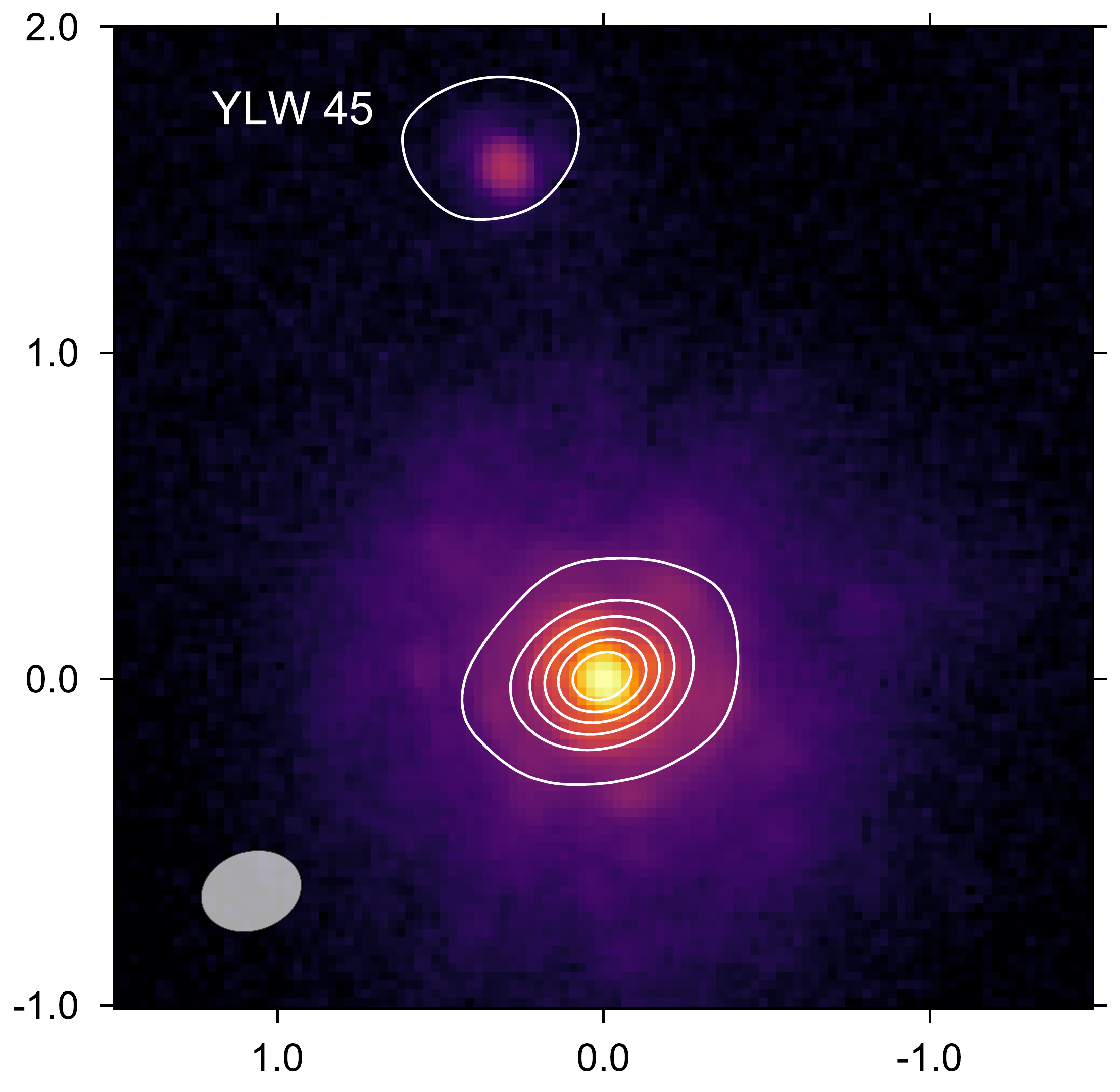}  \\    
  \includegraphics[height=0.3\textwidth]{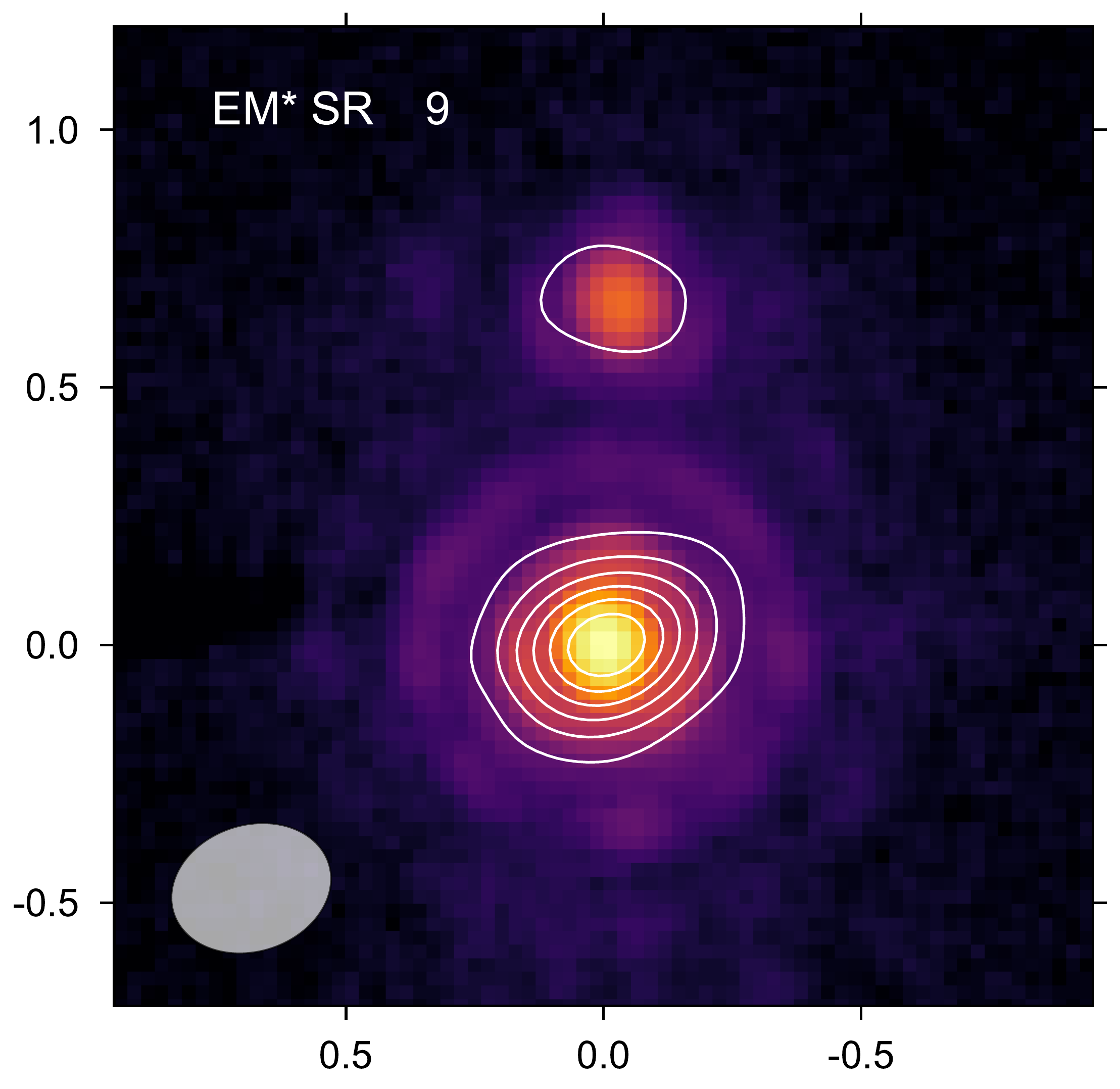}  \hfill
  \includegraphics[height=0.3\textwidth]{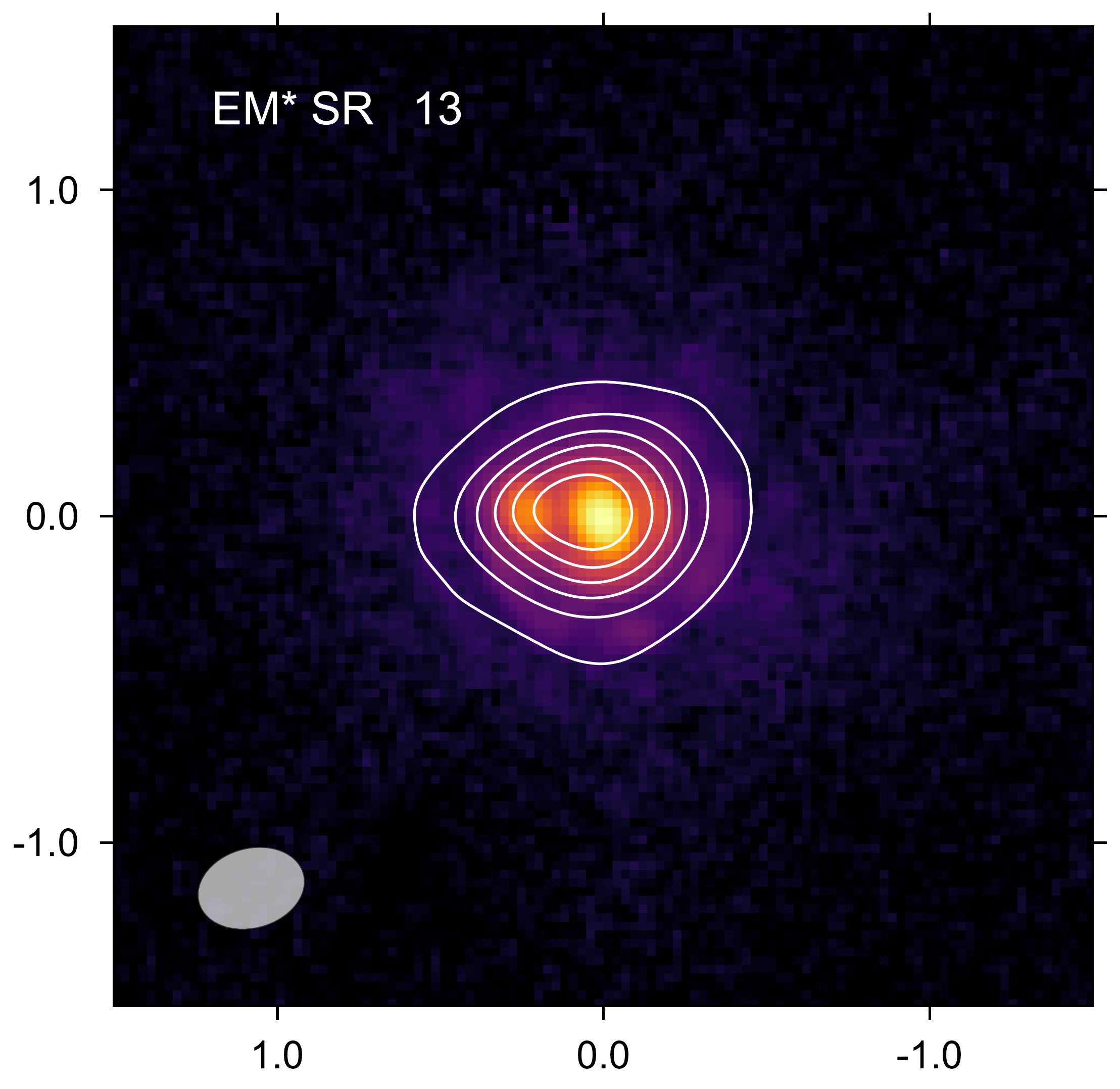}  \hfill
  \includegraphics[height=0.3\textwidth]{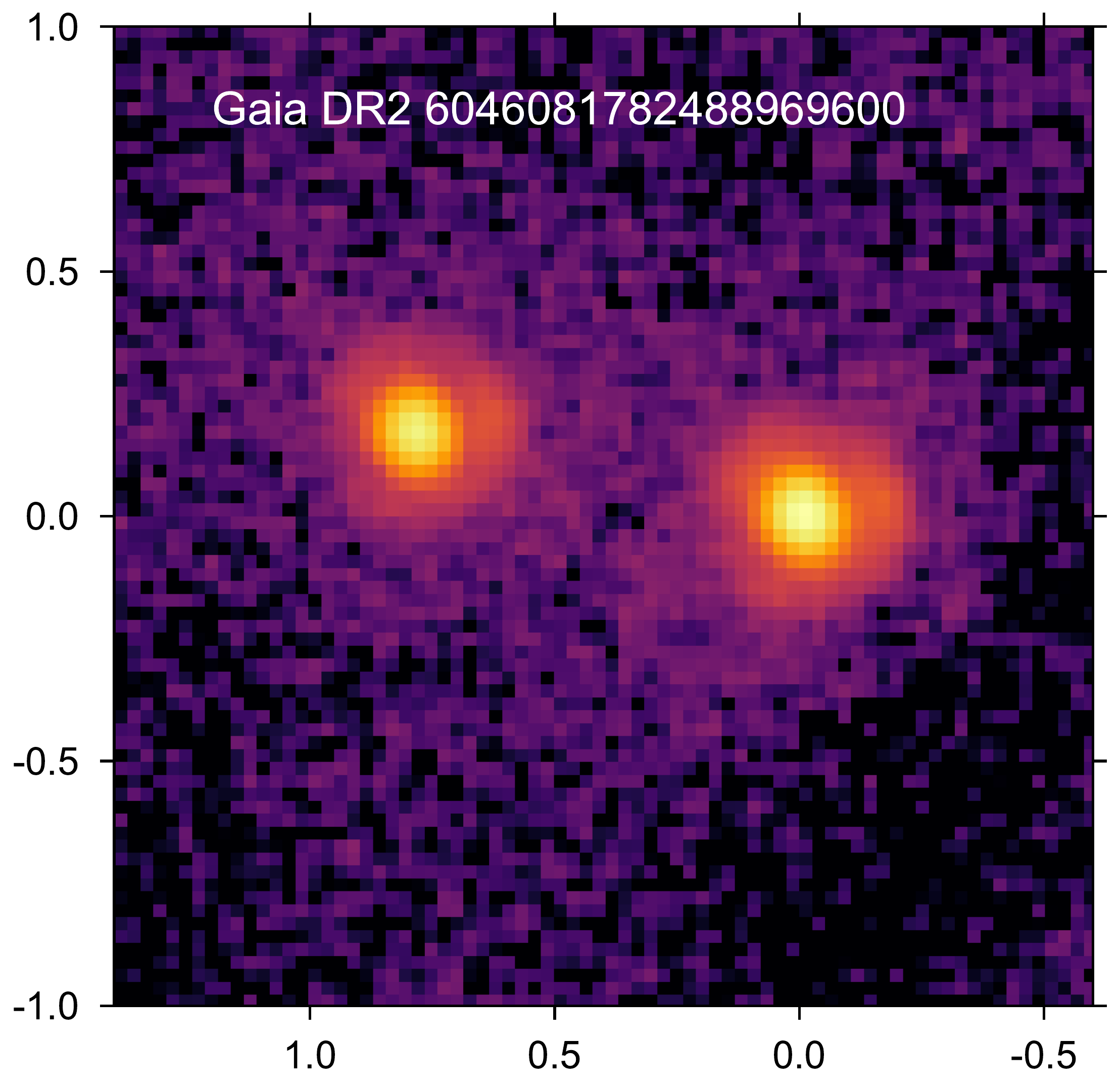}  \\    
  \includegraphics[height=0.3\textwidth]{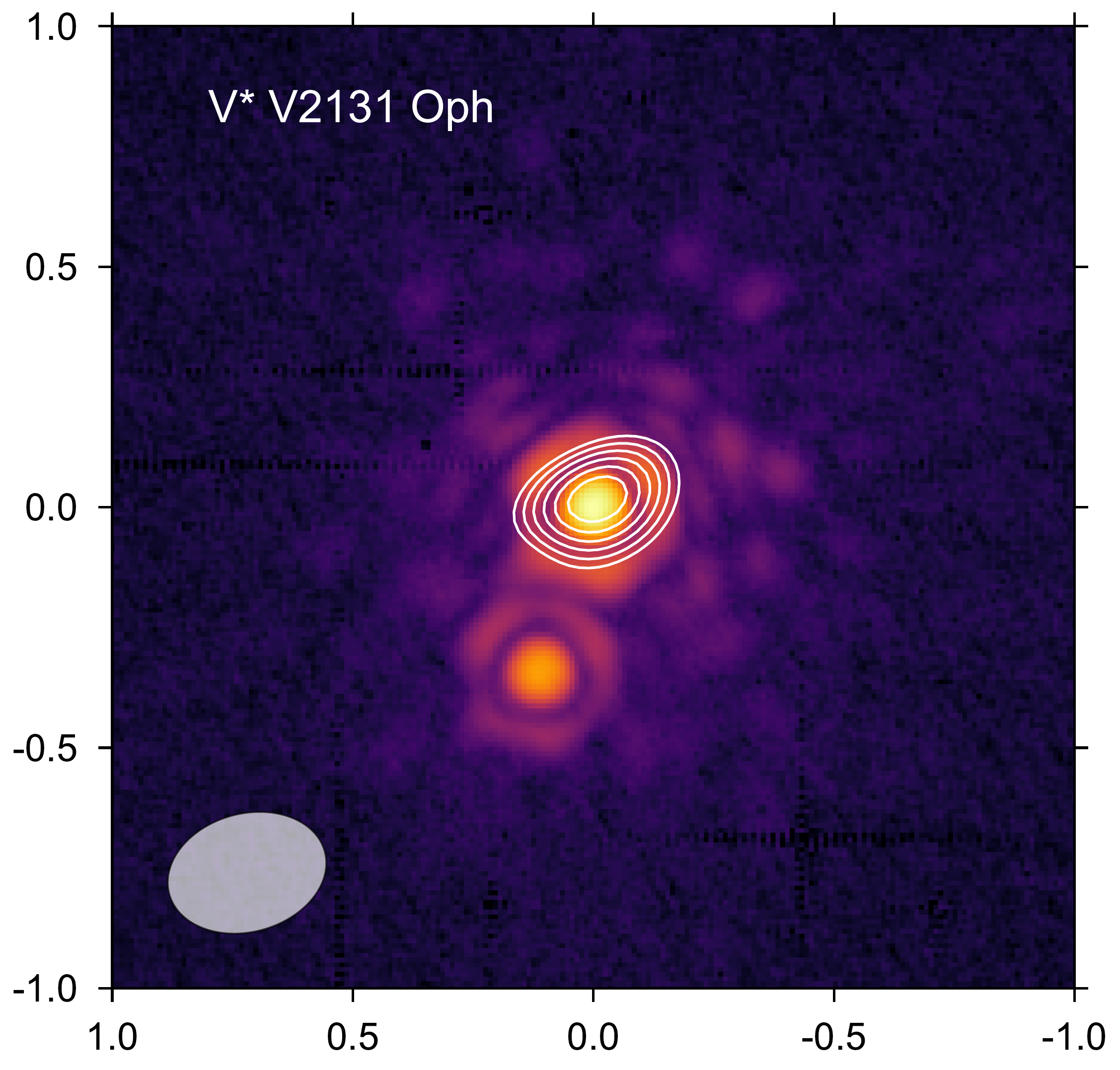}  \hfill
  \includegraphics[height=0.3\textwidth]{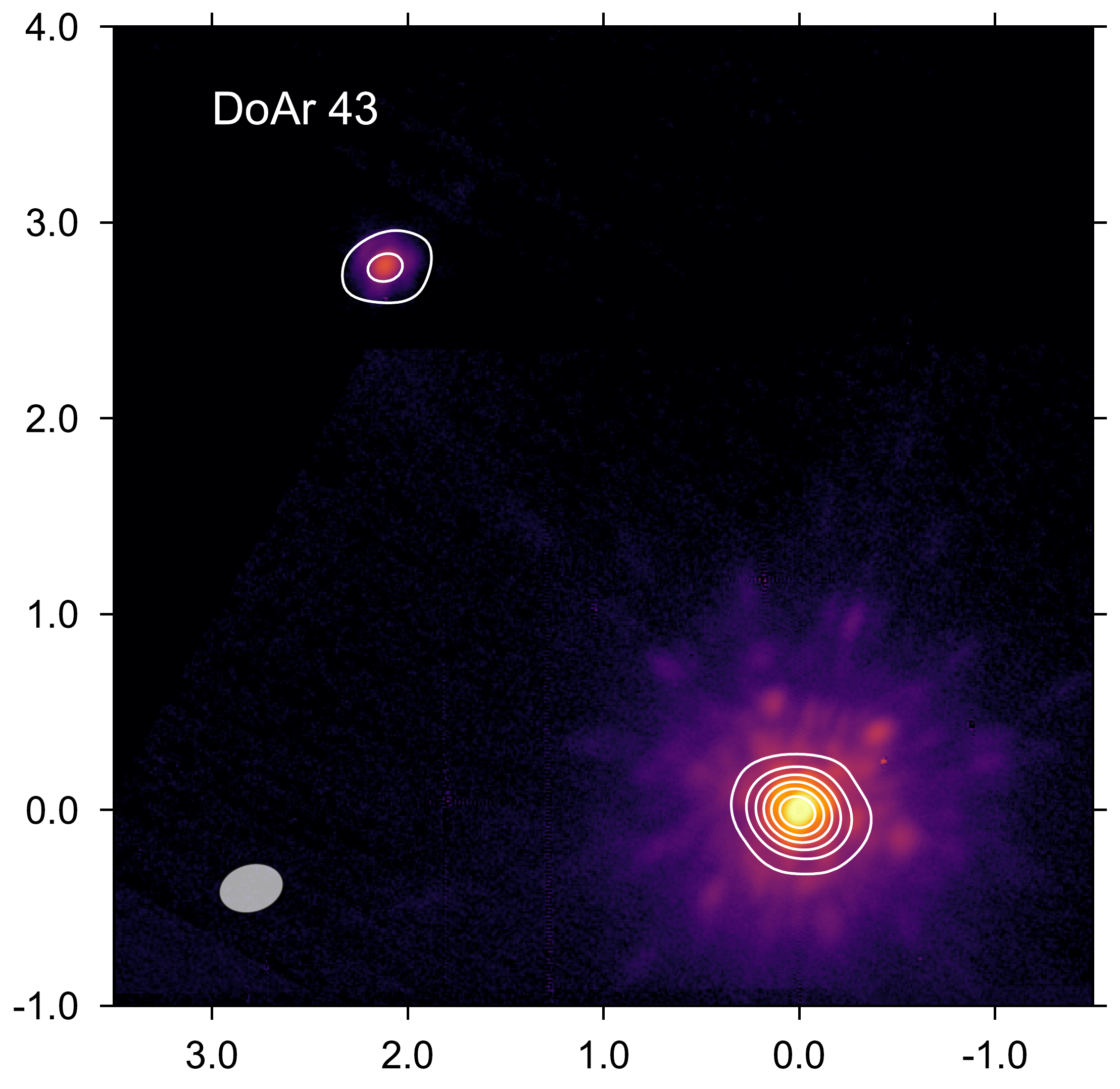}  \hfill
  \includegraphics[height=0.3\textwidth]{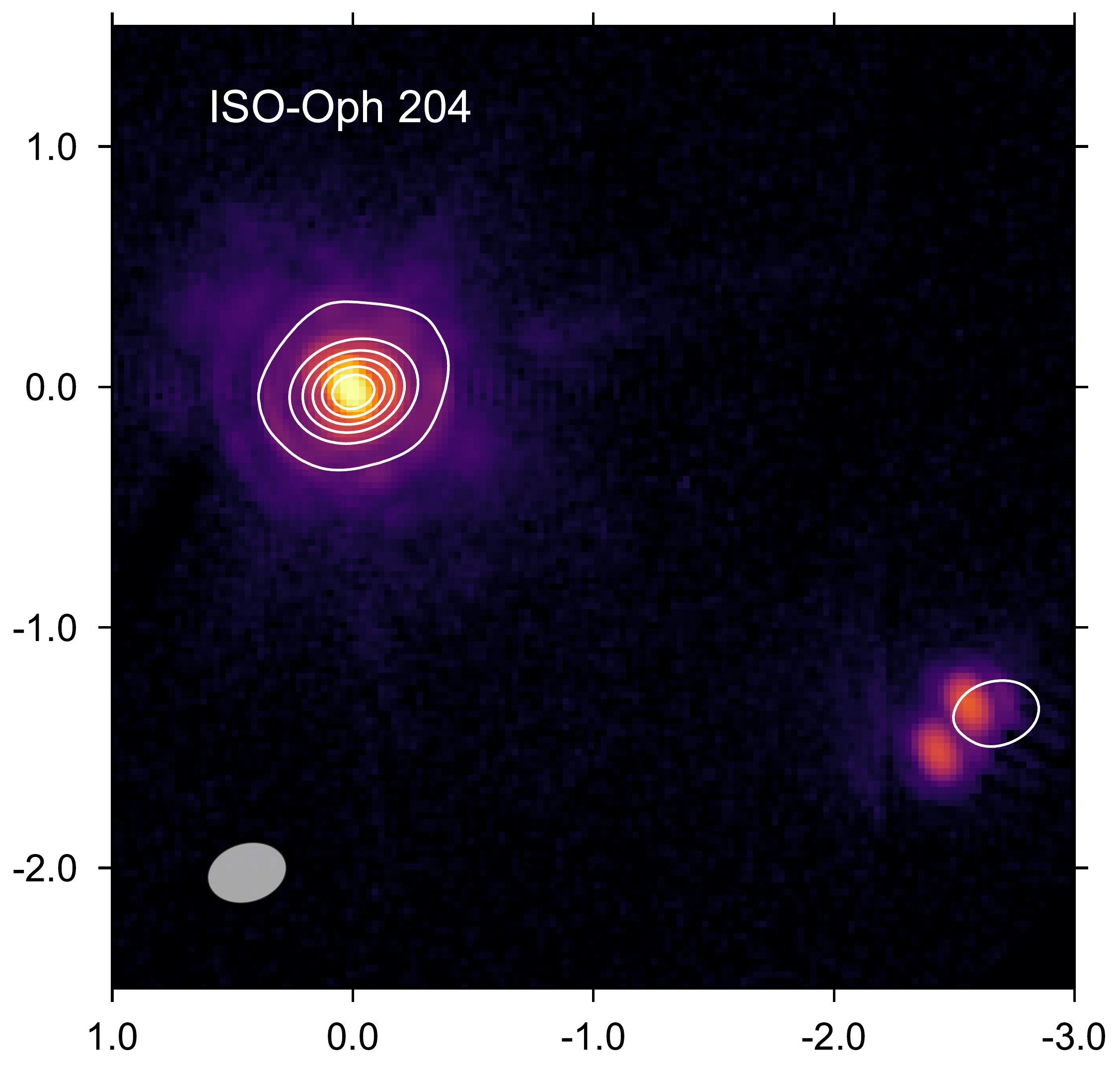}  \\    
  \includegraphics[height=0.3\textwidth]{Odisea035_final.pdf}  \hfill
  \includegraphics[height=0.3\textwidth]{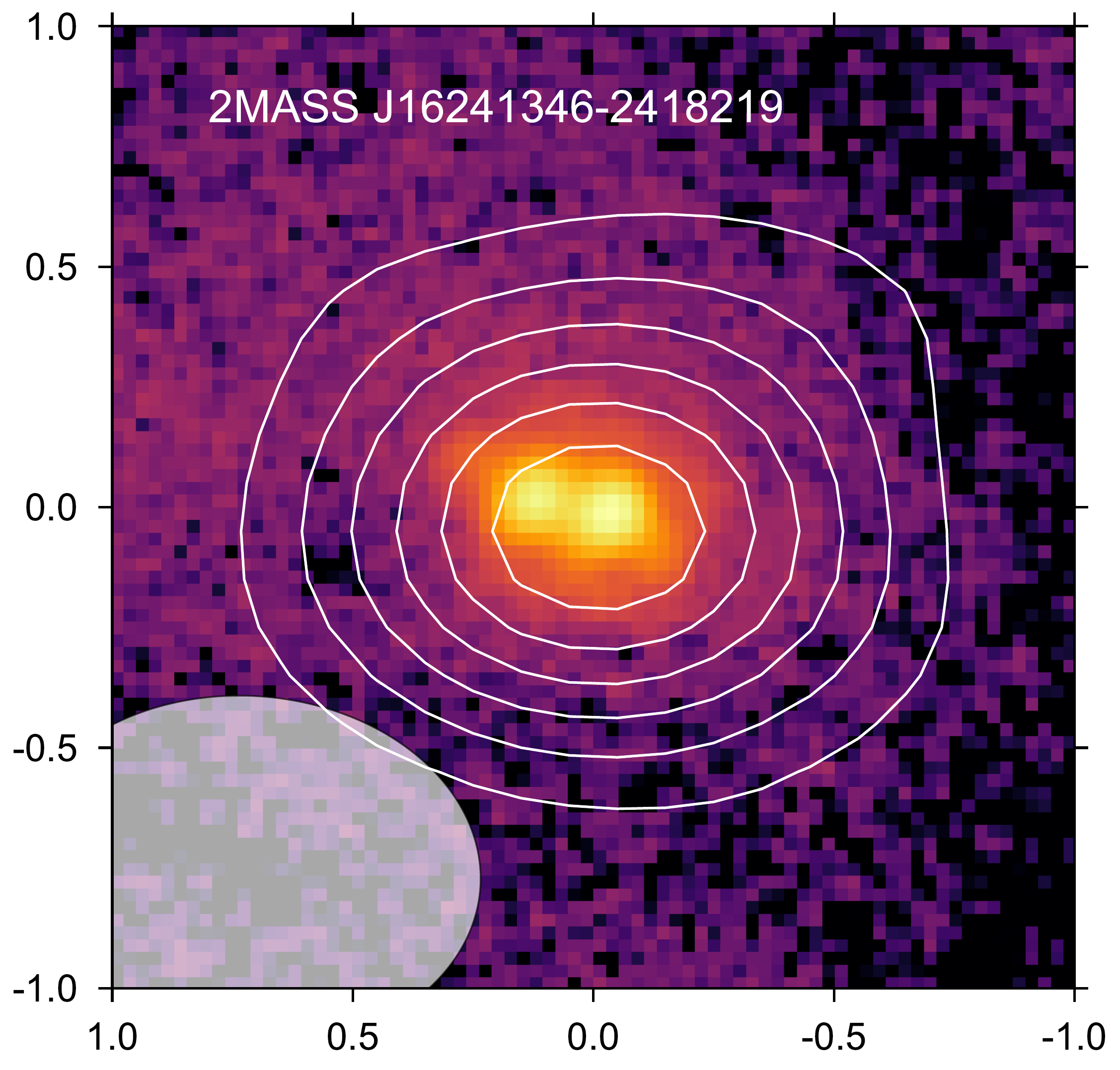}  \hfill
    \includegraphics[height=0.3\textwidth]{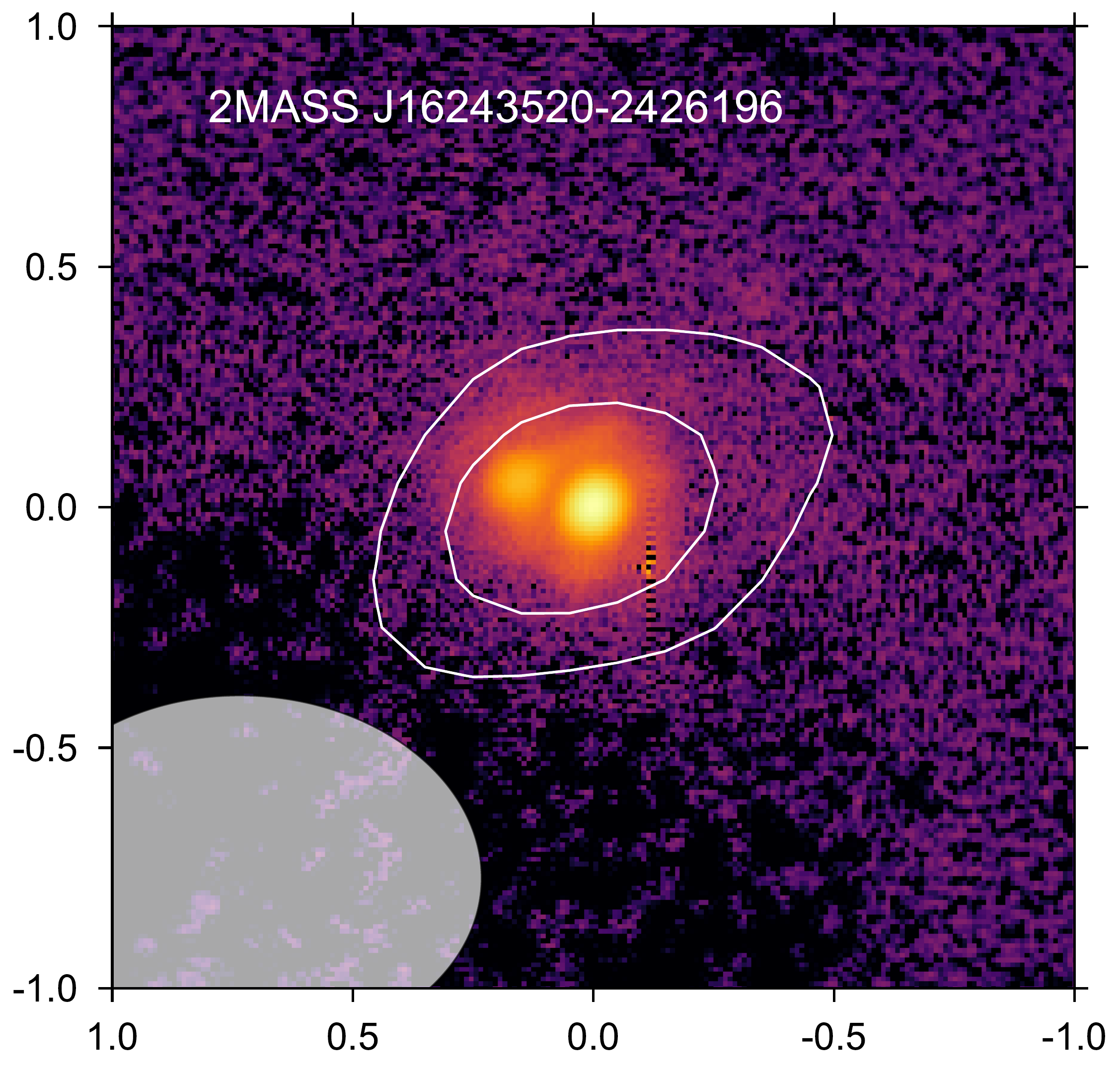}
 
\caption{Same as Figure~\ref{f:mos}. 
}
\label{f:mos2}
\end{center}
\end{figure*}

\begin{figure*}
  \begin{center}
  
  \includegraphics[height=0.3\textwidth]{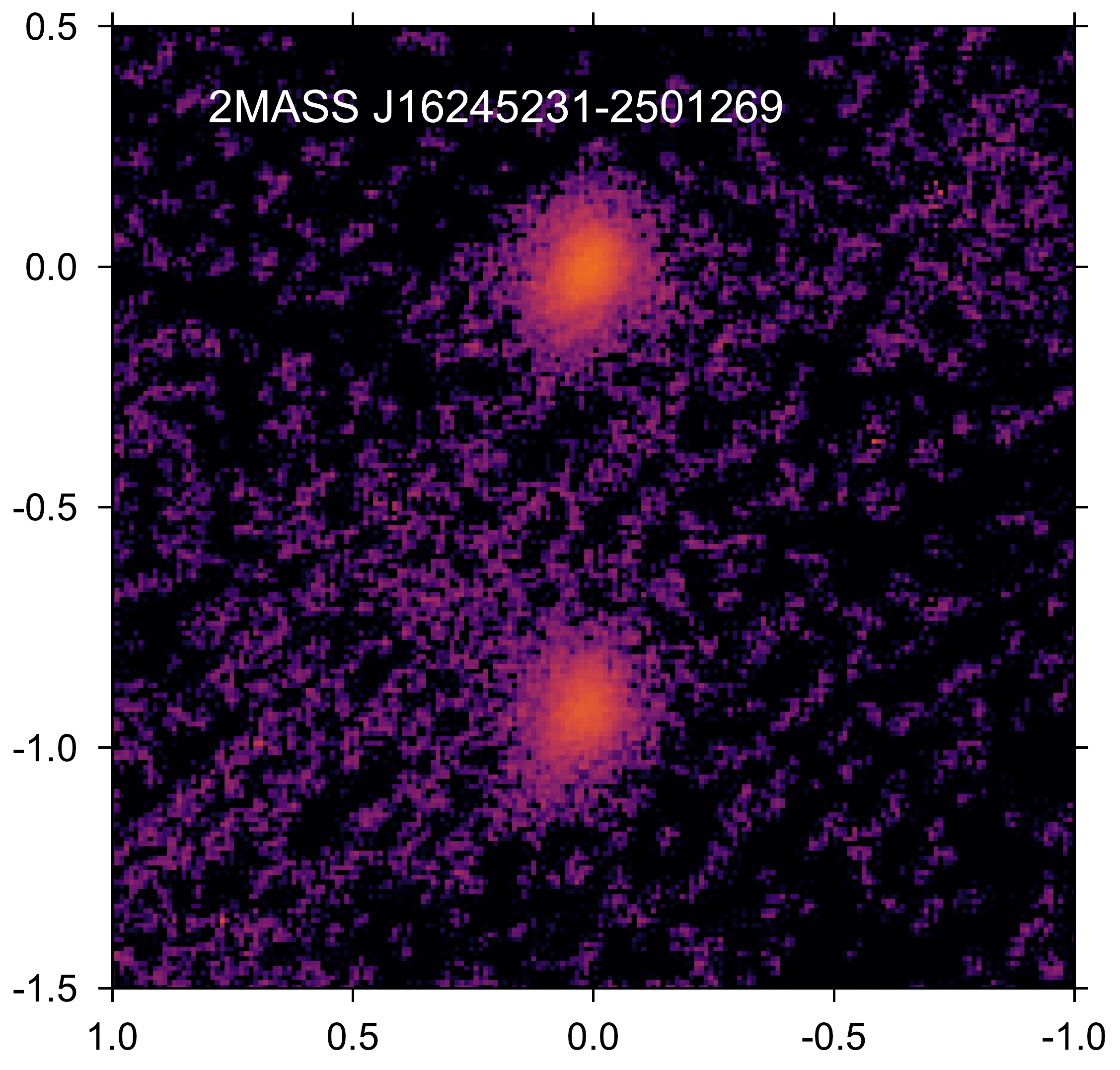}  \hfill
  \includegraphics[height=0.3\textwidth]{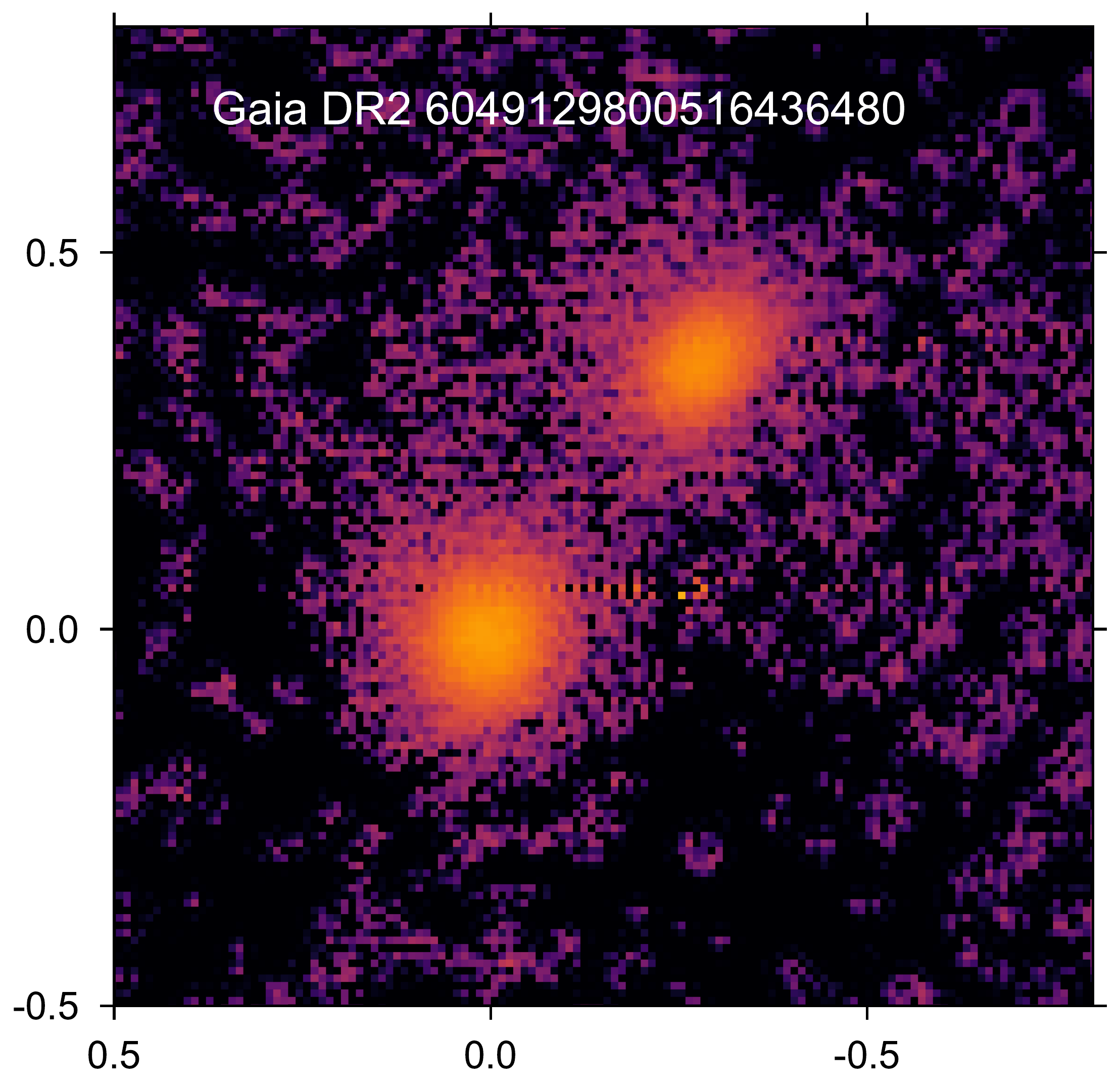}  \hfill
  \includegraphics[height=0.3\textwidth]{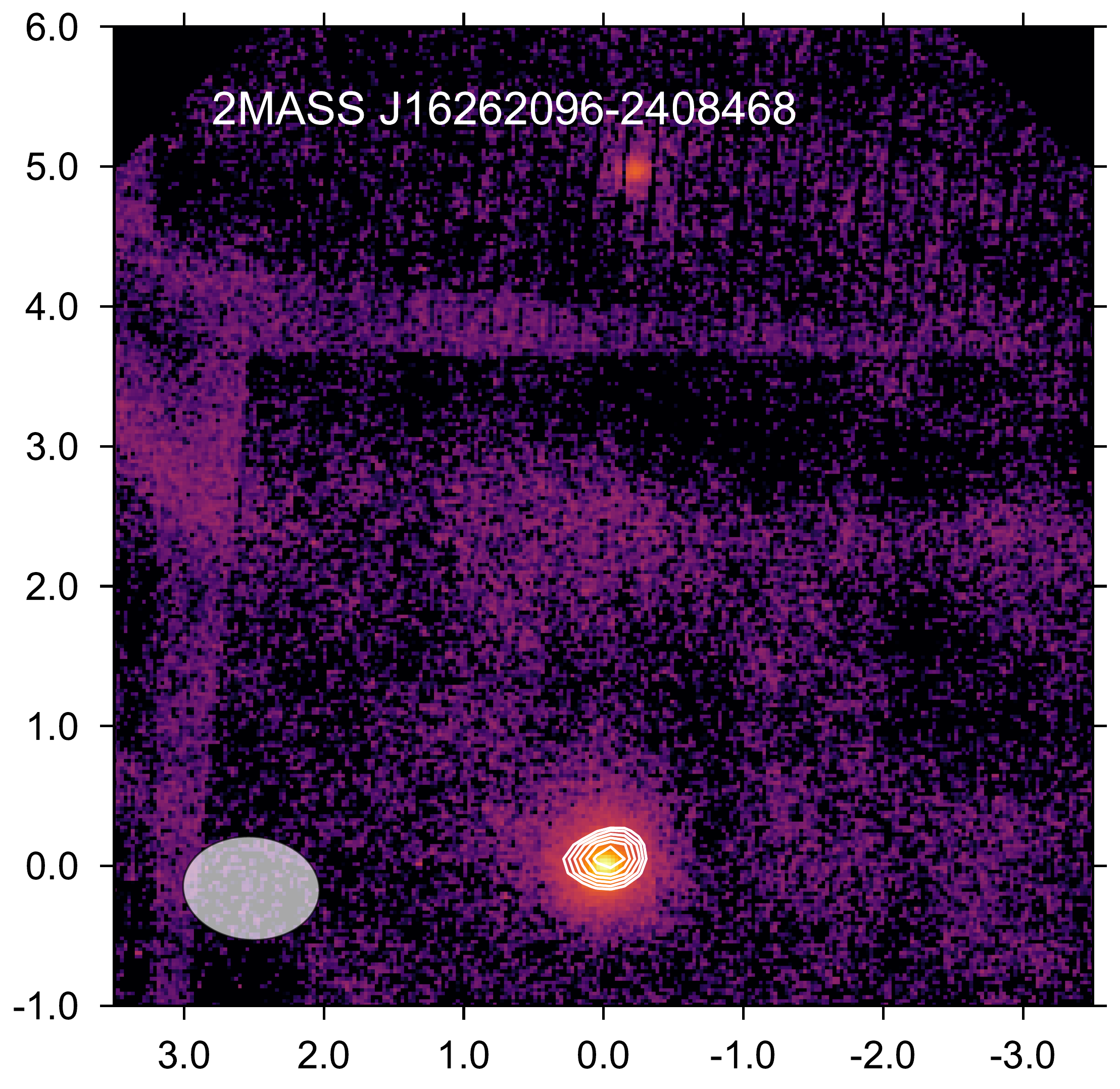}  \\    
  \includegraphics[height=0.3\textwidth]{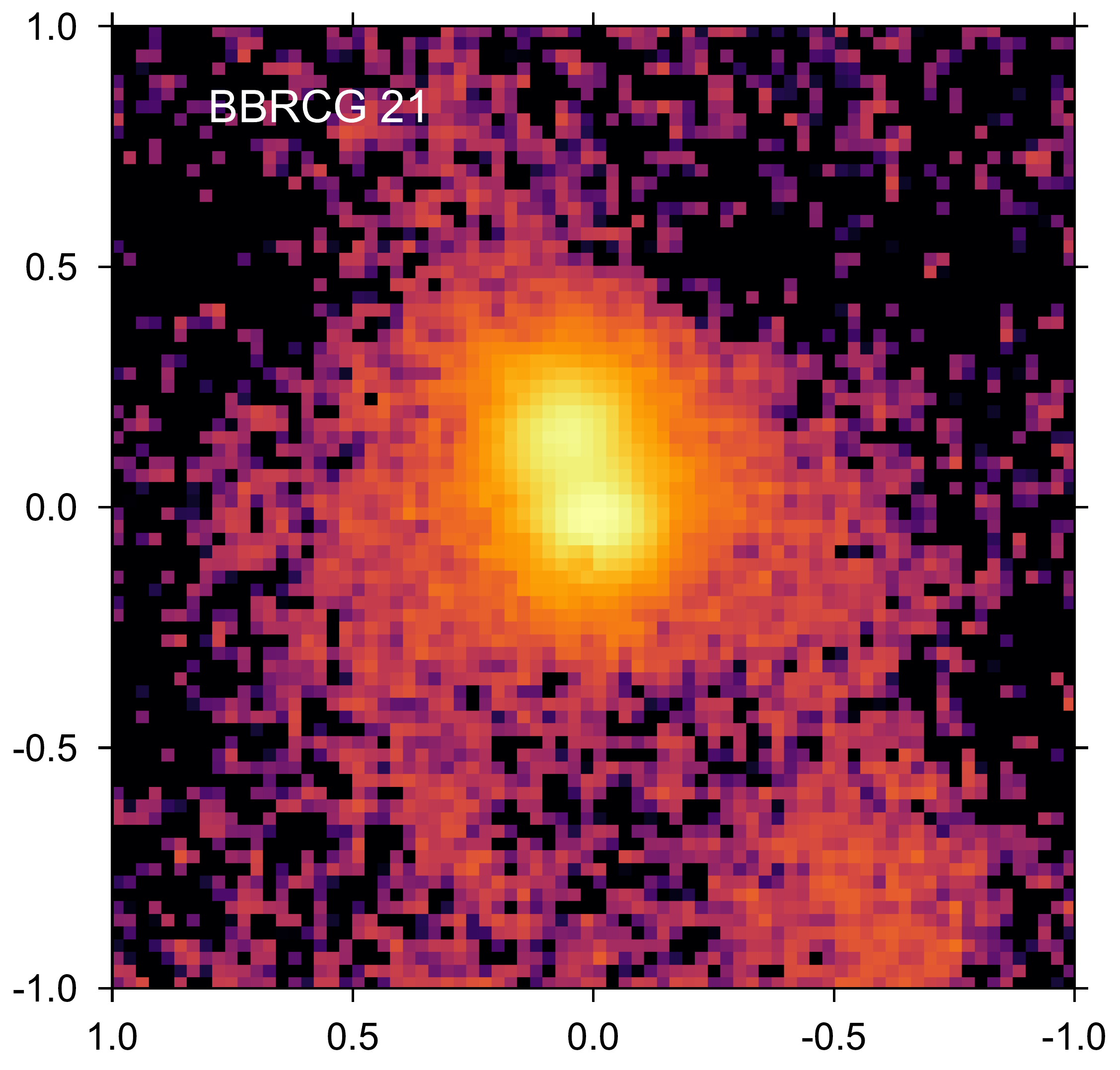}  \hfill
  \includegraphics[height=0.3\textwidth]{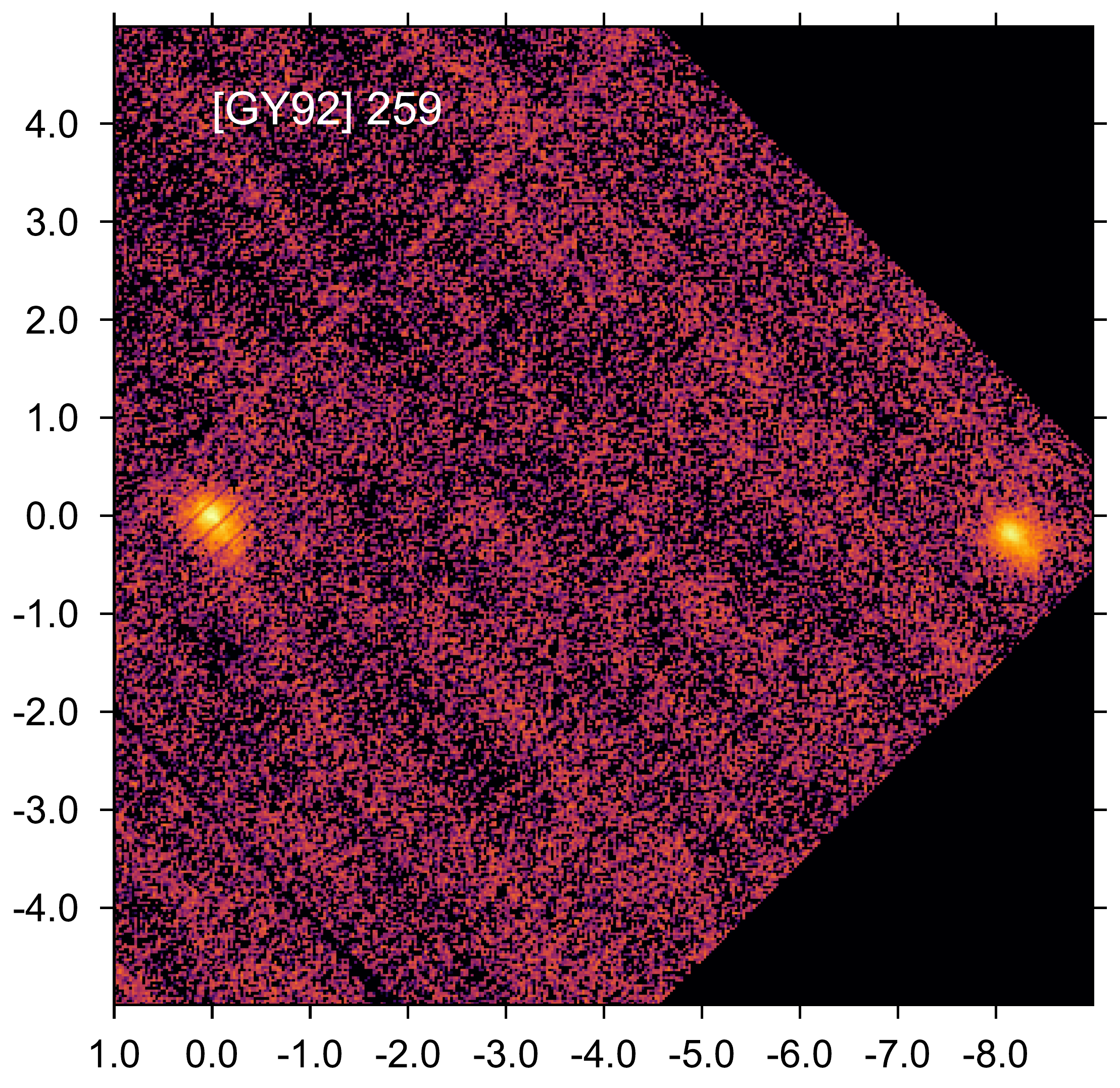}  \hfill
  \includegraphics[height=0.3\textwidth]{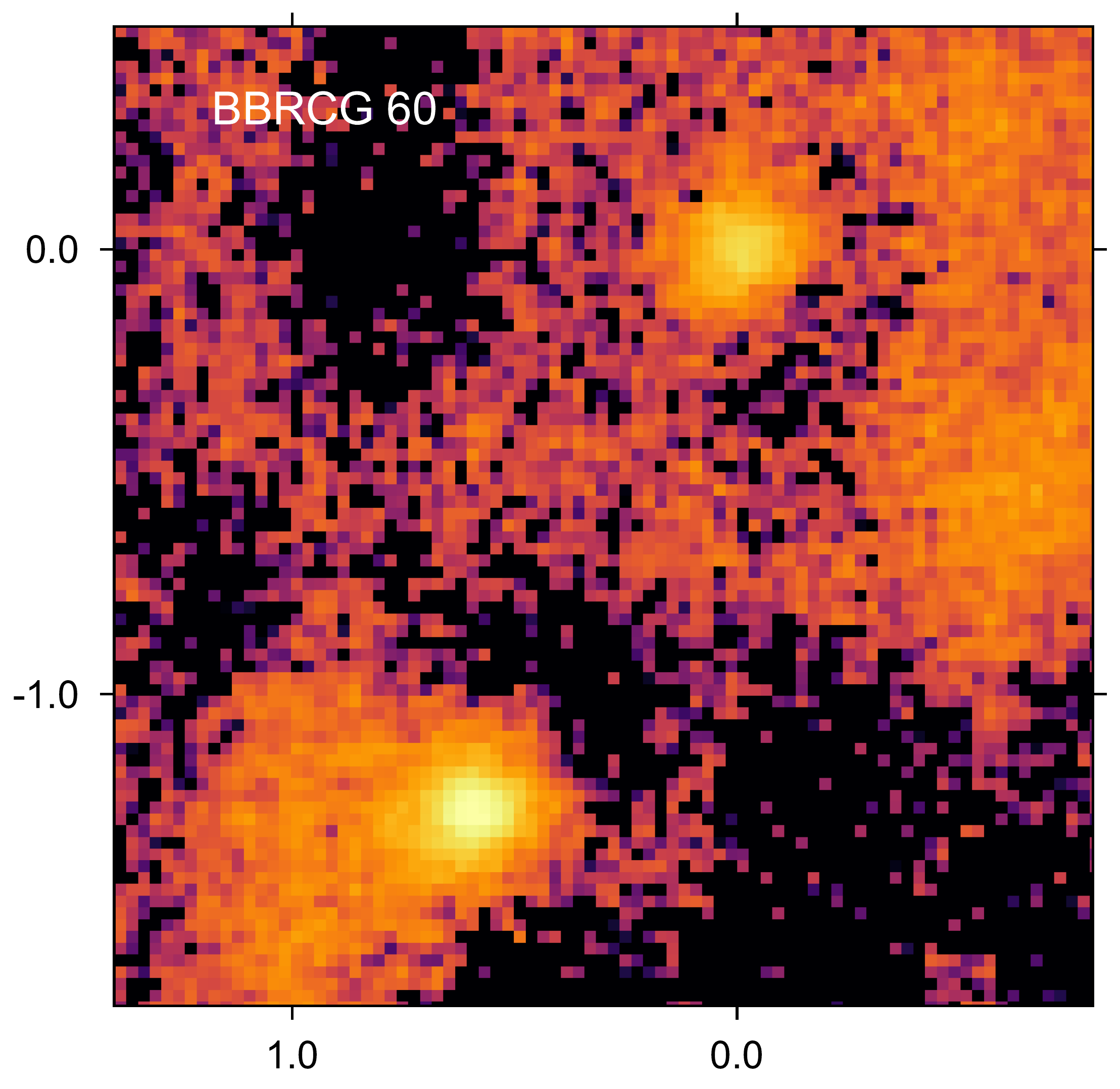}  \\    
  \includegraphics[height=0.3\textwidth]{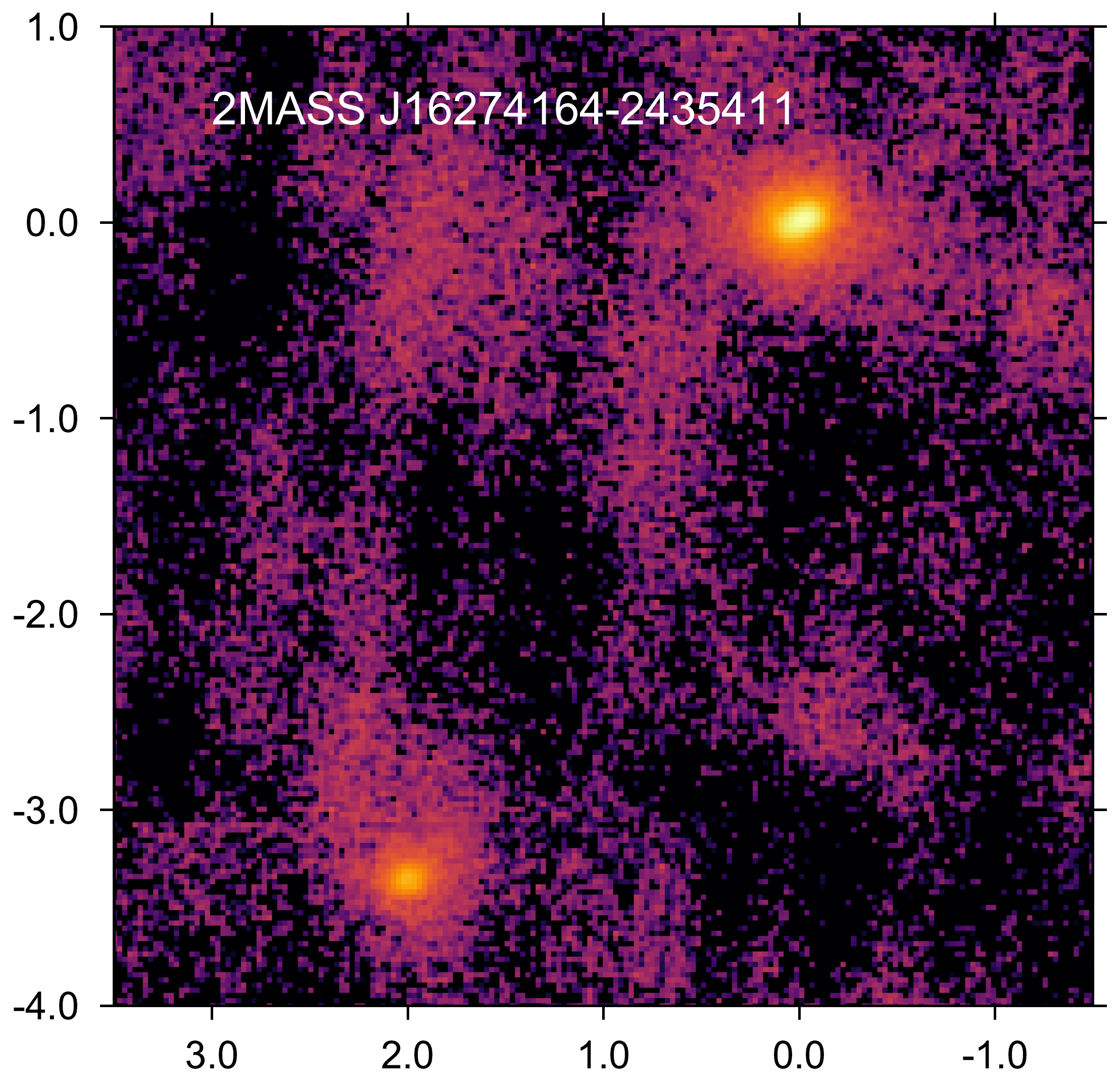}  \hfill
  \includegraphics[height=0.3\textwidth]{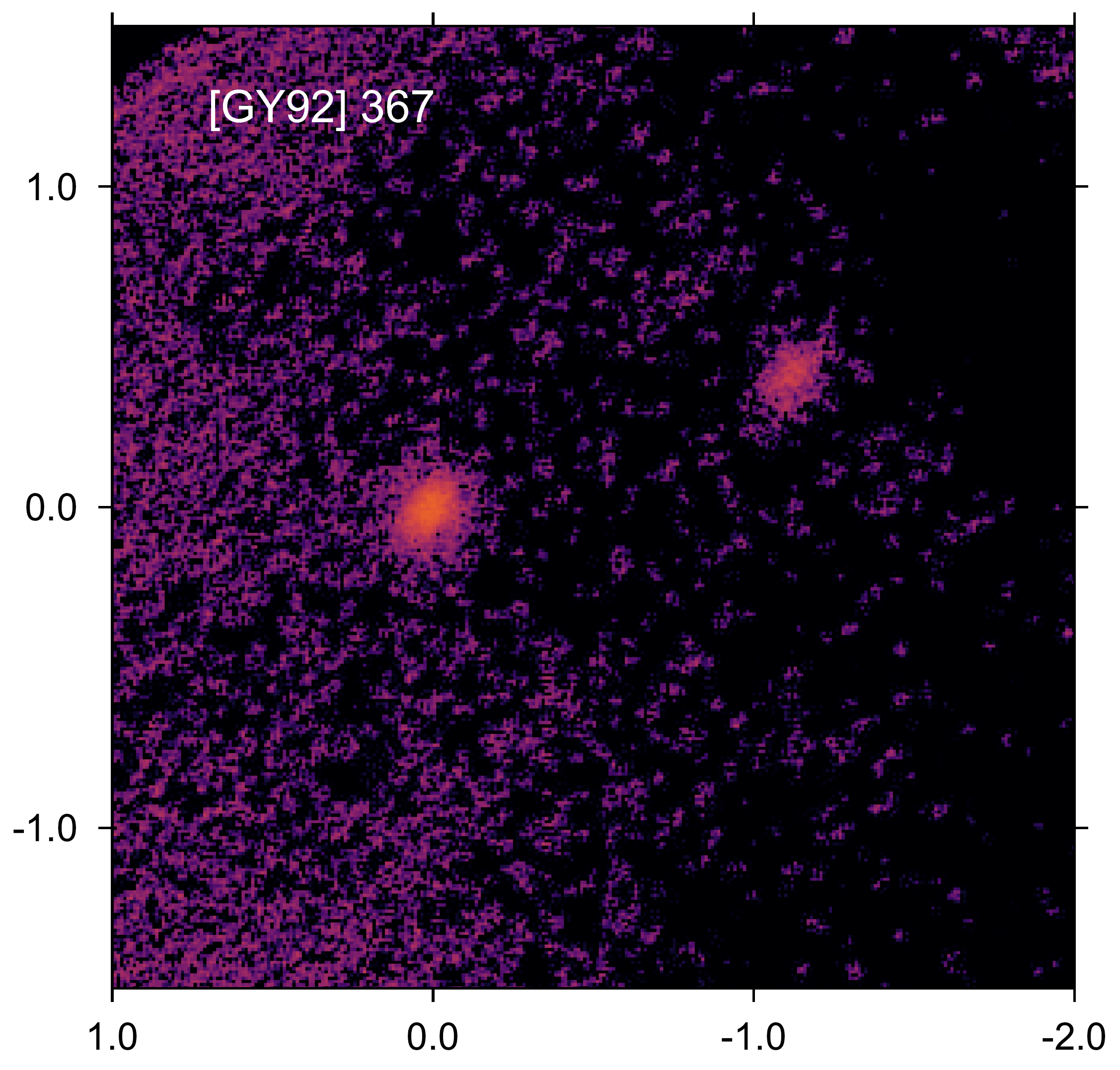}  \hfill
  \includegraphics[height=0.3\textwidth]{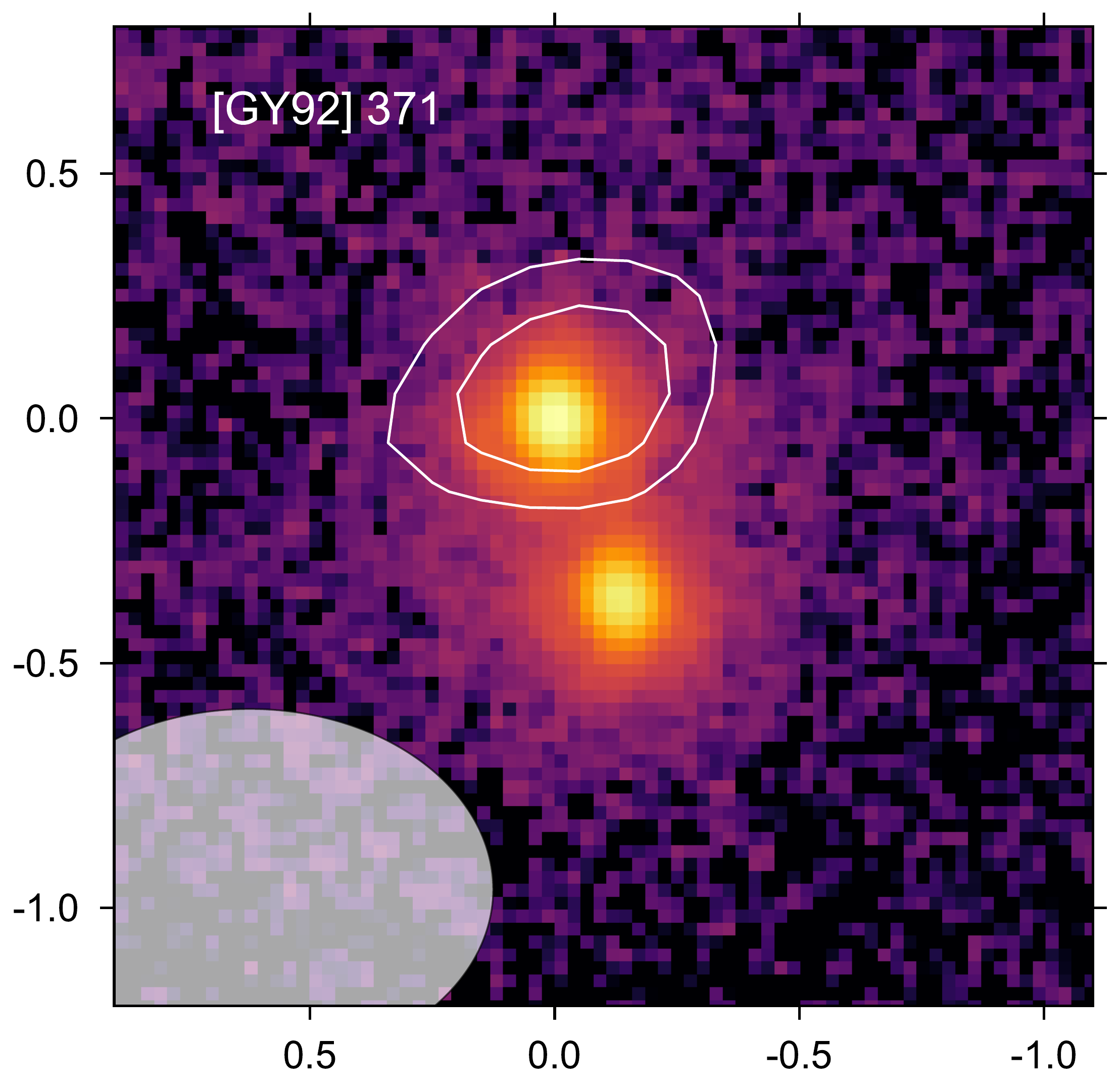}  \\    
  \includegraphics[height=0.3\textwidth]{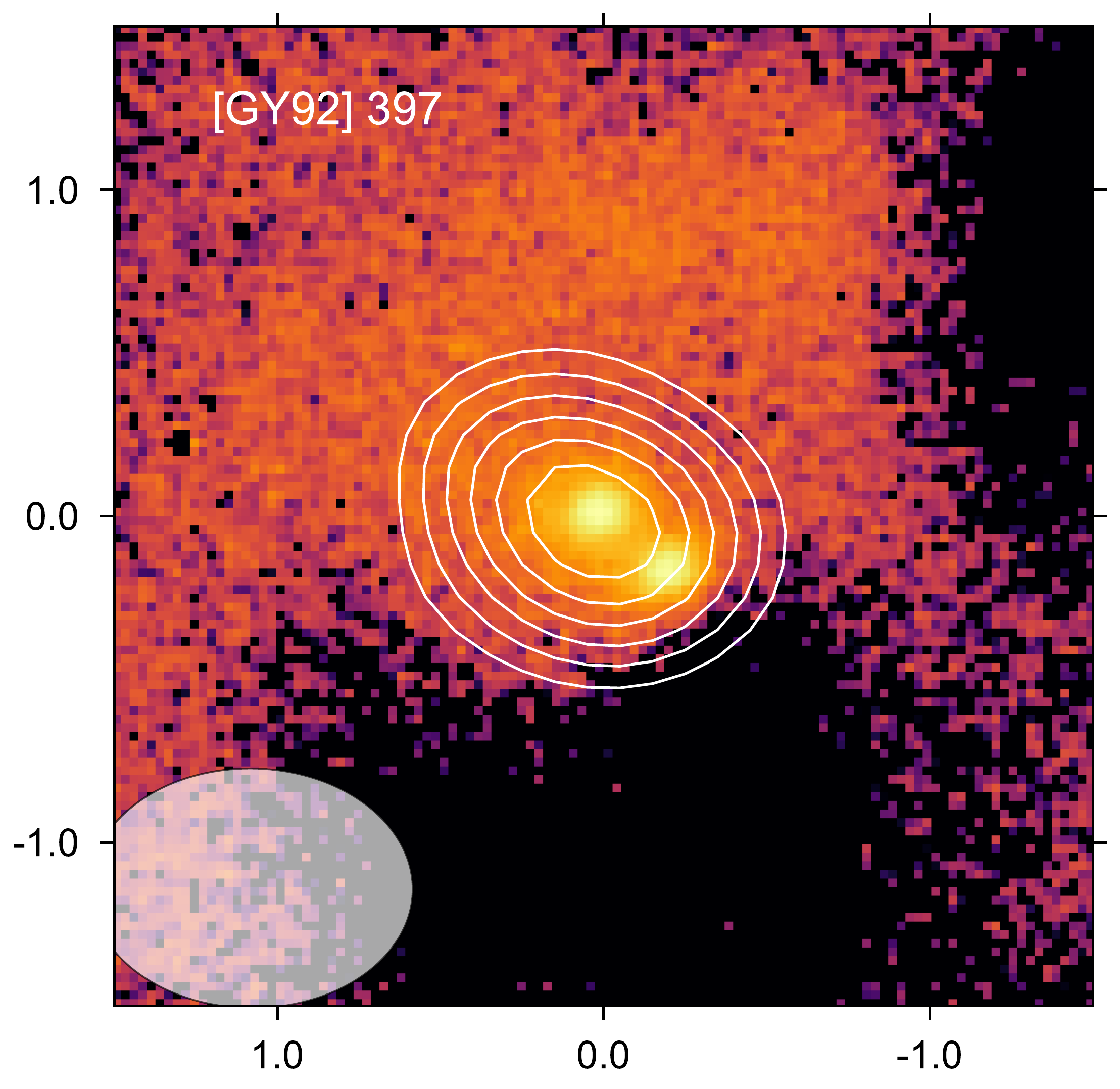}  \hfill
  \includegraphics[height=0.3\textwidth]{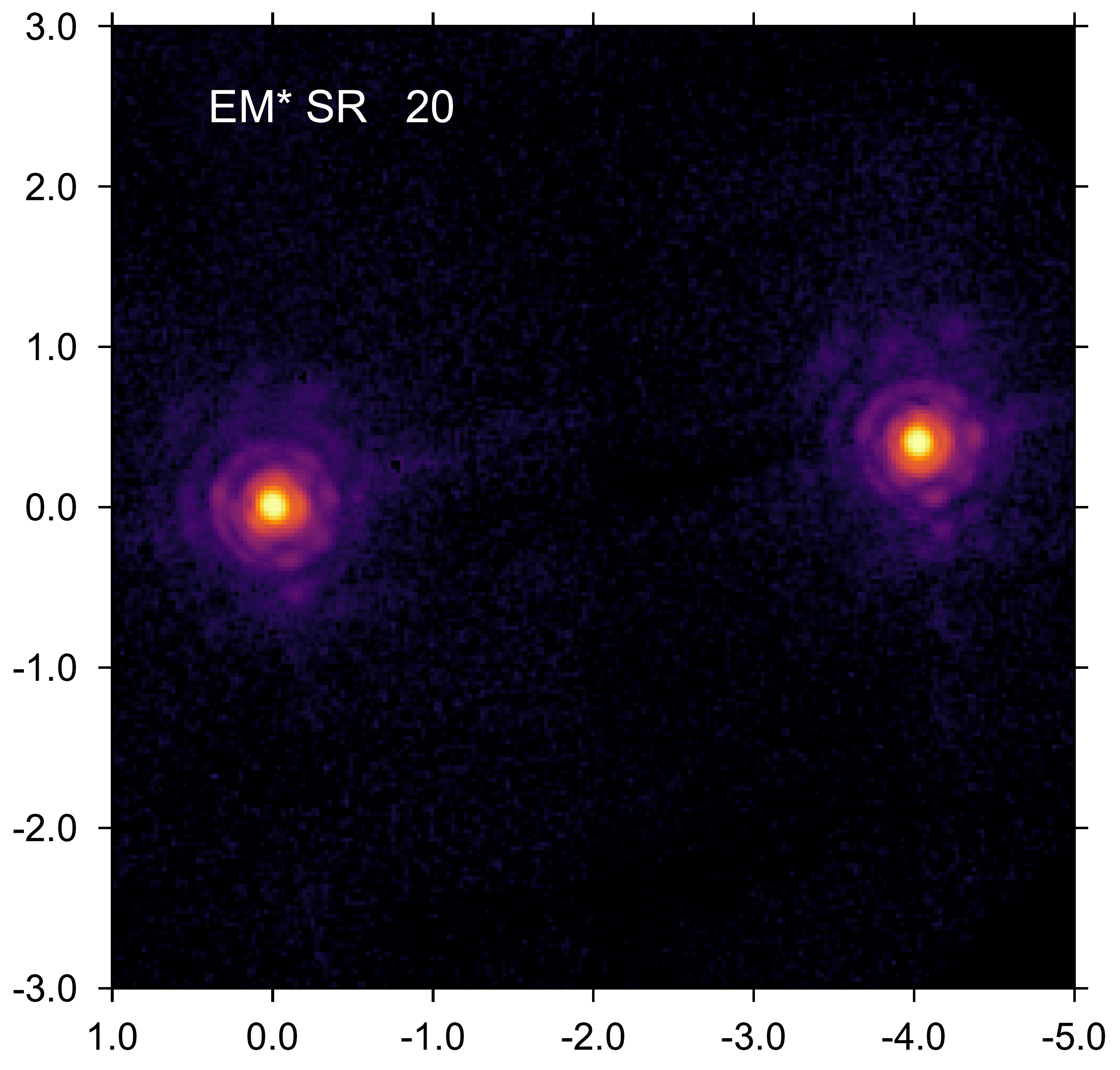}  \hfill
  \includegraphics[height=0.3\textwidth]{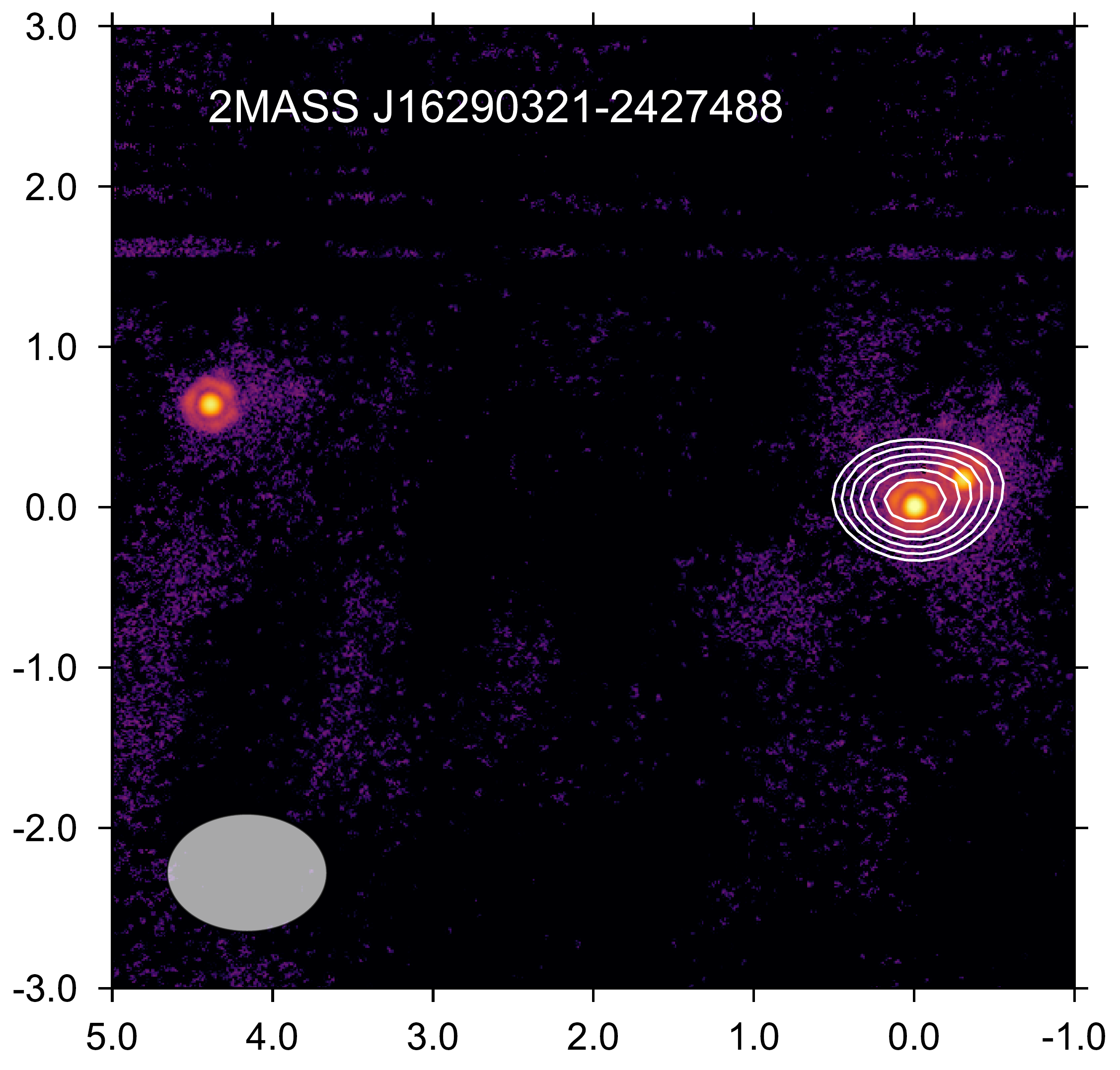}
  \caption{Same as Figure~\ref{f:mos}.}
\label{f:mos3}
\end{center}
\end{figure*}

\begin{figure*}
  \begin{center}

  \includegraphics[width=0.3\textwidth]{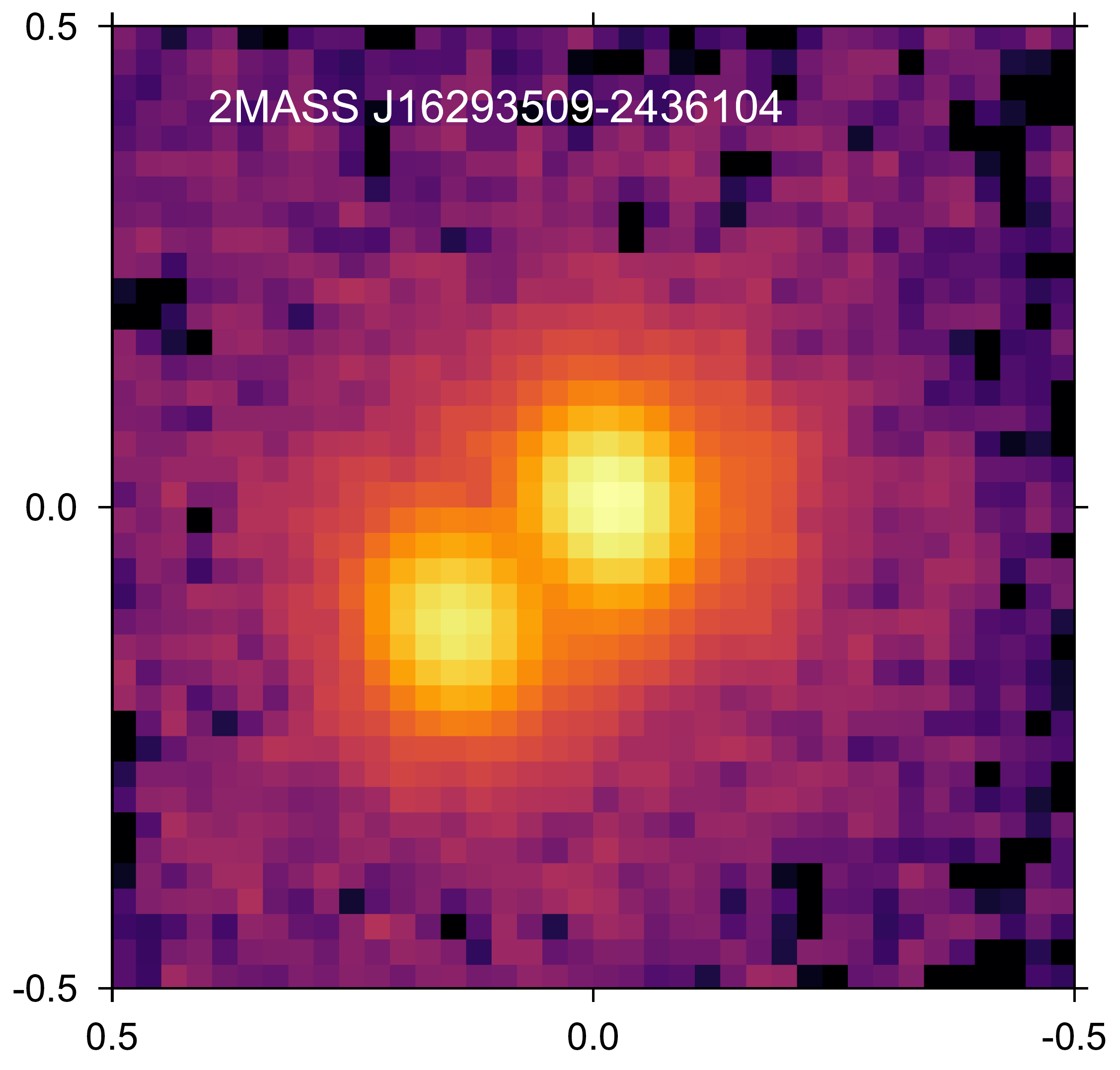}
\caption{Same as Figure~\ref{f:mos}.}
\label{f:mos4}
\end{center}
\end{figure*}

\begin{figure*}
  \begin{center}

    \includegraphics[height=0.3\textwidth]{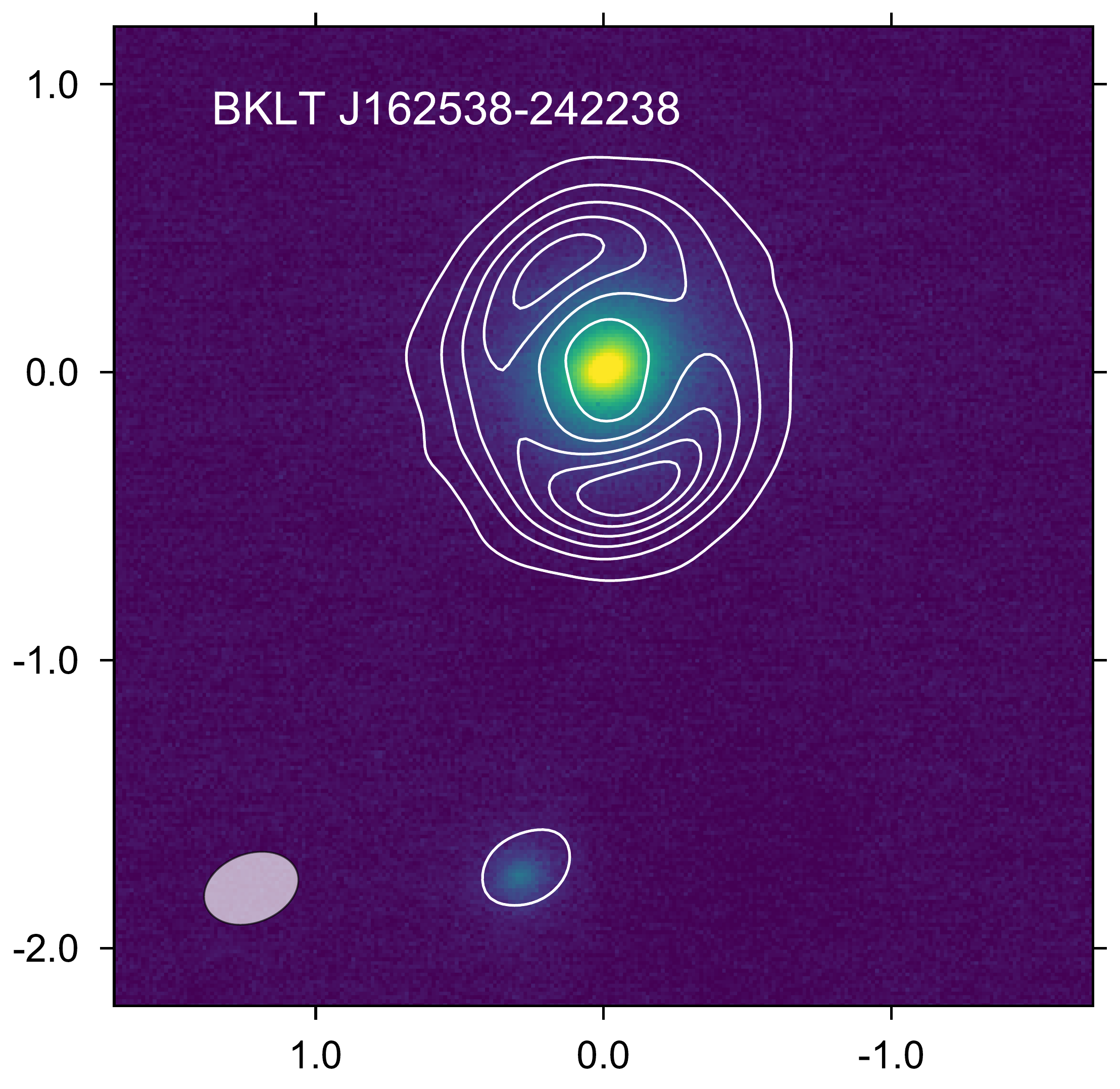}
    \includegraphics[height=0.3\textwidth]{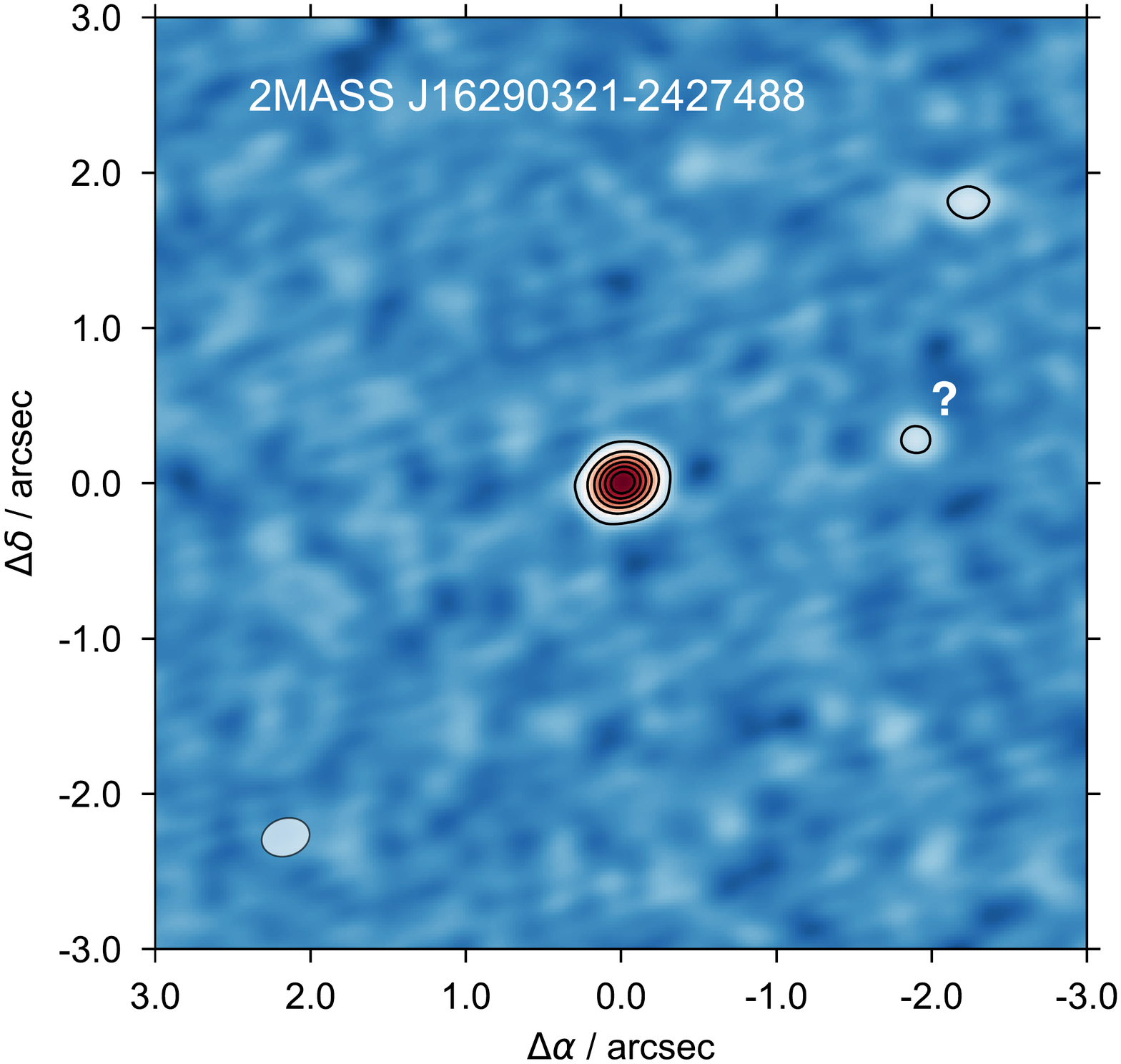}

\caption{{Multiple systems identified in ALMA and not included in NIR-ODISEA. {\it Left:} NACO archive data are shown, overplotted by ALMA contours, as in Figure~\ref{f:mos}.  {\it Right:} In 2MASS J16290321-2427488, the point source identified with a ``?'' is $>$3$\sigma$, only marginally detected. } }
\label{f:onlyalma}
\end{center}
\end{figure*}

Multiple systems were identified and presented in \citep{2019MNRAS.482..698C}: one triple system and 11 binaries were detected in the millimeter. The ALMA images are shown as contours in their NIR counterparts in Fig.~\ref{f:mos}, \ref{f:mos2}, \ref{f:mos3}, and \ref{f:mos4}. The multiple system detected with ALMA for which we do not have NIR AO-imaging is shown in Fig.~\ref{f:onlyalma} {(right panel)}.

\begin{table*}
  \begin{minipage}{1\textwidth}
\caption{{Properties of the stellar companions of the Ophiuchus discs detected during the NIR-ODISEA survey. {Objects marked with a $\blacktriangledown$ are sub-stellar, in the range 30-50 \MJup.} }} 
\label{t:bin}
\centering
\begin{tabular}{lccccccccc}
\hline
\hline
Name & ODISEA ID   & RA & DEC  & Sep  & PA  & NIR  & mm  &  Category\footnote{See Table~\ref{t:cat} and Sec.~\ref{s:res} for more details.} & Instrument \\
 &   &   &  &  (au) & (deg) & flux ratio & flux ratio &  & \\
\hline
2MASSJ16224539-2431237     & C4\_008B  	&	245.689	&	-24.523	&  80  	&       -41 		&	0.17	& <0.01	&	ND &NACO	\\
2MASSJ16230923-2417047     & C4\_012B	&	245.788	&	-24.285	&  226       &	30  	&	0.04	&	<0.01	&ND& NIRC2	\\
IRAS16220-2452             & C4\_016B	&	246.259	&	-24.992	&  209       &	-98      &	0.1	& 0.29 &	Res&NACO	\\
2MASSJ16253253-2326264     & C4\_020B	&	246.386	&	-23.441	&  128       &	-141     &	0.17	&<0.01	&	ND&NACO	\\
YLW19                      & C4\_021B	&	246.403	&	-24.262	&  46        &	59  	&	0.11	&<0.01	&	ND&NACO	\\
$\blacktriangledown$  BKLTJ162538-242238\footnote{For this target K$_s$ filter archive data were used.} & C4\_022B	&246.409   & -24.377	&  236 	&  172		& 0.08 &	0.04	&	Res&NACO	\\
2MASSJ16253943-2326419     & C4\_023B	&	246.414	&	-23.445	&  39        &	-67 	&	0.97	&<0.01	&	ND & NIRC2	\\
IRAS16226-2420             & C4\_024B	&	246.415	&	-24.443	&  151       &	118 	&	0.15	&<0.01	&	ND&NACO	\\
WDSJ16264-2421C            & C4\_037B	&	246.598	&	-24.350	&  185       &	-12 	&	0.09	&  0.79 &	Res& NIRC2	\\
GY9293                   & C4\_048B	&	246.672	&	-24.672	&  9         &	-  	&	0.99	&<0.01	&	ND&NACO	\\
REP40A                   & C4\_050B	&	246.678	&	-24.342	&  198       &	70  	&	0.09	&  0.12 &	Res&NACO	\\
YLW37                      & C4\_052B	&	246.693	&	-24.200	&  57        &	116 	&	0.11	&<0.01	&	ND	&NACO\\
$\blacktriangledown$ WL2                         & C4\_053B	&	246.702	&	-24.478	&  590       &	-18 	&	0.23	&  0.51 &	Res&NACO	\\
VSSG31                     & C4\_065B	&	246.767	&	-24.475	&  127       &	-37 	&	0.77	&<0.01	&	ND&NACO	\\
YLW10B                     & C4\_078B	&	246.814	&	-24.445	&  33        &	-136     &	0.77	&<0.01	&	Res&NACO	\\
YLW45                      & C4\_105B	&	246.916	&	-24.721	&  230       &	10  	&	0.76	& 0.29 &	Res	&NACO\\
EM*SR9                     & C4\_106B	&	246.918	&	-24.368	&  89        &	-4  	&	0.3	& 0.07 &	Res&NACO	\\
EM*SR13                    & C4\_117B	&	247.189	&	-24.472	&  17        &	79  	&	0.16	& 0.14 &	Res&NACO	\\
GaiaDR26046081782488969600 & C4\_120B	&	247.434	&	-24.689	&  116       &	77  	&	0.86	&<0.01	&	ND&NACO	\\
V*V2131Oph                 & C4\_123B	&	247.816	&	-24.567	&  36        &	162 	&	0.25	&<0.01	&	UR& NIRC2	\\
DoAr43                     & C4\_125B	&	247.879	&	-24.411	&  339       &	37  	&	0.06	& 0.02& 	Res& NIRC2	\\
ISO-Oph204                 & C4\_134B	&	247.967	&	-24.938	&  498       &	-118     &	0.15	& 0.11 &	Res&NACO	\\
2MASSJ16233609-2402209     & C5\_021B	&	245.900	&	-24.039	&  251       &	-143     &	0.59	&<0.01	&	ND	&NACO\\
2MASSJ16241346-2418219     & C5\_025B	&	246.056	&	-24.306	&  22        &	77  	&	0.8	&<0.01	&	ND	&NACO\\
$\blacktriangledown$ 2MASSJ16243520-2426196     & C5\_027B	&	246.147	&	-24.439	&  12        &	69  	&	0.33	&<0.01	&	ND	& NIRC2\\
$\blacktriangledown$ 2MASSJ16245231-2501269    & C5\_028B	&	246.218	&	-25.024	&  92        &	-180     &	0.71	&<0.01	&	ND	& NIRC2\\
$\blacktriangledown$ GaiaDR26049129800516436480  & C5\_033B	&	246.357	&	-24.619	&  48        &	-38 	&	0.79	&<0.01	&	ND	& NIRC2\\
$\blacktriangledown$ 2MASSJ16262096-2408468     & C5\_047B	&	246.587	&	-24.146	&  734       &	-3  	&	0.26	&<0.01	&	ND&NACO	\\
BBRCG21                    & C5\_066B	&	246.776	&	-24.477	&  9         &	-  	&	0.86	&<0.01	&	ND&NACO	\\
$\blacktriangledown$ GY92 259                  & C5\_072B	&	246.852	&	-24.493	&  1169      &	-92  	&	0.85	&<0.01	&	ND&NACO	\\
BBRCG60                    & C5\_076B	&	246.886	&	-24.556	&  199       &	-25 	&	0.6	&<0.01	&	ND&NACO	\\
2MASSJ16274164-2435411     & C5\_084B	&	246.924	&	-24.595	&  558       &	149 	&	0.3	&<0.01	&	ND&NACO	\\
$\blacktriangledown$ GY92 367                 & C5\_090B	&	246.955	&	-24.414	&  130       &	-66 	&	0.39	&<0.01	&	ND& NIRC2	\\
GY92 371                  & C5\_091B	&	246.957	&	-24.423	&  54        &	-158     &	0.7	&<0.01	&	ND&NACO	\\
GY92 397                  & C5\_092B	&	246.980	&	-24.478	&  6         &	-122     &	0.62	&<0.01	&	UR&NACO	\\
EM*SR20                    & C5\_106B	&	247.136	&	-24.379	&  566       &	-84      &	0.23	&<0.01	&	ND&NACO	\\
$\blacktriangledown$ 2MASS J16290321-2427488         & C5\_111B	&	247.263	&	-24.464	&  39       &	-60 	&	0.48	&<0.01	&	UR& NIRC2	\\
2MASS J16290321-2427488   & C5\_111C	&	247.263	&	-24.464	&  495       &	82  	&	0.59	&<0.01	&	ND	& NIRC2\\
2MASSJ16293509-2436104  & C5\_114B	&	247.396	&	-24.603	&  15	&	130 	&	0.62	&<0.01	&	ND&NACO	\\

\hline
\end{tabular}
\end{minipage}
\end{table*}

\begin{table*}
\caption{{Properties of the stellar companion of ODISEA detected by ALMA but not observed in the NIR.}} 
\label{t:bin_alma}
\centering
\begin{tabular}{lcccccc}
\hline
\hline
Name   & ODISEA ID & RA & DEC  & Sep (au) & PA (deg) & mm flux ratio \\
\hline

 2MASSJ16290321-2427488 & C4\_082B	&	246.823  & -24.482	&    393   &	-50	&	0.15	\\
\hline
\end{tabular}
\end{table*}

\section{Results and discussion}
\label{s:res}

In this NIR survey we detected 38 multiple systems: 20 new binaries, 1 triple system, and 17 already-known binaries \citep[from][]{2009ApJ...696L..84C, 2013ApJ...762..100C, 2015ApJ...813...83C, 2017ApJ...851...83C, 2018AJ....155..109S}. Two literature binaries were not re-detected {by NIR-ODISEA} because of lower spatial resolution: 2MASS J162603.0-242336 (C4\_028) \citep[which is a very tight binary;][]{2019ApJ...884...13C} and 2MASS J162654.4-242621 (C5\_058) \citep[binary with 0\farcs15 projected separation;][]{2009ApJ...696L..84C}. {Then, an already known multiple system was not re-observed,  the triple hierarchical system EM*SR24 (C4\_062)}. {Also, the object BKLTJ162538-242238 (C4\_022), already identified to be a binary by \citet{2005A&A...437..611R, 2009ApJ...696L..84C} was not re-observed as NACO archive data are available (see Fig.~\ref{f:onlyalma}, left panel). There is only one multiple system detected in the millimeter which does not have NIR AO-imaging: 2MASSJ16290321-2427488 (C4\_082) (see Fig.~\ref{f:onlyalma}, right panel). This latter appears as a triple system and it is too faint to be observed with NACO or NIRC2.}

For each binary system we calculated the position of the companion and its flux ratio with respect to the primary. The projected separation of the companion in au was calculated for each target assuming the distances for each system listed in \cite{2019ApJ...875L...9W}. In Table~\ref{t:bin} we list all these values. Note that the coordinates in RA and Dec correspond to the ALMA coordinates of the primary disc. The properties of the {triple} system without NIR data, as measured in the ALMA image, are presented in Table~\ref{t:bin_alma}.

{We found that some companions are sub-stellar, in the brown dwarf regime. To calculate the mass we used the 2MASS K band photometry, available for all the primaries, then we derived the L$^{\prime}$ magnitude using the color for the different spectral types as in \citet{2015A&A...574A..78S}. The spectral type information of each ODISEA target is presented in Ru\'iz-Rodr\'iguez et al. (in prep). Extinction in K band has also been taken into account. Individual extinction values are listed in \citet{2020AJ....159..282E} for 20 of the multiple systems. To convert the absolute magnitude into mass we used the AMES-COND evolutionary models by \citet{2001ApJ...556..357A}. The range of masses are 30-50 \MJup. One object, GY92 367 b (C5\_090b), is in the planetary regime (14 \MJup), and the primary's mass derived from the evolutionary models is only 30 \MJup. However, we note that the IR spectrum of the primary does not show the features expected in a brown dwarf (Ru\'iz-Rodr\'iguez et al. in prep), suggesting that the binary system might be background contamination, a rare but possible occurrence in the ODISEA sample \citep{2019ApJ...875L...9W}.  } 

The NIR images (colour stretch) and their corresponding ALMA 1.3 mm maps (white contours) are overlaid in Figs.~\ref{f:mos}, \ref{f:mos2}, \ref{f:mos3}, and \ref{f:mos4}. Eight systems are resolved in both the NIR and the millimeter, while five systems are unresolved in the ALMA images. In 11 systems, the companions are undetected in the millimeter at ODISEA sensitivities. 

To compare the disc dust masses of single objects to the ones in multiple systems we divided the sample into five categories: single objects, multiple discs where all components are detected and resolved by ALMA, multiple systems unresolved by ALMA, binaries where {one or both the components are not detected } in the mm, and binaries {without IR excess}. This last group contains binaries taken from \cite{2009ApJ...696L..84C} that do not present IR excesses from {\it Spitzer}.  Although these objects are not included in ODISEA, they are taken into account in the following statistical analysis.  In the case where each component is resolved, the total mass of the system is the sum of the mass of each component. {The five categories of the sample are presented in Table~\ref{t:cat}}.

\begin{table*}
\caption{{Summary of the five categories of the population of Ophiuchus observed in the NIR. }} 
\label{t:cat}
\centering
\begin{tabular}{lcl}
\hline
\hline
Category & N. of objects   & Description  \\
\hline
Single objects    &   252	& Single objects with no IR excess (60) and single objects in NIR-ODISEA (193)	\\
Resolved multiple discs (Res)  & 11	& Multiple systems where all the components are resolved in the mm 	\\
Unresolved multiple discs (UR)          & 3	&Binary systems where the components are unresolved in the mm 	\\
ND of component(s) (ND) & 25	&	Only the primary is detected in ALMA, or none of the components is detected	\\
No IR excess binaries                  & 26	&	Multiple systems excluded from the ALMA sample as they do not present IR excess	\\
\hline
\end{tabular}
\end{table*}

\begin{figure}
\begin{center}
\includegraphics[width=0.5\textwidth]{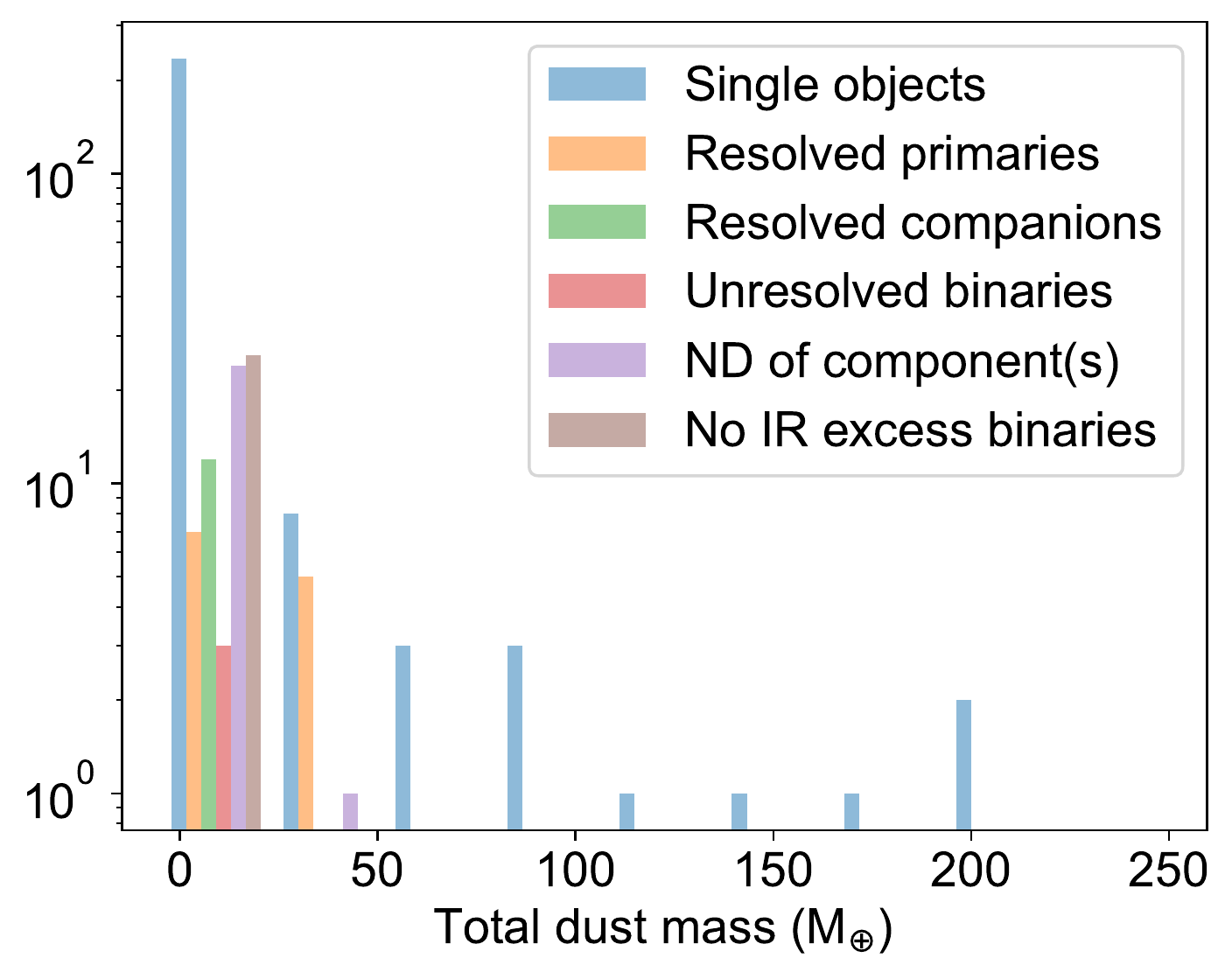}
\caption{Histogram of the total mass of the dust of the discs measured in the ALMA data.  }
\label{f:histo}
\end{center}
\end{figure}

\begin{figure}
\begin{center}
\includegraphics[width=0.5\textwidth]{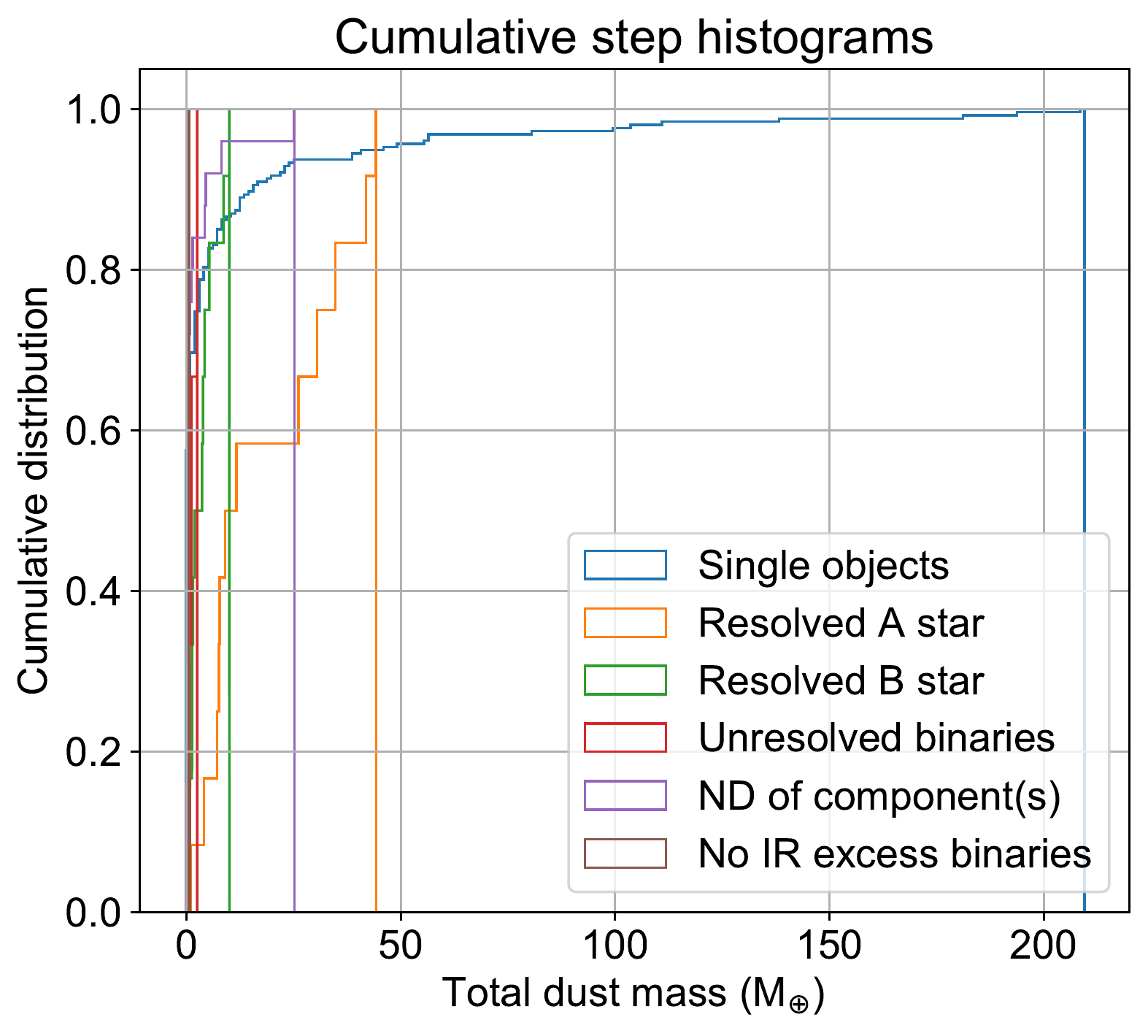}
\caption{Cumulative function of the total mass of the dust in the systems as measured in the ALMA data.}
\label{f:cum}
\end{center}
\end{figure}

\begin{figure*}
\begin{center}
\includegraphics[width=0.45\textwidth]{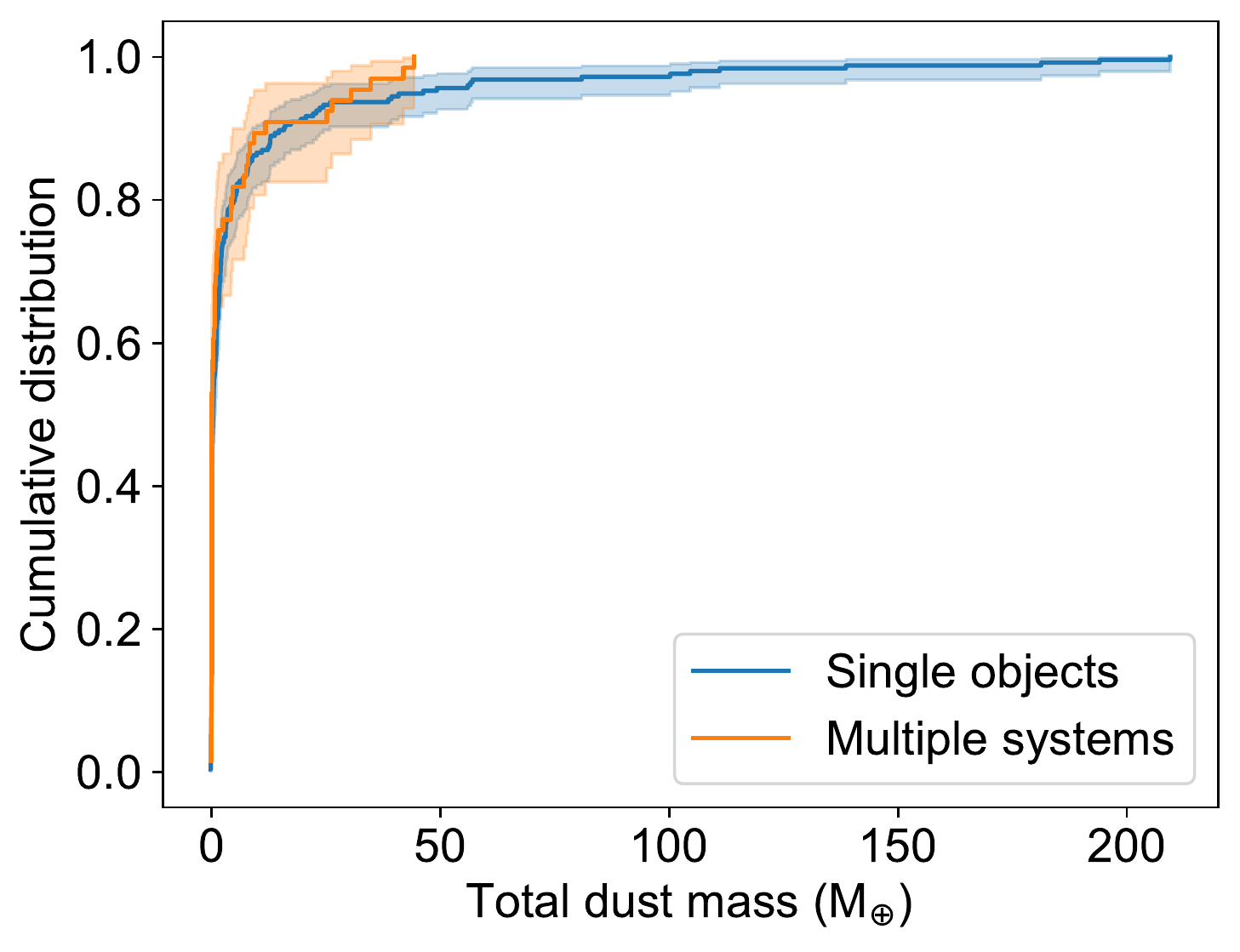}
\includegraphics[width=0.45\textwidth]{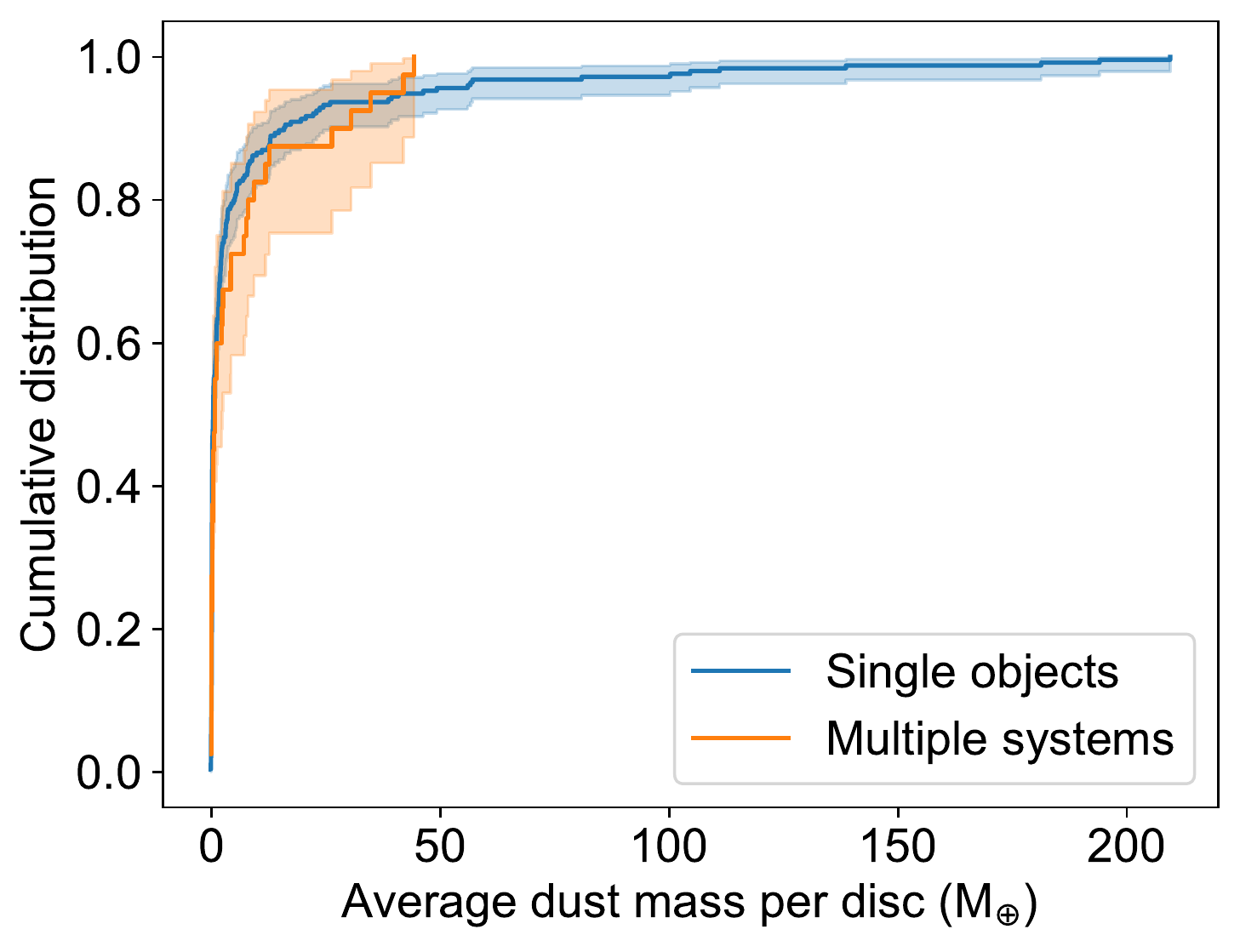}
\caption{ Cumulative function of the total mass of the dust ({\it left}) and of the average mass of the dust around each star ({\it right}) in the systems as measured in the ALMA data, single vs multiple systems are shown.}
\label{f:cum_single}
\end{center}
\end{figure*}

\begin{figure}
\begin{center}
\includegraphics[width=0.5\textwidth]{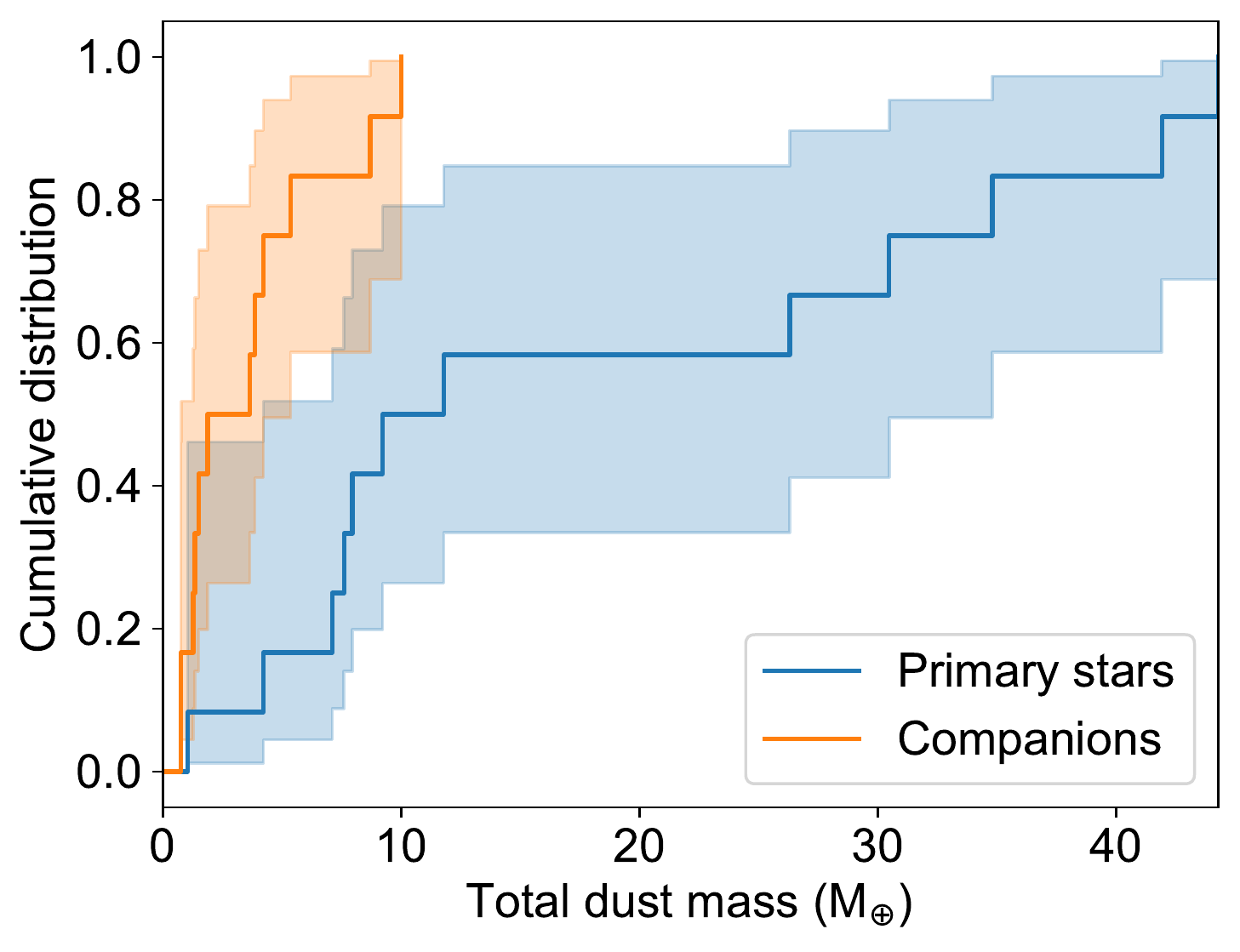}
\caption{Cumulative function of the total mass of the dust in the primaries vs the companions as measured in the ALMA data.}
\label{f:cum_a}
\end{center}
\end{figure}

\begin{figure}
\begin{center}
\includegraphics[width=0.5\textwidth]{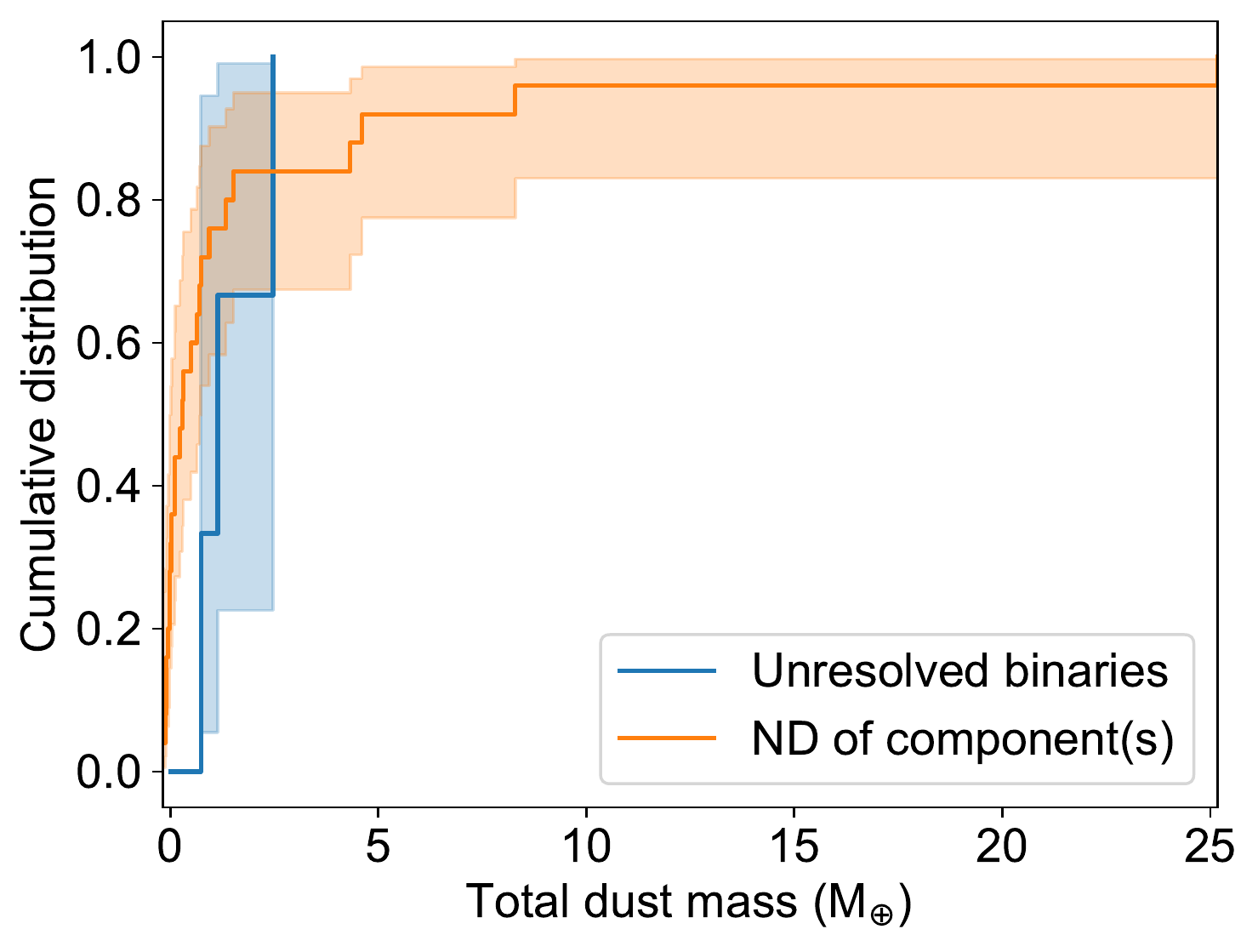}
\caption{Cumulative function of the total mass of the dust in the systems where the tight binaries are not resolved by ALMA vs the ones where the binaries are resolvable in our mm data, but only the primary are detected.}
\label{f:cum_unre}
\end{center}
\end{figure}

To compute the mass in M$_{\oplus}$\xspace we multiplied the millimeter flux in mJy \citep[as listed in][]{2019ApJ...875L...9W} by 0.58 \citep{1990AJ.....99..924B,2005ApJ...631.1134A,2019MNRAS.482..698C}. This conversion from observed mm flux to dust mass assumes a distance of 140 pc, a dust temperature of 20K and suffers from all the uncertainties and caveats discussed in detailed in the references listed above. The histogram of the dust mass of the systems of Ophiuchus is shown in Figure~\ref{f:histo}. The cumulative probability function for all the categories together is displayed in Fig.~\ref{f:cum}. For non-detections, we used the Kaplan-Meier (KM) product estimator. The python package {\sc lifelines} \citep{2019JOSS....4.1317D} was used to estimate the cumulative distribution functions for each category as shown in Figs.~\ref{f:cum_single}, \ref{f:cum_a}, \ref{f:cum_unre}.

Fig.~\ref{f:cum_single} shows the cumulative distribution of disc dust masses for single and binary stars, per system (left panel) and per star (right panel). In the latter case, the total disc dust mass are divided by two. The most massive discs (up to masses of $\sim$200 M$_{\oplus}$\xspace) are found around single stars. Multiple systems reach lower maximum dust masses, with a maximum total mass of $\sim$50 M$_{\oplus}$\xspace.

Note that the distributions of dust masses for single stars and multiple systems are indistinguishable from each other except above M$_{dust}$ $>$ 50 M$_{\oplus}$. The fact that $\lesssim$4$\%$ of singles stars have discs with dust masses $>$ 50 M$_{\oplus}$  suggests that the medium separation binaries (r $>$10 au) only affect a small fraction of the disc population. Fig.~\ref{f:cum_a} shows that, in general, discs around the primary star are more massive than the ones around the companion, {as found for Taurus \citep{1996AJ....111.2431J, 2012ApJ...751..115H,2019ApJ...872..158A}}. This is consistent with the results on the dependance of disc masses on stellar mass in nearby star-forming regions, according to which the dust mass in discs is roughly proportional to M$_{\star}^{1.5}$ \citep{2013ApJ...771..129A, 2016ApJ...831..125P}. { A dedicated publication, Ru\'iz-Rodr\'iguez et al. (in prep.), will include a spectroscopic study of all the objects of the ODISEA sample, with the spectral type derivation and mass. An in-depth discussion on the dependence of the mass will be presented in that publication.} 

\begin{figure}
\begin{center}
\includegraphics[width=0.5\textwidth]{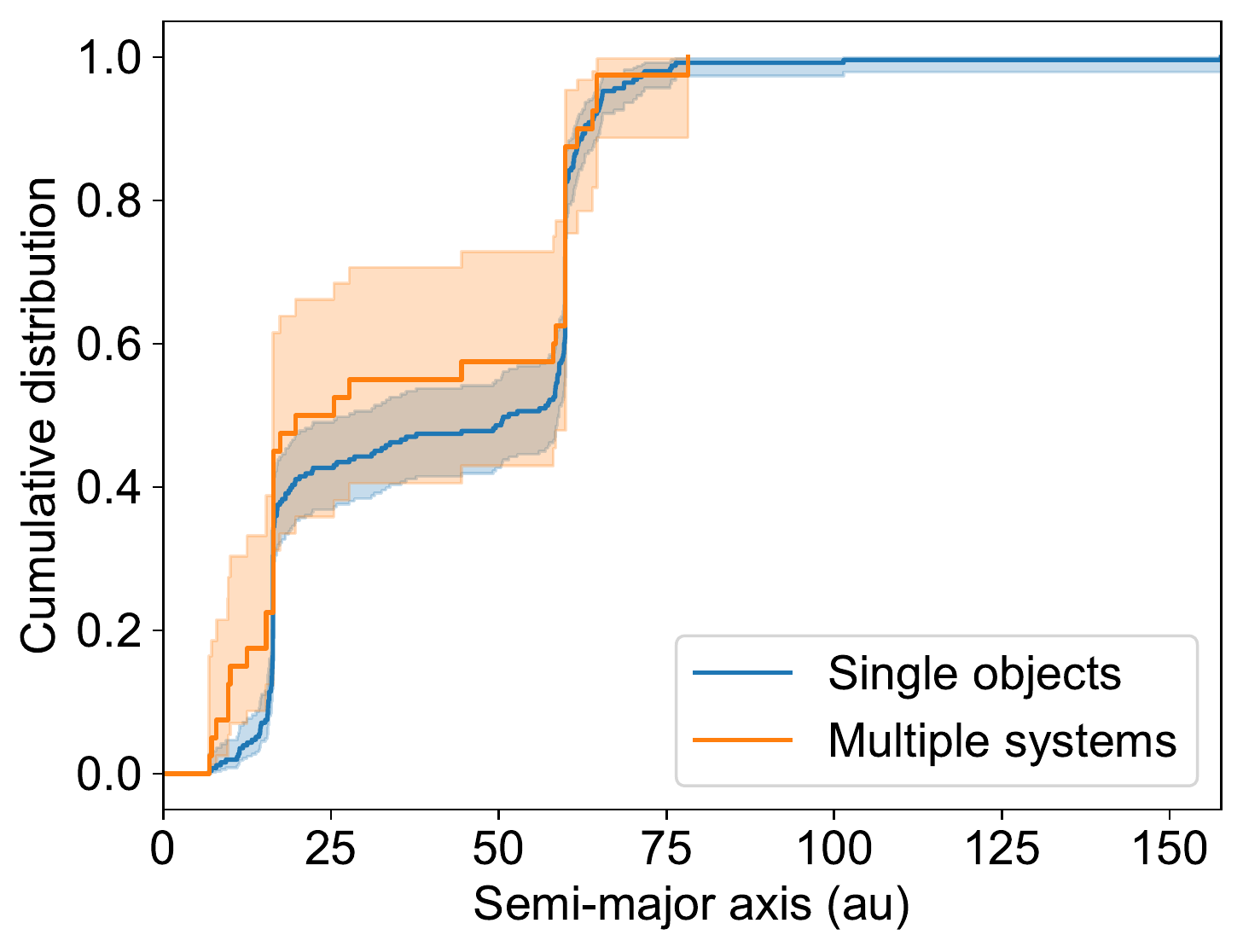}
\caption{Cumulative function of the projected semi-major axis of the discs as measured in the ALMA data.}
\label{f:cum_semi}
\end{center}
\end{figure}

Fig.~\ref{f:cum_unre} shows that tight binaries which are unresolved in the millimeter have total masses M$_{dust}$ $<$ 5 M$_{\oplus}$, while systems where only the primary is detected are in general not very massive. The fact that tight binaries have less massive discs is expected as they should have smaller truncation radii and shorter viscous dissipation timescales \citep{1977MNRAS.181..441P}. Similarly, we speculate that wider binaries where only the primary is detected in the mm are sligthly older or more evolved systems where the discs around the secondaries have already dissipated. In Fig.~\ref{f:cum_semi} the cumulative distribution of the semi-major axis of the discs is shown. As expected, discs in multiple systems reach smaller maximum sizes. However, as seen with the dust masses,  the difference between single stars and binaries is only seen in the tail of the distributions. Notice that the ``ladder shape'' is an effect of the ALMA resolution of the two samples. Also, note that a few sources have disc sizes smaller than the ALMA beam. That is because, for detections with very high signal to noise ratios, the ALMA beam can be deconvolved from the image, which allows us to measure the sizes of sources that are smaller than the beam. However, we emphasize that the deconvolved disc sizes must be interpreted with caution. 

\begin{figure}
\begin{center}
\includegraphics[width=0.5\textwidth]{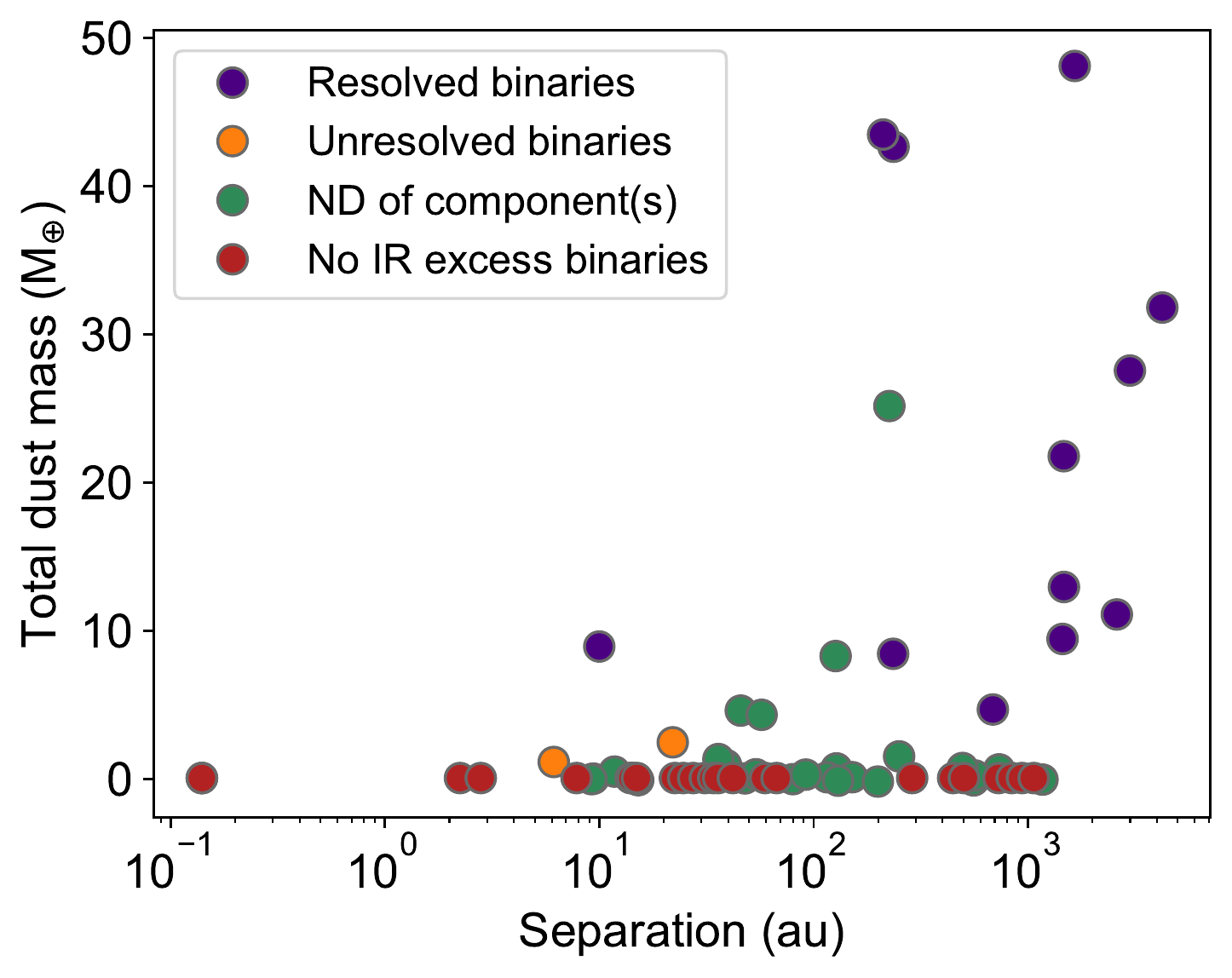}
\caption{Total mass of the system in dust versus projected separation in between each component of the multiple system. }
\label{f:mass_sep}
\end{center}
\end{figure}

\begin{figure}
\begin{center}
\includegraphics[width=0.5\textwidth]{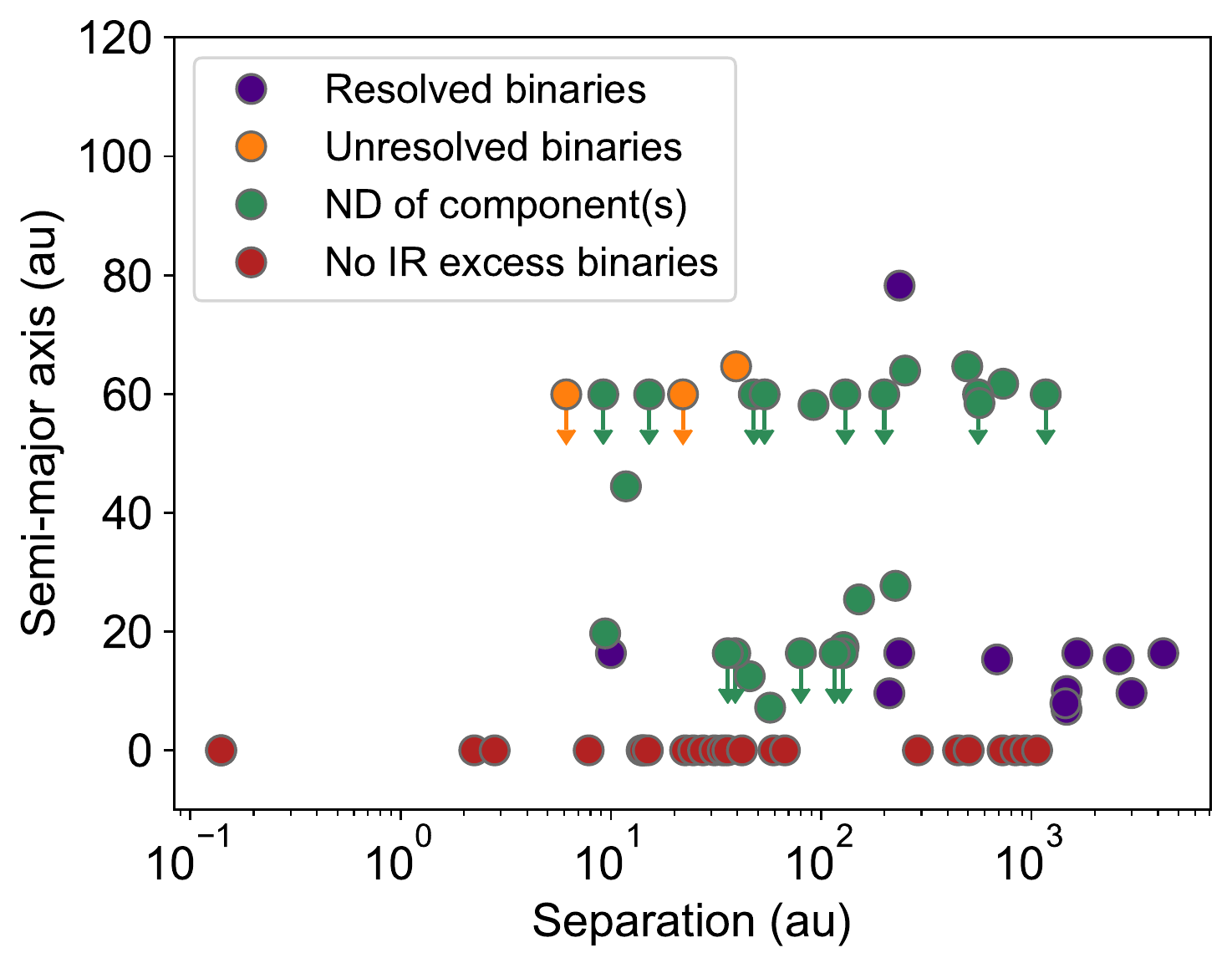}
\caption{Semi-major axis of the discs as measured in the ALMA data versus projected separation in between each component of the multiple system. Unresolved objects are indicated with arrows. }
\label{f:a_sep}
\end{center}
\end{figure}

Figures~\ref{f:mass_sep} and ~\ref{f:a_sep} show the correlation between the dust mass and the size of the disc as a function of the projected separation between the primary and the companion. Massive, big discs are only found around wide ($>$ 100 au) binary systems. Systems with less massive and unresolved discs are part of close binaries with projected separations between components below 150 au. 

{\cite{2012ApJ...751..115H} showed that massive discs (brighter than $>100$ mJy at mm wavelengths) in binaries are only found in Taurus where the stars are very tightly packed and the disc is circumbinary, or around the individual components of  wide separation ($>300$ au) binaries. In particular, they identified massive circumbinary discs around 4 systems: GG Tau Aab, MHO 2 AB, UZ Tau Eab and DQ Tau AB. With a separation of ~35 au, a system like GG Tau Aab, which has one of the most massive discs in all of Taurus \citep[see, e.g.,][]{2020A&A...639A..62K}, would be easily identifiable in the ODISEA sample. Therefore, an analogue system is unlikely to exist in Ophiuchus. However, we note that the other 3 systems with circumbinary discs mentioned above have stellar separations $< 0.1-7$ au and would remain undetectable in our survey.  Multi-epoch high-resolution spectroscopy or optical interferometry would be needed to identify very tight binaries and circumbinary discs. Unfortunately, few such observations are available for ODISEA targets, preventing meaningful comparison to the objects in Taurus.}

\cite{2017ApJ...851...83C} studied the effect of multiplicity on a subsample of ODISEA discs.  In particular, they restricted the {\it Spitzer} sample from the ``cores to disks'' survey \citep{2009ApJS..181..321E} to the 64 targets with 70 \mic detections in order to increase the expected detection rate at mm wavelengths.  This selection criteria, which increased the efficiency of their survey, necessarily introduces a bias towards the largest and most massive discs, where the effect of visual binaries is the strongest. \cite{2017ApJ...851...83C} conclude that discs in binaries are significantly smaller than those around single stars. While our results are very consistent with those conclusions at extremes of the mass and size distributions, we note that most discs in Ophiuchus are small and low-mass and the effect of visual binaries is likely to be much weaker in the general disc population.

\cite{2016AJ....152....8K} show that planets can form and survive even in very tight (projected separations of 2-3 au) binary systems. However, the occurrence rate measured by {\it Kepler} in tight binary systems is 34\% that of the wider binaries or single stars. Wider binaries have very similar occurrence rates to single stars, suggesting that when a companion is separated enough it does not affect the disc of the primary. Following \cite{2016AJ....152....8K}, a fifth of all solar-type stars in our Galaxy are most likely not hosting planets due to the presence of a close binary companion. These results on the occurrence of {\it Kepler} planets could be reconciled  with ours, noting that the planets detected by {\it Kepler} are all within 1 au from their hosts and that our NIR-AO observations are only sensitive to companions with projected separations larger than $\sim$10 au and have been restricted to stars with NIR excess.

The total number of multiple systems of ODISEA is {43}, so the occurrence rate is 18\% (43/236). Notice that this rate is biased by the fact that, in ODISEA, only Ophiuchus members with discs are included.  The objects from \citet{2009ApJ...696L..84C} that are part of Ophiuchus but are not included in the ODISEA sample because they do not display infrared excess are 86 in total, among which 26 are binaries. Therefore, this indicates that the ocurrence rate of visual binaries in the diskless stars in Ophiuchus is 30$\%$, higher than in the ODISEA sample. A similar result is found for Lupus, where the occurrence rate of multiple systems with discs is 12\% \citep{2021MNRAS.501.2305Z}.

The majority of young stars are part of multiple systems, with a frequency twice as high as among solar type main sequence stars \citep{2013ARA&A..51..269D}. In Taurus, for example, the frequency  of multiple systems is $\sim$70\% \citep{2011ApJ...731....8K}. However, the multiplicity census in Ophiuchus is very incomplete for separations smaller than 10 au and for diskless stars.  It is important to mention that \citet{2019A&A...626A..80C} identified $\sim$200 new members of Ophiuchus using GAIA data, for these objects the study of the multiplicity has not been done so far and a significant fraction of them could be multiple systems.  

While some important conclusions can already be derived from the currently available data, additional multiplicity surveys, extending to smaller separations and discless stars, are necessary to obtain a complete picture of the role that companions may have on the planet formation potential of protoplanetary discs.  In particular, muti-epoch radial velocity observations will be necessary to identify the companions that are likely to disrupt planet formation at small separations.

\section{Summary and conclusion}
\label{s:con}

As part of the ODISEA survey, we have obtained NIR AO imaging at 0\farcs08 resolution for a sample of 164 stars in the Ophiuchus molecular clouds. Combining our results with the ODISEA ALMA 1.3 mm data, archival NIR AO data and multiplicity information from the literature, we present the following results and conclusions:
\begin{enumerate}
\item We detect 20 new binary systems and one new triple system.

  \item {Nine companions are in the sub-stellar regime (30-50 \MJup). }

\item The (sub)stellar multiplicity of the ODISEA sample for companions with projected separations in $\sim$ 9 to 1200 au range and flux ratios in the 0.01 to 1 range is 18\%. Since all the ODISEA targets have IR excesses, multiple systems might be underrepresented with respect to the general population of young stars in Ophiuchus.

\item Discs around single stars and stars in multiple systems have similar dust mass cumulative distributions up to 50 M$_{\oplus}$. Disc masses higher than 50 M$_{\oplus}$ and up to 200 M$_{\oplus}$ are only found around single stars. 

\item Primary stars tend to have more massive discs than secondaries.

\item Discs around single stars can be more extended, with semi-major axes up to 150 au. Discs in multiple systems have a smaller maximum size. 

\item Stellar companions with modest projected separations (10-100 au) are likely to affect the formation of massive planets at large radii (> 5-50 au) but still allow the formation of terrestrial planets at small projected separation, like those detected by {\it Kepler}.

\item Our results are consistent with previous claims that discs in visual binaries are significantly smaller and lower mass than  their counterparts around single stars. However, we note that those conclusions only apply to the extremes of the mass and size distributions. Since most discs in Ophiuchus are small and low-mass, the effect of visual binaries seems to be much weaker in the general disc population.
\end{enumerate}

\section*{Acknowledgements}
We wish to thank the anonymous referee for their constructive report. A.Z. acknowledges support from the FONDECYT {\it Iniciaci\'on en investigaci\'on} project number 11190837. L.C. acknowledges support from FONDECYT Regular number 1171246. S.P. acknowledges support from FONDECYT grant number 1191934 and the Joint Committee of ESO and the Government of Chile. This work has been carried out within the framework of the NCCR PlanetS supported by the Swiss National Science Foundation. G.G. acknowledges the financial support of the SNSF. This paper makes use of the following ALMA data: ADS/JAO.ALMA number 2016.1.00545.S. ALMA is a partnership of ESO (representing its member states), NSF (USA) and NINS (Japan), together with NRC (Canada), MOST and ASIAA (Taiwan), and KASI (Republic of Korea), in cooperation with the Republic of Chile. The Joint ALMA Observatory is operated by ESO, AUI/NRAO and NAOJ.  \\

\section*{Data availability}
The data underlying this article are available in the article and in its online supplementary material.




\bibliographystyle{mnras}
\bibliography{bin_oph} 



\appendix

\section{ODISEA objects not observable}

\begin{table}
\caption{{List of the ODISEA objects not observable in the NIR with NACO or NIRC2, then excluded in the statistical analysis.}} 
\label{t:non_o}
\centering
\begin{tabular}{lcc}
\hline
\hline
ODISEA ID & RA & DEC  \\
\hline
ODISEA\_C4\_003 & 16:21:45.13 & -23:42:31.63 \\
ODISEA\_C4\_042 & 16:26:25.46 & -24:23:01.31 \\
ODISEA\_C4\_056 & 16:26:51.95 & -24:30:39.45 \\
ODISEA\_C4\_057 & 16:26:53.47 & -24:32:36.12 \\
ODISEA\_C4\_058 & 16:26:54.29 & -24:24:37.89 \\
ODISEA\_C4\_059 & 16:26:54.76 & -24:27:02.14 \\
ODISEA\_C4\_061 & 16:26:58.27 & -24:37:40.75 \\
ODISEA\_C4\_063 & 16:27:02.99 & -24:26:14.61 \\
ODISEA\_C4\_067 & 16:27:05.24 & -24:36:29.59 \\
ODISEA\_C4\_080 & 16:27:15.87 & -24:25:13.93 \\
ODISEA\_C4\_081 & 16:27:16.39 & -24:31:14.46 \\
ODISEA\_C4\_091 & 16:27:26.27 & -24:42:46.09 \\
ODISEA\_C4\_097 & 16:27:32.12 & -24:29:43.46 \\
ODISEA\_C4\_111 & 16:27:48.23 & -24:42:25.43 \\
ODISEA\_C4\_119 & 16:28:57.85 & -24:40:54.88 \\
ODISEA\_C4\_135 & 16:31:52.45 & -24:55:36.18 \\
ODISEA\_C4\_144 & 16:39:52.91 & -24:19:31.36 \\
ODISEA\_C5\_043 & 16:26:18.57 & -24:29:51.32 \\
ODISEA\_C5\_051 & 16:26:23.81 & -24:18:28.96 \\
ODISEA\_C5\_056 & 16:26:37.79 & -24:39:03.07 \\
ODISEA\_C5\_057 & 16:26:40.83 & -24:30:50.83 \\
ODISEA\_C5\_059 & 16:26:56.35 & -24:41:20.31 \\
ODISEA\_C5\_060 & 16:26:57.32 & -24:35:38.65 \\
ODISEA\_C5\_063 & 16:26:58.65 & -24:24:55.37 \\
ODISEA\_C5\_073 & 16:27:26.21 & -24:19:22.97 \\
ODISEA\_C5\_077 & 16:27:32.71 & -24:45:00.28 \\
ODISEA\_C5\_094 & 16:27:58.89 & -24:35:14.57 \\
ODISEA\_C5\_096 & 16:28:04.53 & -24:34:48.44 \\
ODISEA\_C5\_126 & 16:31:34.08 & -24:00:59.66 \\
\hline
\end{tabular}
\end{table}

\bsp	
\label{lastpage}
\end{document}